\documentclass[aps,prx,twocolumn,showpacs,superscriptaddress, longbibliography]{revtex4-1}
\usepackage{amsmath, amssymb, braket, dsfont, times, units,verbatim}
\usepackage[mathscr]{euscript}
\usepackage{graphicx}
\usepackage{subfigure}     
\usepackage{mathrsfs}
\usepackage{amsthm}
\usepackage{amsfonts}
\usepackage{amssymb}
\usepackage{yfonts}

\usepackage[breaklinks,colorlinks,citecolor=magenta]{hyperref}			

\usepackage[usenames,dvipsnames]{color}         
\definecolor{Zcolour}{rgb}{0.992, 0.588, 0.22}
\definecolor{dkgreen}{rgb}{0,0.5,0}
\definecolor{purple}{rgb}{0.5,0,0.5}

\newcommand{\cohosub}[1]{\scalebox{0.72}{\textswab{#1}}}
\newcommand{\cohosubsub}[1]{\scalebox{0.6}{\textswab{#1}}}
\newcommand{\coho}[1]{\textswab{#1}}

\renewcommand{\ol}[1]{\overline{#1}}
\newcommand{\mb}[1]{\mathbf{#1}}
\newcommand{\Ref}[1]{Ref.~\onlinecite{#1}}

\newcommand{\SG}{\mathcal{G}}
\newcommand{\U}{\textrm{U}}
\newcommand{\slant}{\iota}
\newcommand{\coh}[2]{  \mathcal{H}^{#1}\left[ #2, \U (1) \right]}
\renewcommand{\vr}{\mathbf{r}}

\begin{document}
\title{Translational symmetry and microscopic constraints on symmetry-enriched topological phases: a view from the surface}
\author{Meng Cheng}
\affiliation{Station Q, Microsoft Research, Santa Barbara, California 93106-6105, USA}
\author{Michael Zaletel}
\affiliation{Station Q, Microsoft Research, Santa Barbara, California 93106-6105, USA}
\author{Maissam Barkeshli}
\affiliation{Station Q, Microsoft Research, Santa Barbara, California 93106-6105, USA}
\author{Ashvin Vishwanath}
\affiliation{Department of Physics, University of California, Berkeley, California 94720, USA}
\author{Parsa Bonderson}
\affiliation{Station Q, Microsoft Research, Santa Barbara, California 93106-6105, USA}

\begin{abstract}
The Lieb-Schultz-Mattis theorem and its higher dimensional generalizations by Oshikawa and Hastings require that translationally invariant 2D spin systems with a half-integer spin per unit cell must either have a continuum of low energy excitations, spontaneously break some symmetries, or exhibit topological order with anyonic excitations. We establish a connection between these constraints and a remarkably similar set of constraints at the surface of a 3D interacting topological insulator.
This, combined with recent work on symmetry-enriched topological phases (SETs) with on-site unitary symmetries, enables us to develop a framework for understanding the structure of SETs with both translational and on-site unitary symmetries, including the effective theory of symmetry defects. This framework places stringent constraints on the possible types of symmetry fractionalization that can occur in 2D systems whose unit cell contains fractional spin, fractional charge, or a projective representation of the symmetry group. As a concrete application, we determine when a topological phase must possess a``spinon'' excitation, even in cases when spin rotational invariance is broken down to a discrete subgroup by the crystal structure. We also describe the phenomena of ``anyonic spin-orbit coupling,'' which may arise from the interplay of translational and on-site symmetries. These include the possibility of on-site symmetry defect branch lines carrying topological charge per unit length and lattice dislocations inducing on-site symmetry protected degeneracies.
\end{abstract}

\maketitle

\tableofcontents

\section{Introduction}

The celebrated Lieb-Schultz-Mattis (LSM) theorem~\cite{LSM,AffleckLieb}, including the higher-dimensional generalizations by Oshikawa~\cite{OshikawaLSM} and
Hastings~\cite{HastingsLSM}, shows that translationally invariant spin systems with an odd number of $S=1/2$ moments per unit cell
cannot have a symmetric, gapped, and non-degenerate ground state. Either some symmetries of the system are broken spontaneously,
the phase is gapless, or there is nontrivial topological order with emergent quasiparticle excitations having exotic exchange
statistics. The LSM theorem is remarkable because a microscopic property, i.e. the spin within a unit cell, constrains the universal
long-wavelength physics in a non-trivial way. Moreover, it generates powerful practical implications. For example, experimentally ruling out symmetry breaking can be sufficient to imply the presence of exotic spin-liquid physics~\cite{lee2008,balents2010}.

Spin-liquid folklore informs us that when a spin rotationally invariant system with an odd number of $S = 1/2$ moments per unit cell forms a symmetric gapped spin liquid, it must also possess deconfined ``spinon'' excitations, which are topologically nontrivial quasiparticles that carry half integer-valued spin~\cite{lee2008,balents2010}. In contrast, the local excitations, such as local spin flips, of such systems always carry integer-valued spin. One aim of our work is to put this understanding on more solid formal footing and, thereby, derive the most general relation between the symmetry properties of the unit cell and the fractional quantum numbers of the emergent excitations.

Intriguingly similar constraints as those required by the LSM theorem have recently been discovered
to arise at the surface of symmetry protected  topological (SPT) phases \cite{chen2013,VishwanathPRX2013,wang2013,burnell2013,Fidkowski13,wang2013b,Bonderson13d,wang2014,chen2014b,metlitski2013,cho2014,chen2014,kapustin2014,metlitski2014}.
Such phases are short-range entangled states that cannot be adiabatically connected to a trivial product state while preserving the symmetries of the system. An eminent example is the 3D topological insulator (TI) protected by time-reversal and charge conservation symmetries, which (at least for weak interactions) has a gapless surface with a single Dirac cone~\cite{Moore_PRB07,Fu_PRL07,hasan2010}. In general, the surface phases at
the boundary of an SPT must also be either symmetry broken, gapless, or gapped with nontrivial
``surface topological order,'' mirroring the three options allowed by the LSM theorem.

In the case where the 3D SPT phase possesses surface topological order and preserves the symmetries, the 2D surface states are anomalous
symmetry-enriched topological (SET) phases. Such surface states are anomalous in the sense that they cannot be consistently realized in purely
2D systems when the symmetry is realized in a local manner. Such anomalies are signaled by an object called the ``obstruction class'' having nontrivial values (in the a corresponding cohomology group)~\cite{VishwanathPRX2013, chen2014}.  The anomalous surface SET can provide a non-perturbative characterization of the bulk SPT order, as the symmetry action in the SET must have exactly the right type of ``gauge anomaly'' required by the
nontrivial SPT bulk.

\begin{figure}[t!]
\includegraphics[width=\columnwidth]{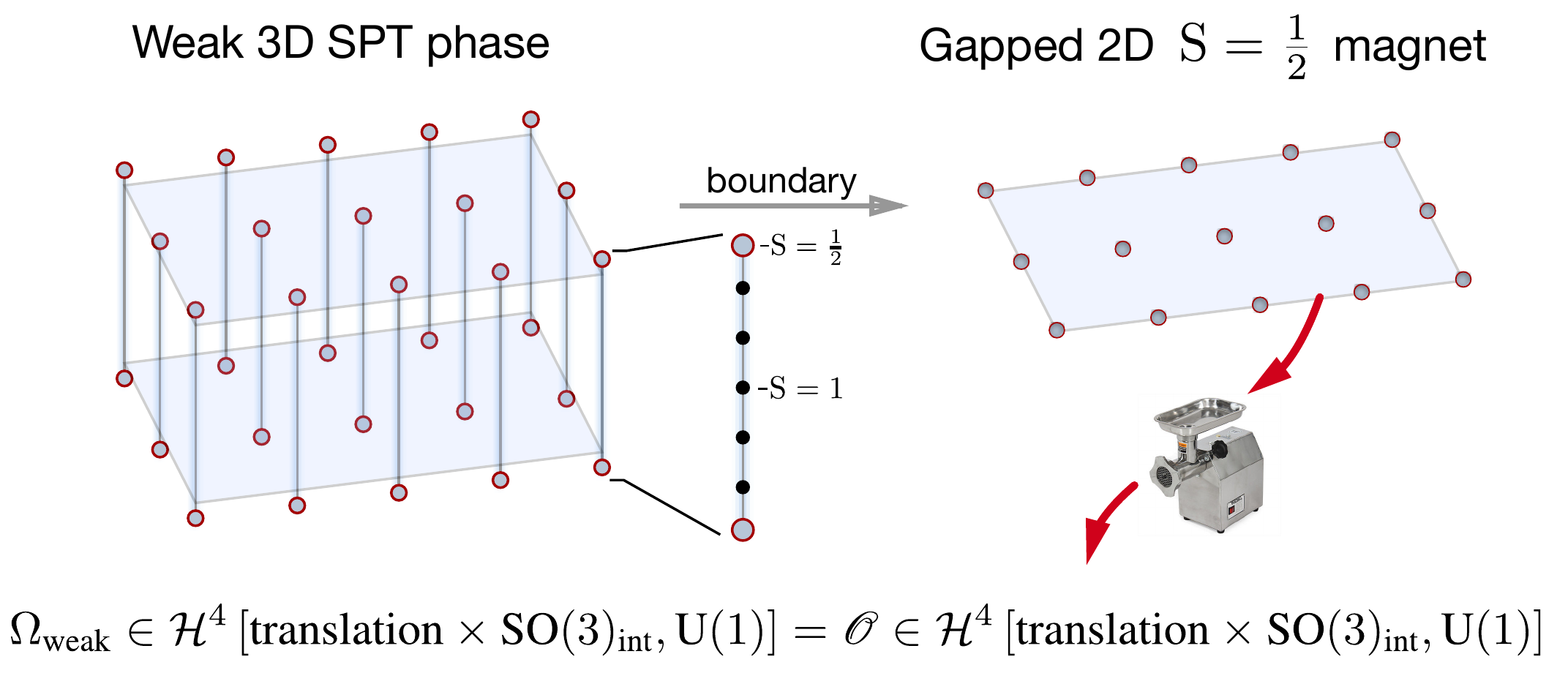}
\caption{A 3D system of integer spins can enter a weak SPT phase equivalent to stacking together 1D AKLT chains (as indicated on the left). The weak SPT phase is characterized by a particular element of the cohomology class $\Omega_{\textrm{weak}} \in \coh{4}{ \mathbb{Z}^3_{\text{trans}} \times \textrm{SO}(3)_{\text{int}}}$. Since each AKLT chain has degenerate emergent $S=1/2$ edge states, the effective Hilbert space of the 2D surface of the 3D SPT phase behaves like a $S=1/2$ magnet; it will have a projective (half integral) representation of SO(3) in each unit cell. Thus, studying surface phases of the 3D weak SPT is equivalent to studying a 2D $S=1/2$ magnet. Given a proposal for the braiding, statistics, and symmetry fractionalization of the 2D magnet, there is a procedure to calculate an obstruction class $\mathscr{O} \in \coh{4}{\mathbb{Z}^2_{\text{trans}} \times \textrm{SO}(3)_{\text{int}}}$, associated with defining an effective theory of the symmetry defects. The bulk-boundary correspondence requires $\mathscr{O}$ of the boundary and $\Omega_{\textrm{weak}}$ of the bulk to be compatible, constraining the allowed 2D SET orders.
}
\label{fig:correspondence}
\end{figure}

In this paper, we develop a new perspective on the LSM theorem by exploiting a relation to the surface of SPT states. A crucial role is played by ``weak'' SPT phases, which are SPT phases protected by translational symmetry~\cite{Fu_PRL07, chen2013}. We argue
that the LSM constraints are, in a precise sense, a special case of the constraints at the surface of weak SPTs.
For example, a 2D spin system with $S=1/2$ per unit cell can be thought of as the boundary of a 3D AKLT model on a
cubic lattice with $S = 3$ per unit cell \cite{AKLT87}, which is an example of a 3D weak SPT.

More specifically, we posit a bulk-boundary ``anomaly matching'' condition when a 2D SET phase is the gapped and symmetric surface state realized on the boundary of a 3D SPT phase. This bulk-boundary correspondence requires that the components of the 2D SET obstruction class match the corresponding bulk 3D SPT invariants. This anomaly matching allows us to precisely formulate a restriction on what kind of SET order can exist at the surface of such a given 3D SPT. A consequence of this bulk-boundary correspondence is that the physical properties associated with the 3D SPT invariants, namely the nature of the emergent boundary modes, imply which SET obstruction classes can be physically realized in purely 2D systems in a local and symmetry preserving manner. In particular, the SET obstruction classes permitted in purely 2D systems correspond to SPT phases whose boundary structure is, at most, a nontrivial projective representation of the on-site symmetry per unit cell. This provides an understanding of the effective theory of symmetry defects in purely 2D SET phases with translational and on-site unitary symmetries, and it requires the SET obstruction class to match the projective representation characterizing the transformation of each unit cell under the on-site symmetry. As such, we obtain constraints on the type of SET order that can be realized in purely 2D systems from the manner in which the microscopic degrees of freedom transform under the symmetry, in particular, from the projective representation per unit cell. These arguments are summarized in Fig.~\ref{fig:correspondence}.

This point of view immediately provides a simple physical ``derivation'' of the LSM theorem.
Namely, the bulk-boundary correspondence tells us that if the boundary of a nontrivial 3D SPT phase is gapped and preserves
the symmetry, then it must also realize nontrivial topological order. This further implies that a purely 2D gapped and symmetric phase must have nontrivial topological order when the microscopic degrees of freedom transform as a nontrivial projective representation of the on-site symmetry for each unit cell.
More significantly, our arguments lead to a far more general and constraining version of the LSM theorem. It applies to arbitrary on-site unitary symmetries, including discrete symmetries, which cannot be probed by the adiabatic flux insertion arguments utilized by Oshikawa and Hastings. Moreover, it also provides sharper restrictions on the \emph{type} of 2D SET order allowed.

As a concrete application, we address the following question: under what general conditions must an SET phase possess
a ``spinon" excitation? We define a spinon to be an excitation which either a) transforms with the same projective representation
as the unit cell, or b) carries the same fractional U(1) charge as the particle filling $\nu = \frac{p}{q}$.
For example, the spinon of a spin-liquid in an $S=1/2$ magnet carries $S=1/2$, while the Laughlin-type quasiparticles (generated by threading a unit quanta of flux)
of the $\nu = \frac{p}{q}$ fractional quantum Hall effect carry the fractional charge $e^\ast = \frac{p}{q} e$.
This question was partially addressed in Ref.~\onlinecite{ZaletelPRL2015}, which left the question open for certain
exotic fractionalization patterns of translation symmetry. We find that a spinon is required if either: (1) the on-site
symmetry group is continuous and connected, or (2)  the anyons are not permuted by the symmetry and there is a point-group symmetry
that can relate the two directions of the lattice, such as $C_3$, $C_4$, or $C_6$ lattice rotations, or certain reflections.
We also examine counterexamples in which either of the conditions is violated.

The work is organized as follows. In Sec.~\ref{sec:overview}, we outline the logical structure of the argument.
In Sec.~\ref{sec:trans_SPTs}, we review the classification of 3D SPT phases and describe the K\"unneth formula decomposition of the SPT class in terms of invariants associated with stacking, packing, and filling of lower dimensional SPT phases.
In Sec.~\ref{sec:symmetry_fractionalization}, we review the theory and classification of symmetry fractionalization in symmetric 2D topological phases, explaining the notion of anyons carrying localized fractional charge and projective representations. We discuss the inclusion of translational symmetry and the decomposition of the symmetry fractionalization classes in terms of anyonic flux per unit cell, anyonic spin orbit coupling, and on-site symmetry fractionalization.
In Sec.~\ref{sec:Symmetry_defects}, we review the algebraic theory of symmetry defects for 2D SET phases with on-site symmetry and their possible obstructions. We discuss the case of topological phases with symmetries that do not permute anyon types in greater detail.
In Sec.~\ref{sec:trans_obstruction}, we discuss the bulk-boundary correspondence and anomaly matching in detail and formulate an understanding of the theory of defects and their obstruction classes for 2D SET phases with translational and on-site unitary symmetries. We consider such 2D SETs in systems with projective representations per unit cell and fractional charge per unit cell in detail. We also describe in detail the phenomena of "anyonic spin-orbit coupling," and a number of resulting physical consequences.
In Sec.~\ref{sec:discussion}, we conclude with a discussion of further directions and open questions.

\section{An overview of the argument}
\label{sec:overview}

Our starting point is a translationally invariant lattice Hamiltonian in two spatial dimensions, with an internal (on-site) symmetry group $G$ and translational symmetry group $\mathbb{Z}^2$. We denote the total symmetry group as $\mathcal{G}=\mathbb{Z}^2 \times G$.
It is not only important to know the symmetry group $G$, but also the particular way in which the microscopic degrees of freedom transform under $G$. For example, a fully spin-rotationally invariant magnet has $G = \text{SO}(3)$, but it may be composed of integer or half-integer moments per unit cell. Likewise, a number conserving system with corresponding symmetry group $G = \U(1)$ has a well-defined charge density.
These microscopic details can constrain the emergent long wavelength physics in important ways.
There are two scenarios (exemplified by these two examples) in which the microscopic degrees of freedom in a unit cell transform ``fractionally'' under the symmetries: 1) when each unit cell transforms as a projective representation of the on-site symmetry, and 2) when the charge per unit cell of a $\mathrm{U}(1)$ symmetry is fractional.
We analyze the consequences of having projective representations per unit cell and fractional filling separately.

Given a symmetry $G$, the microscopic degrees of freedom within a unit cell transform under some representation $R$ of $G$.
$R$ can be either a linear or projective representation of $G$.
For example, spin $S = 1$ corresponds to a linear representation of $\mathrm{SO}(3)$, while $S = 1/2$ corresponds to a projective representation of $\mathrm{SO}(3)$ (a $2\pi$ rotation results in $-\openone$). In general, the possible projective representations of a group $G$ are classified by elements of the cohomology class $\coh{2}{G}$, where linear representations correspond to the trivial element $[1]$. For example, half-integer spins correspond to the nontrivial element in $\coh{2}{\mathrm{SO}(3)} = \mathbb{Z}_{2}$.

For a system with $\text{U}(1)$ charge conservation (such as particle number or magnetization) the average charge $\nu = \frac{p}{q}$ per unit cell defines the filling fraction. In some cases, the representation $R$ of $G$ is enough to specify the fractional part of the filling with respect to a $\mathrm{U}(1)$ subgroup of $G$. For example, in an $\text{SO}(3)$ invariant $S=1/2$ system, the zero magnetization corresponds to $\nu =1/2$ filling of any $\text{U}(1)$ subgroup of $\text{SO}(3)$. In other cases, $\nu$ may be the only relevant quantity, such as when $G = \U(1)$ (because $\coh{2}{\U (1)}$ is trivial).

In order to develop the outline of the argument, we briefly review SETs.
Since anyons are topological objects, they may transform in a fractional manner under $G$, where ``fractional'' means as compared with the quantum numbers carried by local operators. Note that even in an $S=1/2$ magnet, local \emph{operators} (and hence excitations) carry integral representations of $\mathrm{SO}(3)$. For example, in a quantum spin liquid, all local operators correspond to spin flips and carry integer valued spin, whereas the
topologically nontrivial spinon excitations carry half-integer spin.
In fractional quantum Hall states, all local operators are composed of electron operators and therefore carry
integer valued electric charge, while the quasiparticle excitations can carry fractional electric charge.

Thus, in addition to the fusion and braiding statistics specified by the topological order, SET phases are characterized by a pattern of ``symmetry fractionalization.'' We will review this structure later, but for now the reader can take for granted the existence of mathematical objects $\rho$ and $[\coho{w}]$ describing this pattern. Careful examination~\cite{chen2014, SET} reveals that not all choices of $\rho$ and $[\coho{w}]$ allow for a well-defined SET phase in 2D, as certain choices may be anomalous.
Mathematically, the anomaly is encoded in an obstruction class $[\mathscr{O}]$, which is an element of the cohomology class $\coh{4}{G}$.
Given the objects $\rho$ and $[\coho{w}]$ for a proposed symmetry fractionalization pattern, there is a well-defined procedure to obtain
$[\mathscr{O}]$ from the properties of the topological phase~\cite{ENO2009,chen2014,SET}, schematically represented by a grinder in Fig.~\ref{fig:correspondence}. Note that if there are no anyonic excitations, the fractionalization pattern is necessarily trivial  (i.e. $\rho = \openone$ and $[\coho{w}] =[I]$), and whenever the fractionalization pattern is trivial, so is the obstruction class $[\mathscr{O}] =[1]$.

Since 3D SPT phases are also classified by the elements $[\Omega] \in \coh{4}{G}$, this naturally suggests that a fractionalization pattern given by $\rho$ and $[\coho{w}]$ with a nontrivial obstruction $[\mathscr{O}]$ can only occur at the surface of a 3D SPT state with $[\Omega] = [\mathscr{O}]$.
In particular, for a nontrivial SPT phase with $[\Omega] \neq [1]$, we can only find a matching nontrivial obstruction class $[\mathscr{O}] \neq [1]$ if the SET fractionalization class is also nontrivial $[\coho{w}] \neq [I]$, which (as previously mentioned) requires there to be nontrivial anyonic excitations.
Thus, a gapped, symmetric surface of a nontrivial SPT must be topologically ordered.

Thus far in this overview section, the theory of SET phases and their obstructions that we have described assumes that the microscopic symmetry $G$ is an internal (on-site) symmetry. We need to generalize this theory to include translational symmetries. For symmetry fractionalization, the generalization is straightforward and still encoded by $\rho$ and $[\coho{w}]$. The full extension of the SET theory, which includes symmetry defects and their obstructions, is less clear. For this, we use the bulk-boundary correspondence between 3D SPT phases and 2D SET phases that exist on their boundaries. Applying the obstruction theory of 2D SET phases for systems with translational symmetry in the same way as for on-site symmetry, we find that we must have a rather different interpretation of the resulting obstruction class. For this we must first review 3D SPT phases with translational symmetry.

An SPT phase protected by translational symmetry is called a ``weak'' SPT phase.
The canonical example is a weak 3D TI, which arises by vertically stacking 2D TIs. More generally, a $d$-dimensional weak SPT phase protected by the on-site symmetry group $G$ and the $\mathbb{Z}^{d-k}$ translational symmetry group can be obtained by ``stacking'' copies of a $k$-dimensional SPT phase in a $(d-k)$-dimensional lattice, where $0\leq k \leq d$.
As we will show, weak SPT phases \emph{can} be classified by incorporating translational symmetry in the same manner as internal symmetries, that is, by $\coh{d+1}{\mathbb{Z}^d \times G}$.

The most relevant SPT phases for our discussion are 3D weak SPT phases that consist of packing together an array of 1D SPT phases, as illustrated in Fig.~\ref{fig:correspondence}.
At the end of each 1D chain in an SPT phase, there is a degenerate boundary state which transforms as a projective representation under $G$, characterized by $[\nu_{yx}]$.
On a surface perpendicular to the direction of the packed 1D chains of SPT phases, we have precisely a 2D system with the projective representation $[\nu_{yx}]$ per unit cell.
The interpretation is a bit different than the on-site case, since such a surface \emph{can} be physically realized in a strictly 2D system. In particular, one can simply construct the system from microscopic degrees of freedom in such a way that each unit cell transforms as a projective representation $[\nu_{yx}]$.
However, as the obstruction matching condition is still expected to hold, this becomes a powerful constraint on the possible SET orders that can occur in such a 2D system.
To be more precise, we will later show that given $\rho$ and $[\coho{w}]$ for the enlarged symmetry group $\mathcal{G}=\mathbb{Z}^2 \times G$, we can determine the
obstruction class $[\mathscr{O}]$, and, by matching $[\mathscr{O}]$ with the weak 3D SPT phase corresponding to packing together type $[\nu_{yx}]$ 1D chains of SPT phases, we derive stringent constraints on the 2D SET phase. For instance, we can rule out the double-semion model in a time-reversal invariant $S=1/2$ magnet.

Unfortunately, the above argument does not apply to $G = \U(1)$, which does not have projective representations.
Nevertheless, the constraints are rather simple (and well-known) for this case, and we can treat them by formalizing flux-threading type arguments within the SET formalism.

Before presenting our results, we introduce the relevant structures on  either side of the bulk-boundary mapping: 3D SPT phases and 2D SET phases.
While much of this material is discussed elsewhere~\cite{fidkowski2011, turner2011, chen2013, essin2013,SET}, our treatment of translational invariance is novel.

\section{Translationally invariant SPT Phases}
\label{sec:trans_SPTs}

\subsection{Strong SPT Phases}
\label{sec:StrongSPT}	

We first briefly review bosonic SPT phases protected by unitary on-site symmetries (like a spin rotation).
It was proposed that, in the absence of topological order, gapped $G$-symmetric phases in $d$ spatial dimensions are classified by an element of the group cohomology $\coh{d+1}{G}$.~\cite{fidkowski2011, turner2011, chen2013}
By this, we mean that each distinct phase of this type in the parameter space of generic $G$-symmetric Hamiltonians can be uniquely labeled by some element of the Abelian group $\coh{d+1}{G}$. The identity element of this group is identified with the trivial phase, which can be thought of as the universality class of an unentangled product state.

The physics of this classification is most transparent for 1D chains.
While an SPT phase has a unique, gapped ground state on a ring, an open segment may have degenerate states associated with its endpoints.
The degeneracy of these boundary states is protected by their local transformation properties under $\mathbf{g} \in G$.
The action of a global symmetry operation $R_{\bf g}$ on a ground state of the open chain can be localized to the boundaries.
For any ground state $\ket{\Psi}$, this takes the form
\begin{align}
R_{\bf g}\ket{\Psi} = U^{(1)}_{\bf g} U^{(2)}_{\bf g} \ket{\Psi},
\label{eq:1d_sym_loc}
\end{align}
where $U^{(j)}_{\bf g}$ are unitary operators whose nontrivial action is localized near the two ends of the chain, labeled $j=1$ and $2$.
The operators $U^{(j)}_{\bf g}$ do not depend on which of the degenerate ground states they are acting upon.

It is clear from Eq.~\eqref{eq:1d_sym_loc} that, even if $R_{\bf g}$ is a linear representation of $G$ (obeying $R_{\bf g}R_{\bf h}=R_{\bf gh}$), the localized symmetry actions $U^{(j)}_{\bf g}$ may take the form of projective representations of $G$. In particular, they only must respect group multiplication up to $\mathrm{U}(1)$ phases, i.e.
\begin{align}
U^{(j)}_{\bf g} U^{(j)}_{\bf h} = \omega_{j}({\bf g}, {\bf h}) U^{(j)}_{\bf gh}
,
\end{align}
as long as the phases $\omega_{j}({\bf g}, {\bf h}) \in \U (1)$ obey the constraint $\omega_{1}({\bf g}, {\bf h})\omega_{2}({\bf g}, {\bf h})=1$ (assuming the $R_{\bf g}$ is a linear representation). Of course, multiplication of these local operators $U^{(j)}_{\bf g}$ must be associative, which requires the projective phases to satisfy the ``2-cocycle condition''
\begin{equation}
\omega_{j}(\mb{g,h})\omega_{j}(\mb{gh,k})=\omega_{j}(\mb{g,hk})\omega_{j}(\mb{h,k}).
\end{equation}
(The projective phases are also known as the ``factor set'' characterizing a projective representation.)

Additionally, one can always trivially redefine the local operators by a phase $\tilde{U}^{(j)}_{\bf g} = \zeta_{j}({\bf g}) U^{(j)}_{\bf g}$, as long as the phases $\zeta_{j}({\bf g}) \in \U (1)$ satisfy $\zeta_{1}({\bf g})\zeta_{2}({\bf g})=1$. The correspondingly redefined projective phases are, thus, given by $\tilde{\omega}_{j}({\bf g}, {\bf h}) =  \frac{\zeta_{j}({\bf g}) \zeta_{j}({\bf h})}{\zeta_{j}({\bf gh})} \omega_{j}({\bf g}, {\bf h})$. A combination of phase factors of the form $\frac{\zeta_{j}({\bf g}) \zeta_{j}({\bf h})}{\zeta_{j}({\bf gh})}$ is known as a ``2-coboundary.'' Hence, one should view different choices of projective phases that are related by 2-coboundaries as physically equivalent.

Mathematically, taking the quotient of the set of 2-cocycles by the set of 2-coboundaries (i.e. forming equivalence classes of cocycles related by coboundaries) yields the 2nd cohomology group $\coh{2}{G}$. (See Appendix~\ref{app:coh} for a brief review of group cohomology.) We denote the equivalence class that contains the cocycle $\omega$ as the cohomology class $[\omega]$. In this way, $\coh{2}{G}$ classifies the possible types of projective representations of $G$, with the trivial cohomology class $[1] \in \coh{2}{G}$ corresponding to linear representations.

We note that the constraints relating the phase factors associated with the two endpoints, i.e. $\omega_{1}({\bf g}, {\bf h})=\omega_{2}({\bf g}, {\bf h})^{-1}$ and $\zeta_{1}({\bf g})=\zeta_{2}({\bf g})^{-1}$, may be viewed as topological. This is because they are imposed by the topological properties of the system, in this case being that a line segment has precisely two endpoints. The consequence of these constraints is that there is a correlation between the projective representations localized at the endpoints of the chain, specifically $[\omega_{1}] = [\omega_{2}]^{-1}$, and hence the possible gapped phases of the system are classified by a single copy of $\coh{2}{G}$. We emphasize that, when the cohomology class is nontrivial, the corresponding projective representations are necessarily multi-dimensional, and, hence, there is a protected degeneracy of the boundary states.

In higher dimensions, the role of projective representations is played by higher cohomology classes. SPT phases in $d$ spatial dimensions may be characterized by the $(d+1)$th cohomology group $\coh{d+1}{G}$.~\cite{chen2013}

\subsection{Weak SPT Phases}
\label{sec:Weak_SPT}

\begin{figure}[t!]
	\includegraphics[width=80mm]{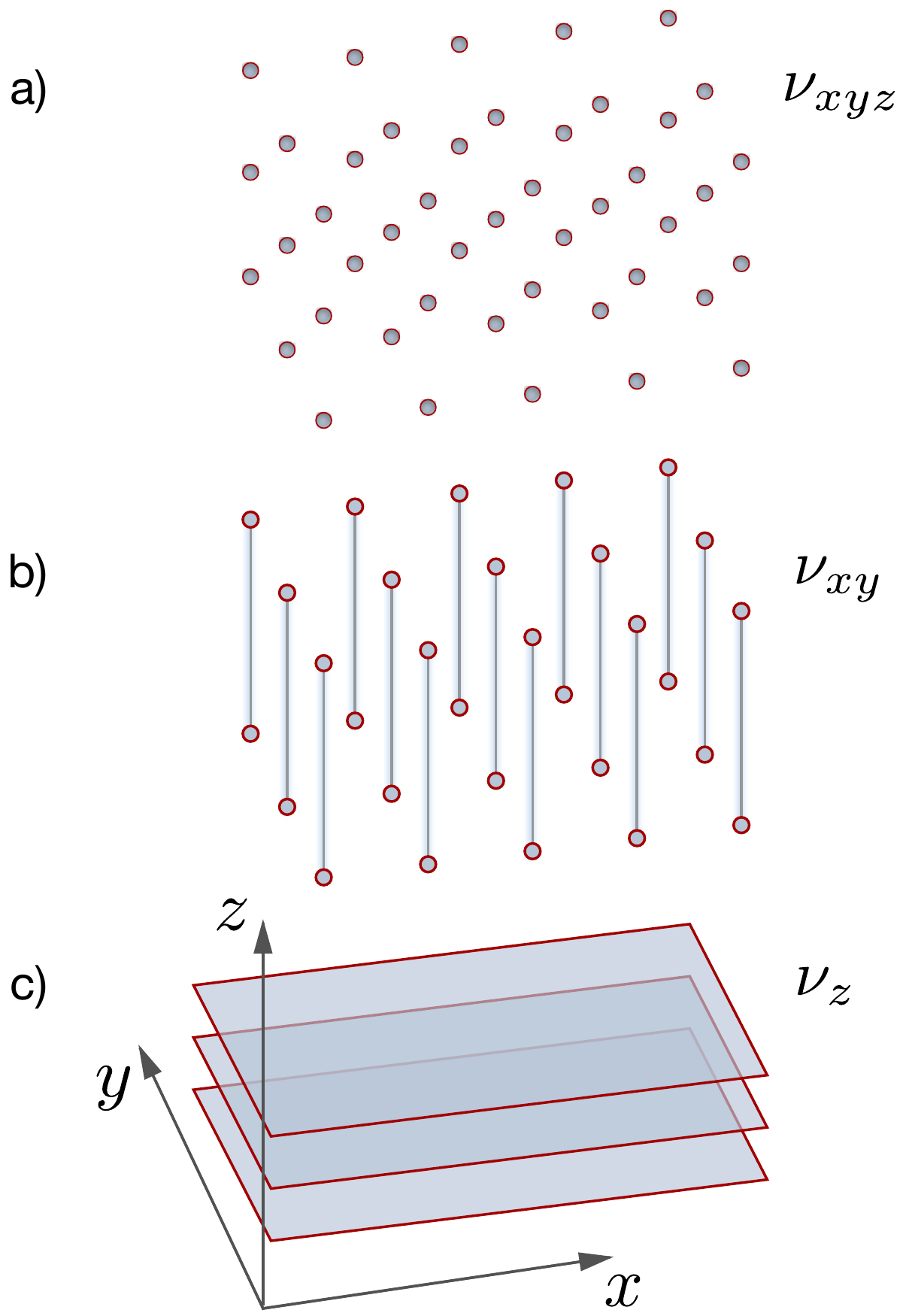}
	\caption{The types of 3D Weak SPT phases. a) Filling 0D SPT phases (charge density), $\nu_{xyz}$.  b) Packing 1D SPT phases, $\nu_{xy}$. c) Stacking 2D SPT phases, $\nu_{z}$. }
	\label{fig:weak_spt}
\end{figure}

A weak SPT phase is a phase protected by translational symmetry in addition to on-site symmetry.
For example, 3D TIs have  three weak indices $\nu_{i} \in \mathbb{Z}_2$ for $i=x,y,z$ corresponding to stacking 2D TIs along the $\hat{i}$-direction.~\cite{Fu_PRL07} Translational invariance is crucial here, otherwise neighboring layers can dimerize and leave the state trivial.
More generally, there are additional weak invariants. For instance, we can also pack together 1D SPT phases. This leads to the following hierarchy of weak invariants in 3D, depicted schematically in Fig.~\ref{fig:weak_spt}.

\emph{a}) $\nu_{ijk}$ the ``0D SPT per unit volume.'' A 0D SPT phase is simply a charge, since $\coh{1}{G}$ is the group of $\U(1)$ (unitary, Abelian) representations of $G$. Filling volume with 0D SPT phases amounts to a charge density given by the index $[\nu_{xyz}] \in  \coh{1}{G}$.
	
\emph{b}) $\nu_{ij}$ the ``1D SPT per unit area.'' Packing 1D SPT phases of index $\nu_{ij}$ perpendicular to the $ij$-plane, for $i,j=x,y,z$ is encoded in three indices $[\nu_{xy}], [\nu_{yz}], [\nu_{zx}] \in  \coh{2}{G}$.

\emph{c})	$\nu_i$ the ``2D SPT per unit length.'' Stacking 2D SPT phases of index $\nu_i$ along the $\hat{i}$-direction for $i=x,y,z$ is encoded in three indices $[\nu_{x}],[\nu_{y}],[\nu_{z}] \in \coh{3}{G}$.

We find that the weak invariants can be elegantly incorporated into the cohomology formulation. Assuming the full symmetry group takes the form $\SG = \mathbb{Z}^d_{\text{trans}} \times G_{\text{int}}$ for a system with $d$ spatial dimensions,  we use the K\"{u}nneth formula~\cite{chen2013, WenPRB2015} to compute $\coh{d+1}{\SG}$ and find
\begin{align}
\coh{d+1}{\mathbb{Z}^d \times G} &= \prod_{r=0}^{d} \left(  \coh{r+1}{G} \right)^{\binom{d}{r}}
.
\end{align}
(See Appendix~\ref{sec:Kunneth} for details.)
Since $\binom{d}{r} = \frac{d !}{ r! (d-r)! }$ is the number of perpendicular $r$ planes in $d$ space, this formula is the obvious generalization of the stacking picture in arbitrary dimensions. In 3D, we have
\begin{align}
\coh{4}{\mathbb{Z}^3 \times G} &=  \coh{4}{G}  \times  \left(\coh{3}{G} \right)^{3}  \notag \\
& \times \left( \coh{2}{G} \right)^{3} \times  \coh{1}{G}.
\end{align}
As such, the SPT classes $[\Omega]\in \coh{4}{\mathbb{Z}^3\times G}$ naturally decompose into octuples as $[\Omega] = \{ [\nu], [\nu_i],[\nu_{ij}],[\nu_{ijk}] \}$, where $[\nu] \in \coh{4}{G}$ is the strong SPT invariant mentioned in Sec.~\ref{sec:StrongSPT} and $[\nu_i]$, $[\nu_{ij}]$, and $[\nu_{ijk}]$ are the weak invariants discussed above.

In this paper, the most significant weak SPT invariant is the 1D SPT per unit area, characterized by $[\nu_{ij}]$.
As discussed, a 1D SPT phase has degenerate boundary states that transform under $G$ with a projective representation encoded by the index $[\nu_{ij}] \in \coh{2}{G}$ that classifies the phase. Consequently, at an boundary surface along the $ij$-plane, there will be a finite density (one per unit area) of low-energy edge states whose fate is determined by the surface interactions. Each unit cell of the effective Hilbert space at the surface will thus transform under $G$ as a projective representation labeled by $[\nu_{ij}]$.

\subsubsection{Example: AKLT spin system}

For concreteness, consider a spin system with $G = \textrm{SO}(3)$. The 2nd cohomology group (which classifies projective representations) is $\coh{2}{\textrm{SO}(3)} = \mathbb{Z}_2$, the trivial element of which corresponds to the integer spin representations and the nontrivial element of which corresponds to the half-integer spin representations.
A well known example of a nontrivial strong 1D SPT phase is the $S=1$ AKLT chain, which has two-fold degenerate $S=1/2$ edge states.~\cite{AKLT87}
Packing AKLT  chains along the $z$-axis at finite density of one per $xy$-unit cell, we trivially obtain a $\nu_{xy} = -1$ weak SPT whose surface theory will behave like an $S=1/2$ square-lattice magnet.
A more familiar model is the 3D cubic $S=3$ AKLT model; in this case $\nu_{xy} = \nu_{yz} = \nu_{zx} = -1$.
However, since we will only analyze one surface at a time, the latter two invariants are irrelevant.

\subsection{Computing the weak invariants from the slant product}

Given a 4-cocycle $\Omega({\bf g}_1, {\bf g}_2, {\bf g}_3, {\bf g}_4)$, there is a computational procedure for finding the invariants $\{ \nu, \nu_i,\nu_{ij},\nu_{ijk} \}$ appearing in the K\"{u}nneth decomposition.
The reader is free to treat the procedure as a ``black box,'' but for completeness, the key tool is the ``slant product''
which we define in Eq.~(\ref{eq:slant_product}). The slant product induces a homomorphism that maps $n$-cohomology classes to $(n-1)$-cohomology classes
\begin{align}
\slant_{\bf g} : \coh{n}{G} & \rightarrow  \coh{n-1}{G}
\notag \\
[ \omega ] & \mapsto  [ \slant_{\bf g} \omega ]
.
\end{align}
More explicitly, for elements ${\bf p}\in Z(G)$ in the center of $G$, the slant product acting on 2, 3, and 4-cocycles gives
\begin{align}
\slant_{\bf p} \omega({\bf g}) &= \frac{\omega({\bf g}, {\bf p})}{\omega({\bf p}, {\bf g})} \\
\slant_{\bf p} \omega({\bf g}, {\bf h}) &= \frac{\omega({\bf p}, {\bf g}, {\bf h}) \omega({\bf g}, {\bf h}, {\bf p})}{\omega({\bf g}, {\bf p}, {\bf h})} \\
\slant_{\bf p} \omega({\bf g}, {\bf h}, {\bf k})&= \frac{\omega({\bf g}, {\bf p}, {\bf h}, {\bf k}) \omega({\bf g}, {\bf h}, {\bf k}, {\bf p})}{\omega({\bf p}, {\bf g}, {\bf h}, {\bf k}) \omega({\bf g}, {\bf h}, {\bf p}, {\bf k})}.
\label{eq:slant_def}
\end{align}

The slant product can also be applied in succession. When ${\bf p}, {\bf q} \in Z(G)$, one finds the (cohomological) equivalence for applying the slant product in different orders to be $[\slant_{\bf p} \slant_{\bf q} \omega]=[ \slant_{\bf q} \slant_{\bf p} \omega]^{-1}$.

If $T_i$ is the group element that generates translations along the $\hat{i}$-direction, the 4-tuple of invariants can be recovered by successive application of the slant products $\slant_{T_i}$ followed by a restriction of the resulting cocycles to the on-site symmetry group $G$:
\begin{align}
\nu({\bf g}, {\bf h}, {\bf k}, {\bf l}) & = \Omega({\bf g}, {\bf h}, {\bf k}, {\bf l})|_{{\bf g}, {\bf h}, {\bf k}, {\bf l} \in G}\\
\nu_{i}({\bf g}, {\bf h}, {\bf k}) &=  \slant_{T_i}\Omega({\bf g}, {\bf h}, {\bf k})|_{{\bf g}, {\bf h}, {\bf k} \in G}\\
\nu_{i j}({\bf g}, {\bf h}) &=  \slant_{T_i} \slant_{T_j} \Omega({\bf g}, {\bf h})|_{{\bf g}, {\bf h} \in G}\\
\nu_{i j k}({\bf g}) &= \slant_{T_i} \slant_{T_j} \slant_{T_k} \Omega({\bf g}) |_{{\bf g} \in G}.
\label{eq:slant_kun}
\end{align}
For more details on the use of slant product in the K\"unneth decomposition of cocycles, see Appendix~\ref{sec:Kunneth_rep}.

\section{Symmetry fractionalization in 2D topological phases}
\label{sec:symmetry_fractionalization}

We now review the theory of symmetry fractionalization in a 2D topologically ordered phase~\cite{essin2013,SET}.
A topological phase supports anyonic quasiparticle excitations whose topologically distinct types are denoted by $a, b, c, \dots$, which we refer to as anyons or topological charges. We assume that the system has total symmetry group is $\mathcal{G}=\mathbb{Z}^2\times G$, where $G$ is the on-site symmetry group and $\mathbb{Z}\times\mathbb{Z}$ is the translational symmetry group generated by $T_x$ and $T_y$, where $x$ and $y$ denote a basis for the Bravais lattice.

\subsection{On-site symmetries}

The structure of symmetry fractionalization is organized according to three successively finer distinctions.

First, symmetry transformations can permute anyon types.
For example, consider a bilayer of two $\nu = \frac{1}{2}$ bosonic quantum Hall states.
There are two types of Laughlin quasiparticles, $s_1, s_2$, corresponding to the two layers, as well as their composite $f= s_1 s_2$.
Exchanging the layers is an on-site symmetry ${\bf g}$ (with ${\bf g}^2 = \openone$) of the model which permutes the two Laughlin quasiparticle types ${\bf g} : s_1 \leftrightarrow s_2$, while leaving their composite invariant, ${\bf g}: f \leftrightarrow f$.
A less trivial example is Wen's plaquette model for $\mathbb{Z}_2$ toric code, where translations by one lattice spacing $T_{i}$ for $i=x,y$ exchange electric and magnetic charges, $T_{i} : e \leftrightarrow m$. ~\cite{Wen2003,you2012}.
In this paper, we will, however, assume that the symmetries $\SG$ do not permute any of the anyon types, as this greatly simplifies the computations.

Second, anyons can carry fractionalized symmetry quantum numbers.
To be more precise, let us prepare a state $\ket{\Psi_{a_1 , \ldots , a_{n} }}$ with $n$ well-separated quasiparticles, where the $j$th quasiparticles carries topological charge $a_j$. (We use the convention where the ``vacuum'' topological charge is denoted $I$ and the ``topological charge conjugate'' of $a$ is denoted $\bar{a}$, which is the unique topological charge that can fuse with $a$ into vacuum.) We assume the overall topological charge is trivial $I$, so that the state can be created from the ground state by applying local operators. (When there are non-Abelian anyonic quasiparticles, one must further specify the fusion channels to specify the state, but this will play no role if our discussion, so we leave it implicit.) We consider the global symmetry transformation $R_\mathbf{g}$ acting on the state $\ket{\Psi_{\{a \}}}$, where we introduce the shorthand $\{a \}$ for the collection of topological charges $a_1 , \ldots , a_{n}$ carried by the quasiparticles. Without loss of generality, we shall assume that the system transforms as a linear representation of $G$ (e.g. if there is a $S=1/2$ spin per site, then we require the number of sites to be even). Although anyons are non-local excitations, the local properties (i.e. local density matrix) of regions away from the positions of the anyons remain the same as those of the ground state.
Therefore the global symmetry transformation $R_\mathbf{g}$ should have the following decomposition (when the symmetries do not permute anyon types):
\begin{equation}
R_\mathbf{g}\ket{\Psi_{\{a\}}}=\prod_{j=1}^{n} U_\mathbf{g}^{(j)}\ket{\Psi_{\{a\}}}.
	\label{eq:sym_loc}
\end{equation}
Here, $U_\mathbf{g}^{(j)}$ is a local unitary operator whose nontrivial action is localized in the neighborhood of the $j$th quasiparticle.
Similar to the localized symmetry actions at endpoints of 1D chains, discussed in Sec.~\ref{sec:StrongSPT}, the localized symmetry transformations $U_\mathbf{g}^{(j)}$ only needs to projectively represent group multiplication:
\begin{equation}
U_\mathbf{g}^{(j)}U_\mathbf{h}^{(j)}\ket{\Psi_{\{a \}}}=\eta_{a_j}(\mathbf{g},\mathbf{h})U_\mathbf{gh}^{(j)}\ket{\Psi_{\{a \}}}
,
\label{eq:eta_def}
\end{equation}
where the projective phase $\eta_{a_j}(\mathbf{g},\mathbf{h}) \in \U (1)$ only depends on the topological properties localized in the neighborhood of the $j$th quasiparticle, which is just the topological charge $a_j$ carried by this quasiparticle. Since $R_{\bf g}$ is a linear representation, the projective phases must satisfy the condition $\prod_j \eta_{a_j}(\mathbf{g}, \mathbf{h})=1$.~\footnote{In writing this condition, we have made a canonical gauge choice for the gauge transformations of symmetry operations on fusion spaces, which is always possible when the symmetry do not permute anyons. In particular, this is the gauge choice in which $U_{\bf g}(a,b;c) = \openone$ and $\eta_{a}({\bf g},{\bf h})= M_{a, \coho{w}({\bf g},{\bf h})}$ in Ref.~\cite{SET}.}

The constraints on the projective phases here provides a richer structure than we observed for 1D SPTs. In particular, this condition is equivalent to the condition that $\eta_a(\mathbf{g},\mathbf{h})\eta_b(\mathbf{g},\mathbf{h})=\eta_c(\mathbf{g},\mathbf{h})$ whenever the topological charge $c$ is a permissible fusion channel of the topological charges $a$ and $b$ (i.e. whenever $N_{ab}^{c} \neq 0$). It follows that $\eta_a(\mathbf{g},\mathbf{h})$ must take the form~\cite{SET}
\begin{equation}
\eta_a(\mathbf{g},\mathbf{h})=M_{a, \cohosub{w}({\bf g},{\bf h})},
\label{eqn:etaasbraiding}
\end{equation}
where $\coho{w}(\mathbf{g},\mathbf{h})$ is an Abelian anyon, and $M_{a, \cohosub{w}({\bf g},{\bf h})}$ is the mutual braiding statistics between anyons $a$ and $\coho{w}({\bf g},{\bf h})$.

Associativity of the localized operators gives the condition
\begin{equation}
\eta_a (\mb{g,h}) \eta_a (\mb{gh,k})= \eta_a (\mb{g,hk}) \eta_a (\mb{h,k})
,
\end{equation}
for all $a$, which translates into
\begin{equation}
\coho{w} (\mb{g,h}) \times \coho{w} (\mb{gh,k}) = \coho{w} (\mb{g,hk}) \times \coho{w} (\mb{h,k})
.
\end{equation}
This is precisely the $2$-cocycle condition, though not for elements in $\U (1)$, but rather for elements in $\mathcal{A}$ the group whose elements are the Abelian anyons, with group multiplication defined by fusion of anyons.

There is also freedom to trivially redefine the localized symmetry transformations by local operators $\tilde{U}^{(j)}_{\bf g} = Z^{(j)}_{\bf g} U^{(j)}_{\bf g}$, whose action on the state is $Z^{(j)}_{\bf g} \ket{\Psi_{\{a \}}} = \zeta_{a_j}({\bf g}) \ket{\Psi_{\{a \}}}$, where the phases $\zeta_{a_j}({\bf g}) \in \U (1)$ satisfy $\prod_j \zeta_{a_j}({\bf g})=1$. This, similarly, generates a map between these phases and anyons, with the relation $\zeta_{a}({\bf g}) = M_{a, \cohosub{z}({\bf g})}$, where $\coho{z}({\bf g})$ is an Abelian anyon. This translates the trivial redefinitions of the local operators into the redefinition $\tilde{\coho{w}} (\mb{g,h}) = \coho{z}({\bf g}) \times \overline{\coho{z}({\bf gh})} \times \coho{z}({\bf h}) \times \coho{w} (\mb{g,h})$. This is precisely redefinition by an $\mathcal{A}$ $2$-coboundary $\coho{z}({\bf g}) \times \overline{\coho{z}({\bf gh})} \times \coho{z}({\bf h})$.

The result is a classification~\cite{SET} by the Abelian anyon-valued 2nd cohomology group $\mathcal{H}^2[G, \mathcal{A}]$. In particular, we form equivalence classes of $\mathcal{A}$ 2-cocycles that are related to each other by $\mathcal{A}$ 2-coboundaries. A given equivalence class $[\coho{w}] \in \mathcal{H}^2[G, \mathcal{A}]$ completely specifies the symmetry fractionalization of the system, i.e. the local projective phases $\eta_a (\mb{g,h})$ are given by Eq. \eqref{eqn:etaasbraiding} for all anyon types. Thus, we call these cohomology classes the ``symmetry fractionalization classes.''

The manifestation of symmetry fractionalization for the anyons may exhibit two {characteristic} properties: 1) anyons can carry a fractional charge (like the Laughlin quasiparticles), and 2) anyons can carry a localized projective representation of the symmetry group $G$, i.e. they have an internal degeneracy (like spin), which transforms projectively under $G$. We can determine these symmetry fractionalization properties for an anyon of topological charge $a$ by examining the corresponding local projective phases $\eta_a (\mb{g,h})$, as we now explain in more detail.

\begin{enumerate}
\item
An anyon $a$ can carry fractional charge under some subgroup $H<G$. We first consider the group $H = \U (1)$. The transformation of an object of charge $Q$ corresponding to $e^{i \theta } \in \U (1)$ is $e^{i \theta Q }$. The total charge in the system (and local excitations) must be an integer, so that the state is left invariant when $\theta$ goes from $0$ to $2 \pi$, i.e. $\ket{\Psi} \rightarrow e^{i 2 \pi Q}\ket{\Psi}=\ket{\Psi}$. Topologically nontrivial quasiparticles may carry fractional charge, as long as the sum of the charges of all quasiparticles in a system adds up to an integer. In other words, the fractional charge assignments must be compatible with the fusion rules, which is the condition imposed by the fractionalization class. One might, thus, write the action of the localized symmetry operation as $U_{\bf \theta}^{(j)} \ket{\Psi_{\{a\}}} = e^{i \theta Q_{a_{j}} } \ket{\Psi_{\{a\}}}$, where $Q_{a}$ is the (possibly) fractional $\U (1)$ charge carried by anyons with topological charge $a$. Of course, this is not gauge invariant for arbitrary $\theta$, but, rather, only when one has wound the angle by $2 \pi$, giving $U_{2 \pi}^{(j)} \ket{\Psi_{\{a\}}} = e^{i 2 \pi Q_{a_{j}} } \ket{\Psi_{\{a\}}}$. Strictly speaking, we have defined $\theta \in [0, 2\pi )$, so this expression is interpreted to mean that we wind $\theta$ by $2 \pi$. We can make this notion well-defined by writing it as
\begin{equation}
\label{eq:fractional_charge_1}
U_{\theta}^{(j)} U_{ 2 \pi - \theta}^{(j)} \ket{\Psi_{\{a\}}} = e^{i 2 \pi Q_{a_{j}} } \ket{\Psi_{\{a\}}}
.
\end{equation}
In this way, the fractional charge is given in terms of the projective phases as
\begin{equation}
\label{eq:fractional_charge_3}
e^{i 2 \pi Q_{a}} = \eta_{a} (\theta , 2\pi - \theta)
.
\end{equation}
Alternatively, we can discretize the process into $N$ steps and write it as
\begin{equation}
\label{eq:fractional_charge_2}
\left[ U_{\frac{2 \pi}{N}}^{(j)} \right]^{N}\ket{\Psi_{\{a\}}} = e^{i 2 \pi Q_{a_{j}}} \ket{\Psi_{\{a\}}}
,
\end{equation}
in which case the fractional charge can be written as $e^{i 2 \pi Q_{a}} = \prod_{k=0}^{N -1} \eta_{a} (\frac{2 \pi}{N} , k \frac{2 \pi}{N} )$.~\footnote{
We can immediately generalize this result to the case of a subgroup $H = \mathbb{Z}_{N}$ of any on-site symmetry group $G$ (assuming it has such a subgroup), even when there is no $\U (1)$ subgroup of $G$. Consider an element ${\bf g }\in G$ of order $N$ (i.e. ${\bf g}^{N} = \openone$), which thus generates a $\mathbb{Z}_{N}$ subgroup. We, similarly, have $[U_{\bf g}^{(j)} ]^{N}\ket{\Psi_{\{a\}}} = e^{i 2 \pi Q_{a_{j}; {\bf g}}} \ket{\Psi_{\{a\}}}$, where $Q_{a; {\bf g}}$ is the fractional charge of anyon $a$ under the symmetry ${\bf g}$. Repeatedly applying Eq.~(\ref{eq:eta_def}), we find $e^{i 2 \pi Q_{a; {\bf g}}} = \prod_{k=0}^{N -1} \eta_{a} ({\bf g},{\bf g}^{k})$, just as we did from Eq.~(\ref{eq:fractional_charge_2}). In contrast with the case of the continuous $\U (1)$ symmetry, in order for this quantity to be gauge invariant, we must have $M_{a, z}^{N}=1$ for any $z \in \mathcal{A}$. If $\mathbb{Z}_{N}$ is being chosen as an arbitrary subset of $\U (1)$, then we can always choose this to be satisfied. For example, this holds for any $a$ by letting $N = |\mathcal{A}|$ (or any multiple of this), since $M_{a, z}^{N}=M_{a, z^{N}} = M_{a, I}=1$ for any $z \in \mathcal{A}$. However, if $N$ is somehow constrained, e.g. when there is no $\U (1)$ subgroup, then the notion of fractional $\mathbb{Z}_{N}$ charge carried by an anyon $a$ may not be well-defined.
}

\item
When the symmetries do not permute anyon types, the local projective phases $\eta_a(\mathbf{g},\mathbf{h})$ satisfy the conditions of a $\mathrm{U}(1)$ $2$-cocycle. Thus, $\eta_a(\mathbf{g},\mathbf{h})$ may be viewed as a representative element of the $\U (1)$ 2-cohomology class $[ \eta_a ]\in \mathcal{H}^2[G, \U (1)]$. As previously discussed, $\mathcal{H}^2[G, \U (1)]$ classifies projective representations of $G$. Hence, when $[ \eta_{a_j} ]$ is a nontrivial element of $\mathcal{H}^2[G, \U (1)]$, the local unitary operators ${U}^{(j)}_\mathbf{g}$ acting in the neighborhood of the $j$th quasiparticle (which has topological charge $a_j$) form a projective representation of $G$.~\footnote{Even when $\eta_a(\mathbf{g},\mathbf{h}) \neq 1$, it may be a $\U(1)$ 2-coboundary, and, hence, be a representative of the trivial $\U (1)$ cohomology class, i.e. $[ \eta_a ]=[1] \in \mathcal{H}^2[G, \U (1)]$.} In this case, the ${U}^{(j)}_\mathbf{g}$ must be acting on a multi-dimensional Hilbert space. In other words, there is a symmetry-protected local degeneracy associated with anyons of topological charge $a_j$. Thus, when $[ \eta_{a} ]$ is a nontrivial element of $\mathcal{H}^2[G, \U (1)]$, we say that $a$ carries a localized projective representation of $G$.

\end{enumerate}

Famously, property 1) is exhibited by symmetry fractionalization in the $\nu=\frac{1}{m}$ Laughlin fractional quantum Hall states, where the Laughlin quasiparticles carry $1/m$ electric charge of the $\mathrm{U}(1)$ symmetry. When $G= \U(1)$ and $\mathbb{Z}_m < \mathcal{A}$, the possibility of charge fractionalization is captured by the symmetry fractionalization class in $\mathcal{H}^2[\mathrm{U}(1), \mathbb{Z}_m]=\mathbb{Z}_m$. On the other hand, for $\mathrm{U}(1)$ cohomology, we have $\coh{2}{\mathrm{U}(1)}=\mathbb{Z}_1$, indicating that nontrivial projective representations do not even exist. Hence, quasiparticles in the Laughlin FQH state can have no internal degeneracy.

A well-known example of property 2) is a $\mathbb{Z}_2$ spin liquid with $G=\mathrm{SO}(3)$, where spinon excitations carry spin-$1/2$ projective representations of $G$. Mathematically, we have $\mathcal{H}^2[\mathrm{SO}(3), \mathbb{Z}_2]=\mathcal{H}^2[\mathrm{SO}(3), \mathrm{U}(1)]=\mathbb{Z}_2$, for which the trivial group element corresponds to integer representations and the nontrivial group element corresponds to half-integer representations of SO(3).

The third level of SET structure is that of the extrinsic symmetry defects (fluxes)~\cite{SET}. In fact, given the symmetry action $\rho$ and a symmetry fractionalization class $[\coho{w}]$, it is not necessarily the case that a consistent structure can be defined for the defects in purely 2D systems. There is an object $[\mathscr{O}] \in \coh{4}{G}$, which is uniquely defined in terms of the topological order, $\rho$, and $[\coho{w}]$. This object $[\mathscr{O}]$ indicates whether or not a consistent 2D defect theory is possible, depending on whether or not it is a trivial element of $\coh{4}{G}$, and so it is called the obstruction class. When $[\mathscr{O}]$ is a nontrivial, the symmetry fractionalization class is anomalous and it can only appear as the surface termination of a 3D SPT phase~\cite{VishwanathPRX2013, chen2014}. We will discuss this in more detail later in the paper.

\subsection{Incorporating translational symmetry}
\label{sec:set_translation}

The analysis of symmetry fractionalization can be generalized to translational symmetry.
In 2D, lattice translations form a $\mathbb{Z}\times\mathbb{Z}$ group. For the generators $T_i$ of translation in the $i=x$ and $y$ directions, the decomposition of the action of the symmetry (again assuming the symmetries do not permute anyon types) is generalized to~\cite{SET}
\begin{equation}
R_{T_{i} }\ket{\Psi_{\{a\}}}=\prod_{j=1}^{n} U_{T_{i}}^{T_{i} (j)}\ket{\Psi_{T_{i} \{a \}}},
\label{}
\end{equation}
where $T_{i} \{a \}$ is used here to indicate that the location of the quasiparticles have all been translated by $T_{i}$, and $T_{i} (j)$ is used to indicate that the neighborhood in which the corresponding local operator
$U_{T_{i}}^{T_{i}(j)}$ acts nontrivially has also been translated by $T_{i}$.~\footnote{This is slightly different notation that was used in Ref.~\cite{SET}, where it is also explained that, more generally, the notion of symmetry fractionalization is sensible when the symmetry is ``quasi-on-site,'' meaning the symmetry $R_{\bf g}$ can have nonlocal action, as long as it preserves the relative locality of all operators when acting upon them by conjugation.}
We can then repeat the entire analysis and arrive at essentially the same classification, but now including the translational symmetries.

It is very instructive to study the fractionalization classes in $\mathcal{H}^2[\mathbb{Z}^2\times G, \mathcal{A}]$ in greater detail. Using the K\"unneth formula for group cohomology, we have
\begin{equation}
\mathcal{H}^2[\mathbb{Z}^2\times G, \mathcal{A}]= \mathcal{H}^2[\mathbb{Z}^2, \mathcal{A}]\times \left(\mathcal{H}^1[G, \mathcal{A}]\right)^2
\times \mathcal{H}^2[G, \mathcal{A}]
,
\label{eq:trans_frac_decomp}
\end{equation}
where we have used the fact that $\mathcal{H}^1\left[G, \mathcal{H}^1[\mathbb{Z}^2, \mathcal{A}]\right]=\mathcal{H}^1 [G, \mathcal{A}^{2} ]=\left(\mathcal{H}^1[G, \mathcal{A}]\right)^2$. We note that $\mathcal{H}^2[\mathbb{Z}^2, \mathcal{A}]=\mathcal{A}$ and that $\mathcal{H}^1[G, \mathcal{A}]$ is just the (group formed by the) homomorphisms from $G$ to $\mathcal{A}$.

In order to understand the terms in this decomposition, it is useful to consider the symmetry localization of the operations of the form $R_{\bf g}^{-1} R_{\bf h}^{-1} R_{\bf g} R_{\bf h}$,  where the group elements ${\bf g}$ and ${\bf h}$ commute. In this case, the localized operators gives
\begin{eqnarray}
\rho_{\bf h} [ U_\mathbf{g}^{{\bf \bar{h}}(j)}]^{-1} \rho_{\bf h}^{-1} [U_\mathbf{h}^{(j)}]^{-1} U_\mathbf{g}^{(j)} \rho_{\bf g} U_\mathbf{h}^{{\bf \bar{g}}(j)} \rho_{\bf g}^{-1} \ket{\Psi_{\{a\}}} &&
\notag \\
= \frac{\eta_{a_j}(\mathbf{g},\mathbf{h})}{\eta_{a_j}(\mathbf{h},\mathbf{g})}\ket{\Psi_{\{a\}}} &&
,
\label{eq:local_commutator}
\end{eqnarray}
where $\rho_{\bf g}$ here is an operator on the physical Hilbert space that acts precisely as the symmetry action does on the topological state space, i.e. it acts on the topological quantum numbers of the physical states. (See Ref.~\onlinecite{SET} for more details.) For our purposes here, it only has the effect of translating the locations of quasiparticles when ${\bf g}$ is a translational symmetry group element.

We define
\begin{equation}
\coho{b}(\mathbf{g},\mathbf{h}) = \overline{ \coho{b}(\mathbf{h} ,\mathbf{g})} = \coho{w}(\mathbf{g},\mathbf{h}) \times  \overline{ \coho{w}(\mathbf{h},\mathbf{g})}
,
\end{equation}
for $\mathbf{g},\mathbf{h} \in \mathcal{G} = \mathbb{Z}^{2} \times G$, which takes values in the set of Abelian anyons $\mathcal{A}$. We note that, when ${\bf gh} = {\bf hg}$, the quantity $\coho{b}(\mathbf{g},\mathbf{h})$ is gauge invariant, i.e. it does not change when $\coho{w}(\mathbf{g},\mathbf{h})$ is modified by a 2-coboundary.

Then, in the gauge where $\eta_{a}(\mathbf{g},\mathbf{h}) = M_{a , \cohosub{w}(\mathbf{g},\mathbf{h})}$, we can write the ratio of projective phases as
\begin{equation}
\frac{\eta_{a}(\mathbf{g},\mathbf{h})}{\eta_{a}(\mathbf{h},\mathbf{g})} = M_{ a , \cohosub{b}(\mathbf{g},\mathbf{h})}
.
\end{equation}

A representative $2$-cocycle of a fractionalization class in $\mathcal{H}^2[\mathbb{Z}^2\times G, \mathcal{A}]$ can be chosen to take the form
\begin{equation}
\begin{split}
\coho{w}(T_x^{m_x} T_y^{m_y} {\bf g}, T_x^{n_x} T_y^{n_y} {\bf h} )=
[\coho{b}(T_y,T_x)]^{m_y n_x} \\
\times [\coho{b}(T_x ,{\bf h})]^{m_x}
\times [\coho{b}(T_y , {\bf h})]^{m_y}
\times \coho{w}({\bf g} , {\bf h})
,
\end{split}
\label{eq:decomposed_2cocycle}
\end{equation}
for $\mathbf{g},\mathbf{h} \in \mathcal{G} = G$.
It it straightforward to see that this is indeed a $2$-cocycle, as long as
\begin{equation}
\coho{b}(T_i , {\bf g}) \times \coho{b}(T_i, {\bf h} ) = \coho{b}(T_i , {\bf gh}),
\end{equation}
for $i=x,y$. These conditions are just the requirement that $\coho{b}(T_i , {\bf g})$ are group homomorphism from $G$ to $\mathcal{A}$, i.e. that $\coho{b}(T_i , {\bf g}) \in \mathcal{H}^1[G, \mathcal{A}]$.

\begin{figure}[t!]
	\centering
	\includegraphics[width=0.6\columnwidth]{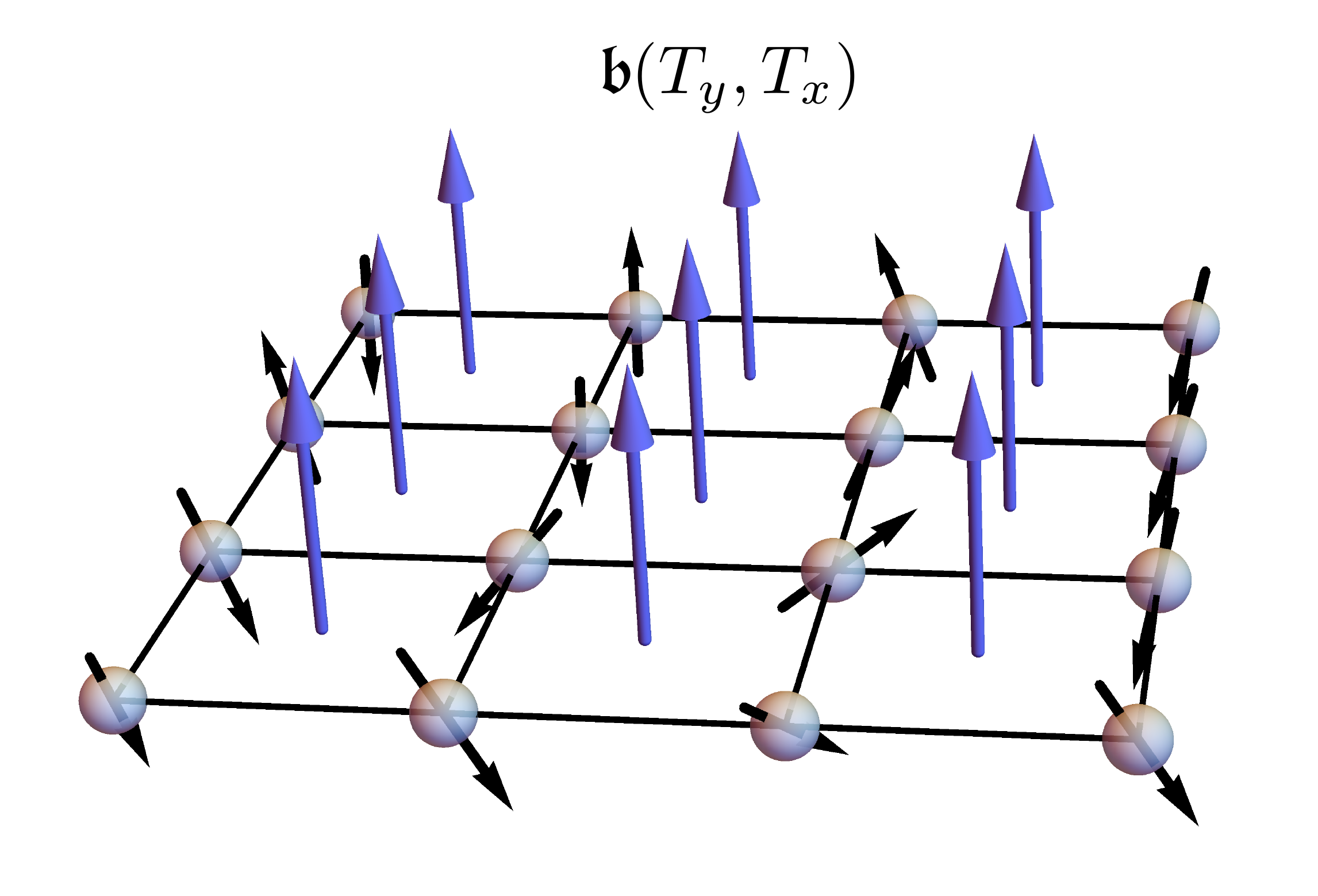}
	\caption{Anyonic flux per unit cell $\coho{b}(T_y,T_x)$. }
	\label{fig:flux}
\end{figure}

The terms in the decompositions of Eqs.~(\ref{eq:trans_frac_decomp}) and (\ref{eq:decomposed_2cocycle}) can now be understood physically as follows:
\begin{enumerate}
\item
$\coho{b}(T_y,T_x)$ the ``anyonic flux per unit cell.'' The first term, $\mathcal{H}^2[\mathbb{Z}^2, \mathcal{A}]=\mathcal{A}$, in the K\"unneth formula decomposition of Eq.~(\ref{eq:trans_frac_decomp}) is characterized by the gauge invariant quantity $\coho{b}(T_y,T_x) \in \mathcal{A}$, which we can interpret as the background topological flux through each unit cell (as depicted in Fig.~\ref{fig:flux}) in the following way. The symmetry operation $T^{-1}_y T^{-1}_x T_y T_x$ is a sequence of translations corresponding to a path that encloses one unit cell in a counterclockwise fashion. From Eq.~(\ref{eq:local_commutator}), we see that this operation has the corresponding local projective phase factor of $M_{ a , \cohosub{b}(T_y,T_x)}$ for quasiparticles of topological charge $a$. (See also Ref.~\onlinecite{ZaletelPRL2015} for an operational definition.) This phase
$M_{ a , \cohosub{b}(T_y,T_x)}$ is the mutual braiding statistics associated with an anyon $a$ encircling an anyon $\coho{b}(T_y,T_x)$ in a counterclockwise fashion. Thus, we can picture this type of symmetry fractionalization as being generated by an Abelian anyon $\coho{b}(T_y,T_x)$ sitting in each unit cell, as shown in Fig.~\ref{fig:flux}. This behavior is familiar from ``odd'' gauge theories~\cite{MoessnerZ2Gauge2001, MoessnerFFIM2001}.

\item
$\coho{b}(T_i, \mathbf{g})$ the ``anyonic spin orbit coupling'' in the $\hat{i}$-direction for ${\bf g} \in G$. The two factors in the second term, $\left(\mathcal{H}^1[G, \mathcal{A}]\right)^2$, in Eq.~(\ref{eq:trans_frac_decomp}) are characterized by the two gauge invariant quantities $\coho{b}(T_x , \mathbf{g})$ and $\coho{b}(T_y , \mathbf{g})$, respectively. These encode the possible non-commutativity between the localized operators for translations and on-site symmetries acting on anyons. In other words, moving an anyon in the $\hat{i}$-direction can change its symmetry charge under ${\bf g} \in G$. For this reason, we think of it as a sort of anyonic spin orbit coupling, which couples the internal quantum numbers of an anyon with its motion.

An operational definition of $\coho{b}(T_i , \mathbf{g})$ may be given as follows. Consider a state $\ket{\Psi_{a,\bar{a}}}$ with a pair of well separated anyons $a$ and $\bar{a}$ localized at $\mathbf{r}$ and $\mathbf{r}'$. We now move the anyon $a$ by one lattice spacing $\mathbf{e}_i$ while holding the other anyon fixed, and denote the resulting state by $\ket{\Psi'_{a, \bar{a}}}$.
By translational invariance, the local density matrices in a region centered at $a$ (but far away from $\bar{a}$) for $\ket{\Psi_{a,\bar{a}}}$ and $\ket{\Psi'_{a,\bar{a}}}$ should be identical, as they are related by a spatial translation. We then compute the $\mb{g}$ charge on the two states
\begin{equation}
\frac{\langle{\Psi_{a,\bar{a}}}|R_\mb{g}|{\Psi_{a,\bar{a}}}\rangle}{\langle{\Psi'_{a,\bar{a}}}|R_\mb{g}|{\Psi'_{a,\bar{a}}}\rangle} \approx \frac{\eta_a(T_i, \bf g)}{\eta_a(\mb{g}, T_i)} = M_{a,\cohosub{b}(T_i , \mathbf{g})}
.
\end{equation}
When $M_{a,\cohosub{b}(T_i , \mathbf{g})} \neq 1$, moving the anyon $a$ changes the $\bf g$-charge it carries. Thus, the local operators and processes which move anyons by one lattice spacing are necessarily ``charged'' under the on-site symmetry. Equivalently, one can view this as there being an integral symmetry charge (linear representation) per directed unit length of the quasiparticle string operators (Wilson lines), potentially with a different value of symmetry charge per directed length in the two different directions. A consequence of this property is that, when $M_{a,\cohosub{b}(T_i , \mathbf{g})} \neq 1$, it is not possible to transport a quasiparticle carrying topological charge $a$ in the $\hat{i}$-direction via a process that is both adiabatic and $\mathbf{g}$-symmetric.

Note that if $G$ is continuous and connected (e.g. $G=\mathrm{U}(1)$ or $G=\mathrm{SO}(3)$), the spin-orbit fractionalization $\mathcal{H}^1[G, \mathcal{A}]$ is always trivial.
This is because $\mathcal{H}^1[G, \mathcal{A}]$ consists of all group homomorphisms from $G$ to $\mathcal{A}$, so the image of the identity element $\openone$ in $G$ is the trivial anyon $I$ in $\mathcal{A}$.
When $G$ is continuous, continuity is also required of the cochain $\mathcal{C}^{n}(G,\mathcal{A})$. This forces the image of every element in the connected component of the identity of $G$ to equal $I$, since $\mathcal{A}$ is discrete. Thus, if $G$ is continuous and connected, $\mathcal{H}^1[G, \mathcal{A}]=\mathbb{Z}_1$ and $\coho{b}(T_i , \mathbf{g}) = I$. On the other hand, if $G$ has multiple connected components, e.g. all discrete nontrivial $G$ or $G=\mathrm{O}(2)=\mathrm{U}(1)\rtimes\mathbb{Z}_2$, it can have nontrivial anyonic spin orbit coupling.

\item
$\coho{w}(\mathbf{g},\mathbf{h})$ the on-site fractionalization class, for ${\bf g},{\bf h} \in G$. The third term, $\mathcal{H}^2[G, \mathcal{A}]$, in Eq.~(\ref{eq:trans_frac_decomp}) is just the previously discussed symmetry fractionalization for the on-site symmetries $G$.

\end{enumerate}

\subsection{Examples}

We now illustrate symmetry fractionalization for systems with translational symmetry using three concrete examples.

\subsubsection{The Laughlin states}
\label{sec:Laughlin_frac}

The first example is the $\nu=\frac{1}{m}$ Laughlin fractional quantum Hall states~\cite{Laughlin83}
with (magnetic) translational symmetry and $\mathrm{U}(1)$ charge conservation.
It has been known since Laughlin's seminal work that the fundamental quasiholes (quasiparticles corresponding to a $2 \pi$ vortex) in these states carry fractional electric charge $1/m$ (where the electron carries charge $-1$).

These states support $m$ anyon types, all of which are Abelian.~\footnote{The topological properties of fermionic quantum Hall states are more correctly described using a fermionic $\mathbb{Z}_{2}$ grading, in which topological charges related by fusion with the electron are grouped into doublets, e.g. the the (bosonic) vacuum and (fermionic) electron form a doublet. For our interests, we can simply ignore this, as it enters the symmetry fractionalization in a trivial way.} Let us associate the label $\phi$ with the anyon type of the fundamental quasihole, i.e. a vortex obtained by threading $2 \pi$ flux. The anyons are then given by $\mathcal{A} = \{ I, \phi, \phi^2, \cdots \phi^{m - 1} \}$, where the corresponding fusion rules are given by $\phi^{j} \times \phi^{k} = \phi^{[j+k]_{m}}$. In other words, $\mathcal{A} = \mathbb{Z}_{m}$. The mutual braiding statistics of anyons $\phi^{j}$ and $\phi^{k}$ (corresponding to taking one around another in a counterclockwise fashion) is given by the phase $M_{\phi^{j}, \phi^{k}} = e^{i 2 \pi jk / m}$.~\cite{Arovas84}

In the presence of a magnetic field, translations obey a magnetic algebra: $T_{-\delta y} T_{-\delta x}  T_{\delta y}  T_{\delta x} = \exp [- i \frac{\delta  x \delta y}{\ell_B^2 } \hat{N} ]$.
It would be interesting to study this more general symmetry group involving continuous translations. However, for our purposes, we consider only a discrete subset of the translations generated by the ``unit'' translations $T_x$ and $T_y$, which are chosen to define a unit cell with area $2 \pi \ell_B^2$. This corresponds to one flux quanta enclosed in each unit cell and $T_x T_y = T_y T_x$.
The corresponding symmetry group is thus $\mathcal{G} = \mathbb{Z}^2 \times \textrm{U}(1)$, as discussed. No anyons are permuted by symmetry, i.e. $\rho_{\bf g}(a) = a$).
The cohomology classes decomposing the classification of the possible symmetry fractionalization [as in Eq.~(\ref{eq:trans_frac_decomp})] for this topological order and symmetry group are $\mathcal{H}^2[\mathbb{Z}^{2}, \mathcal{A}] = \mathbb{Z}_{m}$, $\mathcal{H}^1[ \U (1), \mathcal{A}] = \mathbb{Z}_{1}$, and  $\mathcal{H}^2[ \U (1) , \mathcal{A}] = \mathbb{Z}_{m}$.

The symmetry fractionalization of the Laughlin states are described by
\begin{align}
\coho{w}(T_{x}^{j}, T_{y}^{k}) &= \phi^{[jk]_{m} } \\
\coho{w}(T_{y}^{j} , T_{x}^{k}) &= I \\
\coho{w}(T_{i}^{j}, \theta) &= \coho{w}( \theta , T_{i}^{j}) = I \\
\coho{w}(\theta, \theta') &= \phi^{ (\theta + \theta' - [\theta + \theta']_{2 \pi}) / 2 \pi }.
\end{align}
Here $\theta , \theta' \in [0, 2 \pi)$ label the elements of $\textrm{U}(1)$, and $[\,]_x$ denotes modulo $x$. Let us unravel the physics of this fractionalization class.

In order to determine the fractional $\U (1)$ charge of $\phi$, consider the localized action of $\U(1)$ rotations on $\phi$. From Eq.~(\ref{eq:fractional_charge_1}), we find that
$e^{i 2 \pi Q_{\phi}} = M_{\phi, \phi} = e^{i 2 \pi / m}$, indicating a fractional charge of $Q_\phi = \frac{1}{m}$.
Similarly, the anyons $\phi^k$ have fractional $\U (1)$ charge $Q_{\phi^{k}} = \frac{k}{m}$.

The anyonic flux per unit cell for this fractionalization class is $\coho{b}(T_y, T_x) = \overline{\phi} = \phi^{m-1}$.
The system is at filling $\nu = \frac{1}{m}$ electrons per unit cell.
Since anyons of topological charge $\overline{\phi}$ carry fractional $\U (1)$ charge $Q_{\overline{\phi}} = - \frac{1}{m}$, it appears as if the electric charge density of the quantum Hall liquid arises from the finite density of such anyons $\overline{\phi}$ (one per unit cell). Furthermore, it is known that if a quasiparticle of charge $Q$ adiabatically moves along a path that encloses a region of area $A$ (containing no quasiparticles), it will accumulate an Aharanov-Bohm phase of $\gamma = e^{-i Q A  / \ell_B^2}$.~\cite{Arovas84} Restricting to areas enclosing $n$ flux quanta, i.e. $A = n \cdot 2 \pi \ell_B^2$, we have $\gamma = e^{- i 2 \pi Q n}$. Local excitations have charge $Q \in \mathbb{Z}$ and so accumulate no phase from encircling $n$ flux quanta. However, a quasiparticle excitation with topological charge $\phi^k$ will accumulate an Aharonov-Bohm phase of $\gamma = e^{- i 2 \pi n k / m}$ when encircling $n$ flux quanta. On the other hand, the mutual statistics between $\coho{b}(T_y,T_x)=\overline{\phi}$ and $\phi^k$ is $M_{\overline{\phi}, \phi^k} = e^{- i 2 \pi k / m}$.
Thus, the Aharonov-Bohm phase $\gamma$ can equivalently be interpreted as the statistics between quasiparticles and the $n$ anyons per unit cell $\overline{\phi}$ within the encircled region (containing $n$ flux quanta).

\subsubsection{The $\mathbb{Z}_2$ Spin Liquid}

Our second example is translationally invariant $\mathbb{Z}_2$ spin liquids, with $G=\mathrm{SO}(3)$.
For clarity, we only explicitly write the cocycles for the $\mathbb{Z}_2\times\mathbb{Z}_2=\{\openone,X,Y,Z\}$ subgroup of $\mathrm{SO}(3)$, consisting of the three $\pi$ rotations about the $x$, $y$ and $z$ axes.
We will consider phases with topological order of the $\mathbb{Z}_2$ toric code type, i.e. $\mathbb{Z}_2$ spin-liquids.
Trial wavefunctions of such states can be constructed by Gutzwiller projection of non-interacting Schwinger fermion/boson mean-field ansatz~\cite{sstri, wen2002psg, Wang2007}.
We denote the bosonic spinons by $e$, the fermionic spinons by $\psi$, and the $\mathbb{Z}_2$ flux~\cite{Read1989a,Read1991z} (also known as a ``vison''~\cite{MoessnerZ2Gauge2001}) by $m$.
All anyons are Abelian, defining the group $\mathcal{A} = \{I, e, m, \psi\}$ with fusion algebra $\mathbb{Z}_2 \times \mathbb{Z}_2$.

In the familiar $\mathbb{Z}_2$ spin-liquid, the spinons $e$ and $\psi$  carry $S=1/2$ under SO$(3)$ while the vison $m$ does not carry spin.
This fixes the fractionalization of the on-site symmetry $G$ to be the following, in a gauge where $\coho{w}(\openone , \mb{g}) = I$:
\begin{equation}
\begin{gathered}
\coho{w}(X,X)=\coho{w}(Y,Y)=\coho{w}(Z,Z)=m \\
\coho{b}(X,Y)=\coho{b}(Y,Z)=\coho{b}(X,Z)=m
\end{gathered}
\label{}
\end{equation}
To see why, consider, for example, the local operation of two $\pi$ rotations about the $x$ axis for some anyon $a$. The corresponding projective phase is
\begin{align}
\eta_a (X, X) = M_{a, \cohosub{w}(X,X)} = \pm 1
,
\end{align}
where $+1$ corresponds to $a$ carrying trivial spin and $- 1$ corresponds to $a$ carrying spin $S=1/2$. Since $M_{e, m} =M_{\psi, m} = -1$ and $M_{m, m} =1$, and the theory is modular, the unique choice consistent with our assignment of spin is $\coho{w}(X,X) = m$.
Similar arguments yield the other terms of the fractionalization class.

If we consider any $\U (1)$ subgroup of the $\mathrm{SO}(3)$, for example rotations about the $x$ axis, we can determine the corresponding fractional charge carried by quasiparticles. From Eq.~(\ref{eq:fractional_charge_1}), we find that $e^{i 2 \pi Q_{a}} = M_{a, m}$, which indicates that the quasiparticle types have rotational $\U (1)$ fractional charge values of $Q_{I}=Q_{m} = 0$ and $Q_{e} = Q_{\psi} = \frac{1}{2}$. These, of course, simply match the corresponding $\mathrm{SO}(3)$ projective representations (integer or half-integer spin values) carried by the quasiparticles, in this case.

For the translational symmetry, the fact that there is a slave particle sitting at each site in the mean-field ansatz leads to the following fractionalization
\begin{equation}
\coho{b}(T_y, T_x)=e \text{ or } \psi.
\label{}
\end{equation}
Which of these two options is actually realized depends on whether one uses a Schwinger boson or fermion construction, as well as the chosen mean-field ansatz ($0$ or $\pi$ flux per plaquette in the band structure). Note that in the PSG language, $\coho{b}(T_y, T_x)$ determines the PSG of the anyons via $\frac{\eta_{a} (T_y, T_x)}{\eta_{a} (T_x, T_y)} = M_{a, \cohosub{b}(T_y, T_x)} = \pm 1$, for the phase factor in Eq.~(\ref{eq:local_commutator}).

Since $G$ is connected, the spin-orbit fractionalization is trivial, $\coho{b}(T_i , \mb{g}) = I$.

\subsubsection{Unconventional symmetry for the toric code}
\label{sec:toriccode}

An example which realizes a nontrivial anyonic spin-orbit fractionalization class is due to M. Hermele~\cite{Hermele}.
We begin with Kitaev's $\mathbb{Z}_2$ toric code model~\cite{Kitaev97}, where qubits live on the bonds of a square lattice. The Hamiltonian of the model is
\begin{equation}
H=-\sum_v A_v -\sum_p B_p.
\label{}
\end{equation}
Here, $A_v=\prod_{j\in v}\sigma^x_j$ and $B_p=\prod_{j\in p}\sigma^z_j$, where the products are over the four bonds meeting at vertex $v$ or forming the plaquette $p$, respectively. Notice that there are two qubits per unit cell.

\begin{figure}[t!]
	\centering
	\includegraphics[width=0.9\columnwidth]{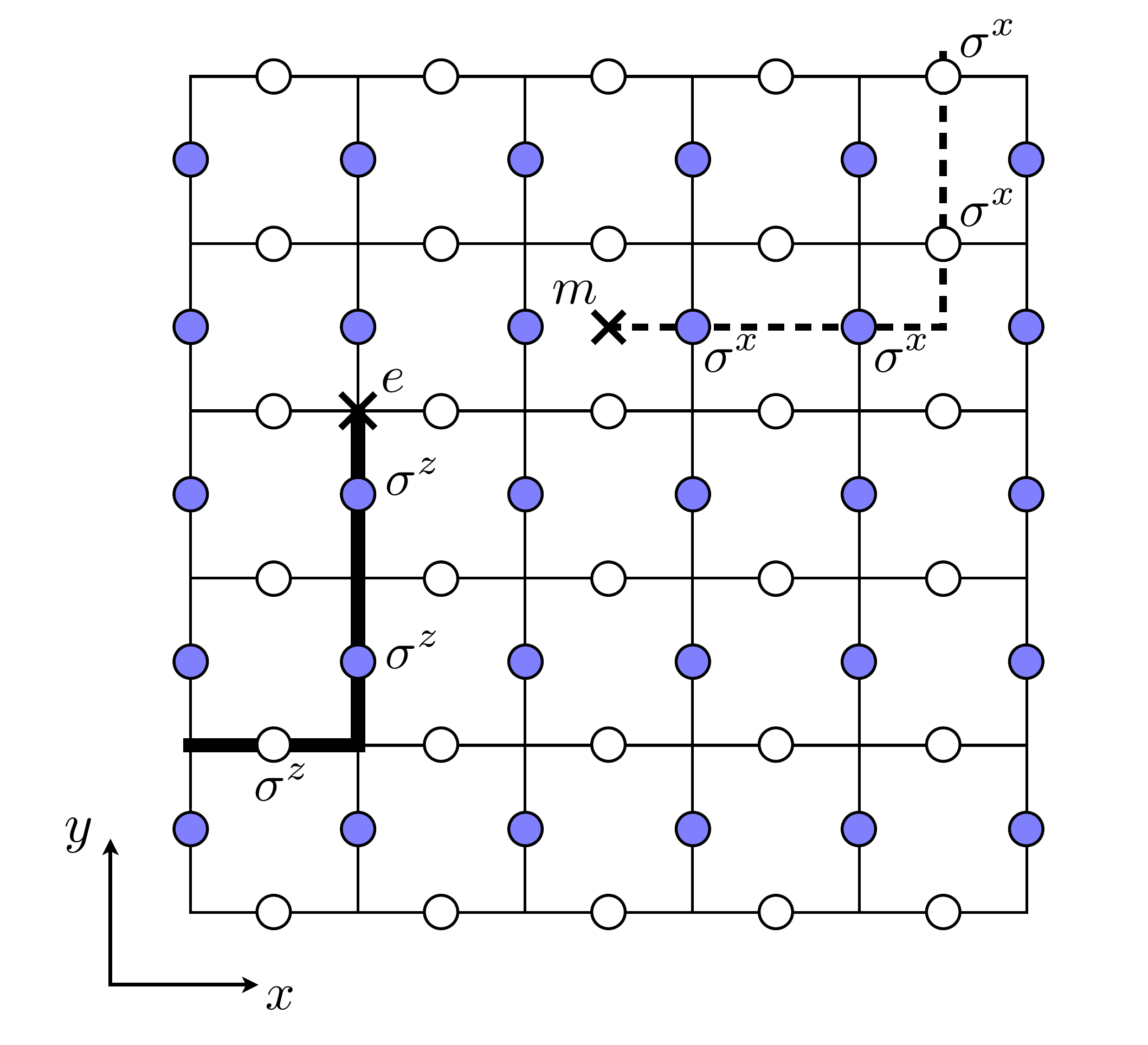}
	\caption{The toric code with an unconventional $\mathbb{Z}_2\times\mathbb{Z}_2$ symmetry that exhibits ``anyonic spin-orbit coupling.'' Dots represent qubits on each bond; the dark blue ones transform as projective representations of the global $\mathbb{Z}_2\times\mathbb{Z}_2$ symmetry, while the light dots do not.
The string operators which move the $e$-particles vertically (thick black) and $m$-particles horizontally (dashed black) involve operators on vertical links, so are charged under the symmetry operations. }
	\label{fig:tc}
\end{figure}

We now define a global $G = \mathbb{Z}_2\times\mathbb{Z}_2 = \{\openone, X, Z, XZ \} $ symmetry with a peculiar representation that only acts on the vertical ($y$) bonds. Let the set of all vertical bonds in the system be denoted $C_{y}$ and define the operators
\begin{equation}
R_X=\prod_{j \in C_{y}}\sigma_j^x, \quad R_Z=\prod_{j \in C_{y}}\sigma_j^z
.
\label{}
\end{equation}
It is straightforward to check that $[H, R_X]=[H,R_Z]=0$, so these operators are symmetries of the system.
However, this peculiar choice of operators forms a projective representation of $\mathbb{Z}_2\times\mathbb{Z}_2$, since only the vertical links transform under the symmetry. In particular, we have $R_X R_Z = (-1)^{L_x L_y} R_Z R_X$, where $L_x$ and $L_y$ is the linear size of the system in the $\hat{x}$ and $\hat{y}$ directions, respectively.
Moreover, each unit cell locally transforms under the $\mathbb{Z}_2\times\mathbb{Z}_2$ symmetry as a projective representation, since there is one vertical bond in each unit cell.

As usual, the $e$ and $m$ excitations correspond to violations of the vertex terms (i.e. $A_v$) and plaquette terms (i.e. $B_p$), respectively.
The usual definition of the toric-code string operators implies neither anyon carries a projective representation of $G$, so $\coho{w}({\bf g}, {\bf h}) = I$ for ${\bf g},{\bf h} \in G$. Furthermore, our choice of Hamiltonian implies trivial anyonic flux per unit cell, so $\coho{b}(T_y, T_x) = I$.

On the other hand, the interplay of $G$ and translational symmetries is more interesting and gives nontrivial anyonic spin orbit coupling.
We can move an $e$ anyon by applying a string of $\sigma^z$ operators along a path in the lattice. Similarly, we can move an $m$ anyon by applying a string of $\sigma^x$ along a path in the dual lattice. As can be seen from Fig.~\ref{fig:tc}, when we move $e$ along the $\hat{y}$-direction by one lattice spacing, the extra $\sigma^z$ that extends the string operator does not commute with $R_X$. Nothing interesting happens, however, when we move $m$ along the $\hat{y}$-direction. Thus, we find
\begin{equation}
\coho{b}(T_y , X)=\coho{b}(T_y , XZ)=m, \quad \coho{b}(T_y , Z)=I
.
\end{equation}
Similarly, when we move $m$ along the $\hat{x}$-direction, the string operator is extended by an extra $\sigma^x$ factor for each step, which implies
\begin{equation}
\coho{b}(T_x , X)=I, \quad \coho{b}(T_x , Z)=\coho{b}(T_x , XZ)=e
.
\end{equation}
An interesting feature of the model is that the definition of the symmetry breaks the $C_4$ lattice rotation symmetry. We will see later that this is unavoidable in such cases.

We can interpret the anyonic spin orbit coupling another way. Recall that an $e$-string operator applies $\prod_{j \in P}\sigma_j^z$ to the sites contained in a path $P$ running along links between vertices, and that such an operator creates $e$-quasiparticles at the endpoints of an open path $P$, while it commutes with the Hamiltonian away from the endpoints (or it commutes everywhere if $P$ is a closed path). Moreover, applying an open $e$-string operator with one of the path's endpoints at the site of an $e$-quasiparticle will move that quasiparticle to the position of the other endpoint of $P$.
The global symmetry operator $R_Z$ is precisely the product of densely-packed $e$-string operators running in the $\hat{y}$-direction.
This is why $\coho{b}(T_x , Z)=e$; growing the lattice by one unit of $T_x$ adds one more of these $e$-string operators, so $Z T_x$ and $T_x Z$ differ by the addition of one $e$-string.
Similarly, $R_X$ can be interpreted as the product of densely-packed $m$-strings running in the $\hat{x}$-direction.

\section{Symmetry defects and $\mathcal{H}^4$ obstruction}
\label{sec:Symmetry_defects}

Not every fractionalization class $[ \coho{w} ]$ of a given topological phase and symmetry can be realizable in a strictly 2D system, as some of them may be anomalous.
For an on-site,  unitary symmetry $G$, one can determine whether $[ \coho{w} ]$ is anomalous by considering the inclusion of symmetry defects, which are extrinsic objects carrying symmetry fluxes. When anyons are transported around the defects they are transformed by the corresponding local group actions.
There is an algebraic theory of the fusion and braiding of defects~\cite{SET} captured by topological data similar to the $F$-symbols and $R$-symbols of anyon models.
Importantly, there are consistency conditions that this topological data must satisfy.
It was shown mathematically in \Ref{ENO2009} that the sufficient and necessary condition for the solvability of the consistency equations is that a certain object $[\mathscr{O}]$ be the trivial cohomology class in $\coh{4}{G}$.
If this object is nontrivial, it indicates the presence of an obstruction, meaning that a consistent 2D theory of symmetry defects does not exist. This can be interpreted as an indicator of an anomaly, since symmetry defects can always be introduced explicitly for any 2D theory realized in a lattice model. Given the topological order (i.e. the $F$ and $R$ symbols of the anyons), the symmetry group $G$, the symmetry action $\rho$ on the topological degrees of freedom, and the symmetry fractionalization class $[ \coho{w} ]$, the corresponding obstruction class $[\mathscr{O}] \in \coh{4}{G}$ is uniquely determined and there is a concrete procedure for computing it.

We now briefly review the theory of defects in a topological phase $\mathcal{C}$ with symmetry group $G$ that does not permute anyon types.~\cite{SET}
We first define an Abelian group $\mathcal{A}$ whose elements are the Abelian topological charges of $\mathcal{C}$ with multiplication given by the fusion rules.
Given a fractionalization class $[\coho{w}] \in \mathcal{H}^2[G, \mathcal{A}]$ of the symmetry $G$ acting on the topological phase $\mathcal{C}$, we choose a representative $\coho{w}$ of the class. One can then proceed to write down the different types of $G$ defects and their fusion rules.  As established in \Ref{SET}, for each $\mathbf{g}\in G$ there are $|\mathcal{C}|$ topologically distinct types of $\mathbf{g}$-defects (i.e. defects that carry a symmetry flux of $\mathbf{g}$), and they are all related to each other by fusion with anyons. We can label the defect types by arbitrarily picking one Abelian defect type for each ${\bf g}$ that we label $I_{\bf g}$ and treat as a reference point within each ${\bf g}$ sector. It follows that all other ${\bf g}$-defect types can be labeled as $a_{\bf g}=a \times I_{\bf g}$, where $a \in \mathcal{C}$ are the labels of all the anyon types. The fusion rules then read
\begin{equation}
a_\mb{g}\times b_\mb{h}= \sum_{c \in \mathcal{C}} N_{ab}^{c} \left[c\times\coho{w}(\mathbf{g},\mathbf{h}) \right]_\mb{gh},
\label{eqn:defect-fusion}
\end{equation}
where $N_{ab}^{c}$ are the fusion coefficients of the anyons in $\mathcal{C}$.

We notice that the fusion rules of defects also encodes the symmetry actions on defect labels by the so-called $G$-crossed braiding.  Let us denote the $\mb{h}\in G$ action on the defect charge $a_\mb{g}$ as $\rho_\mb{h}(a_\mb{g})$. The action can be physically realized by transporting $a_\mb{g}$ across the branch cut of a $\mb{h}$ defect, say $I_{\bf h}$, which amounts to a braiding exchange. Since the total topological charge in the region containing both defects cannot change, and the relative positions of the defects is interchanged, we must have
\begin{equation}
I_\mb{h}\times a_\mb{g}=\rho_\mb{h}(a_\mb{g})\times I_\mb{h}.
	\label{}
\end{equation}
This implies
\begin{equation}
\rho_\mb{h}(a_\mb{g})= [\coho{w}(\mb{h,g})\times \ol{\coho{w}(\mb{hgh^{-1},h})} \times a]_{\mb{hgh^{-1}}}.
\label{eqn:action_on_defect}
\end{equation}
In particular, if $\mb{gh}=\mb{hg}$ and $\coho{w}(\mb{g,h})\neq \coho{w}(\mb{h,g})$, then the two different types of $\mb{g}$ defects, $a_\mb{g}$ and $\coho{b}(\mb{h,g})\times a_\mb{g}$, are related to each other by the symmetry action of $\mb{h}$. As such, $a_\mb{g}$ and $[ \coho{b}(\mb{h,g})\times a]_\mb{g}$ must be energetically degenerate.

Given the topological data of the anyon model (i.e. $F$ and $R$ symbols), the symmetry fractionalization class $[\coho{w}]$, and the fusion rules in Eq. \eqref{eqn:defect-fusion}, one can then solve the $G$-crossed consistency equations for the topological data (i.e. $F$, $R$, $U$, and $\eta$ symbols) of the $G$-crossed defect theory. In doing so, one finds that consistent solutions exist if and only if the obstruction class $[\mathscr{O}] = [1]$ is trivial, where (a representative of) the obstruction class is found to be
\begin{widetext}
\begin{equation}
\label{eq:obstruction}
\mathscr{O} ({\bf g},{\bf h},{\bf k},{\bf l})
=
\frac{ F^{\cohosub{w}({\bf g},{\bf h}) \cohosub{w}({\bf k},{\bf l}) \cohosub{w}({\bf gh},{\bf kl})}
F^{ \cohosub{w}({\bf k},{\bf l})  \cohosub{w}({\bf h},{\bf kl})  \cohosub{w}({\bf g},{\bf hkl}) }
F^{\cohosub{w}({\bf h},{\bf k})  \cohosub{w}({\bf g},{\bf hk})  \cohosub{w}({\bf ghk},{\bf l}) }}
{F^{\cohosub{w}({\bf g},{\bf h})  \cohosub{w}({\bf gh},{\bf k})  \cohosub{w}({\bf ghk},{\bf l}) }
F^{\cohosub{w}({\bf k},{\bf l})  \cohosub{w}({\bf g},{\bf h})  \cohosub{w}({\bf gh},{\bf kl}) }
F^{ \cohosub{w}({\bf h},{\bf k})  \cohosub{w}({\bf hk},{\bf l})  \cohosub{w}({\bf g},{\bf hkl}) }}
R^{\cohosub{w}({\bf g},{\bf h}), \cohosub{w}({\bf k},{\bf l})}
,
\end{equation}
\end{widetext}
which can be shown to be a 4-cocycle. (In this paper, the braiding $R$-symbols are written with two superscript topological charge labels, in contrast to the global symmetry operators $R_{\bf g}$ which are written with a subscript group element.) We emphasize that Eq.~(\ref{eq:obstruction}) only depends on the symmetry fractionalization class, and the $F$ and $R$ symbols of the Abelian anyons in $\mathcal{C}$. For a derivation of the result, we refer the readers to Refs.~\onlinecite{chen2014,SET}.

In the case of topological orders in which the (non-vanishing) $F$-symbols involving only Abelian anyons can be set to $1$ by a choice of gauge (using vertex basis gauge freedom), the expression for the obstruction significantly simplifies to
\begin{equation}
\label{eq:obstruction_simplified}
[\mathscr{O} ({\bf g},{\bf h},{\bf k},{\bf l})] =[R^{\cohosub{w}({\bf g},{\bf h}), \cohosub{w}({\bf k},{\bf l})}]
.
\end{equation}
In other words, $\mathscr{O} ({\bf g},{\bf h},{\bf k},{\bf l}) =R^{\cohosub{w}({\bf g},{\bf h}), \cohosub{w}({\bf k},{\bf l})}$ is a representative of the obstruction class for such topological orders.

Generally speaking, if a $d$-dimensional system exhibits gauge anomaly, it can only exist at the boundary of a $(d+1)$-dimensional system, where there is an ``anomaly inflow'' from the bulk to cancel the gauge anomaly on the boundary.~\cite{Callan1985} This also shows that the bulk cannot be adiabatically connected to a trivial product state while the symmetry is preserved and it is therefore a nontrivial SPT phase. Recently, it has been shown that in low dimensions ($d\leq 3$), SPT phases with a unitary symmetry $G$ are completely classified by $\coh{d+1}{G}$. In particular, 3D SPT phases are classified by $\coh{4}{G}$, which is the same cohomology group to which the obstruction classes belong. This is not a coincidence. In fact, the obstruction class precisely determines the bulk SPT phase that is needed to cancel the gauge anomaly of the boundary SET phase, a novel kind of bulk-boundary correspondence. This bulk-boundary correspondence has been established for Abelian topological phases with on-site symmetries that do not permute the particle types~\cite{chen2014}, and is expected to hold more generally, as well.

\section{Incorporating translational symmetry in SET phases}
\label{sec:trans_obstruction}

We would like to consider the theory of defects and obstructions for 2D SET phases in which there is translational symmetry, as well as on-site symmetry, i.e. with symmetry group $\mathcal{G} = \mathbb{Z}^{2} \times G$. Strictly speaking, the theory of the symmetry defects described by $G$-crossed MTCs in Ref.~\onlinecite{SET} is understood to apply to on-site symmetries $G$. For translational symmetry, we understand how symmetry fractionalization occurs~\cite{SET}, as discussed in Sec.~\ref{sec:set_translation}. However, it is not clear, in general, how to properly define and interpret the algebraic theory of translational symmetry defects (i.e. dislocations or ``movons''). The na\"ive application of the $\mathcal{G}$-crossed consistency conditions of Ref.~\onlinecite{SET} to examples of purely 2D systems with projective representations of the on-site symmetry per unit cell, such as certain spin liquid states, gives rise to nontrivial obstruction classes $[\mathscr{O}]$. This means that, in contrast to the case of purely on-site symmetries, the inclusion of translational symmetry for systems with projective representations of the on-site symmetry per unit cell requires a refined interpretation of these obstruction classes. In the following, we will develop such a refined interpretation.

In order to make progress, we develop our understanding from the following conjecture:

\emph{Given a 3D SPT with translational and on-site symmetry $\mathbb{Z}^{3}_{\text{trans}} \times G_{\text{int}}$ and linear representations of $G_{\text{int}}$ per unit cell, if the boundary is gapped and symmetric, it manifests a 2D boundary SET with symmetry $\mathcal{G} = \mathbb{Z}^{2}_{\text{trans}} \times G_{\text{int}}$ whose defects are described by the $\mathcal{G}$-crossed theory, as described in Ref.~\onlinecite{SET}, with anomaly matching between the 3D bulk SPT indices and the 2D boundary SET obstruction class.}

In order to understand the statement of anomaly matching in this conjecture, we first notice that the cohomology group $\coh{4}{\mathbb{Z}^2 \times G}$, to which the 2D SET
obstruction class $[\mathscr{O}]$ belongs, can be decomposed using the K\"unneth formula to give
\begin{align}
\coh{4}{\mathbb{Z}^2 \times G} =  \coh{4}{G}  & \times  \left(\coh{3}{G} \right)^{2}  \notag \\
& \times \coh{2}{G}
.
\end{align}
The elements of this cohomology group can be similarly decomposed using the slant product (as we have done before).
In particular, we can write the 2D SET obstruction class in this decomposition as
$[\mathscr{O}] = \{ [o] , [o_{x}], [o_{y}] , [o_{yx}]  \}$, where (for a given representative $\mathscr{O}$ of the cohomology class) we have defined
\begin{eqnarray}
o({\bf g}, {\bf h}, {\bf k}, {\bf l}) & =& \mathscr{O}({\bf g}, {\bf h}, {\bf k}, {\bf l})|_{{\bf g}, {\bf h}, {\bf k}, {\bf l} \in G} \\
o_{x}({\bf g}, {\bf h}, {\bf k}) &=&  \slant_{T_x}\mathscr{O}({\bf g}, {\bf h}, {\bf k})|_{{\bf g}, {\bf h}, {\bf k} \in G} \\
o_{y}({\bf g}, {\bf h}, {\bf k}) &=&  \slant_{T_y}\mathscr{O}({\bf g}, {\bf h}, {\bf k})|_{{\bf g}, {\bf h}, {\bf k} \in G} \\
o_{yx}({\bf g}, {\bf h}) &=&  \slant_{T_y} \slant_{T_x} \mathscr{O}({\bf g}, {\bf h})|_{{\bf g}, {\bf h} \in G}
.
\end{eqnarray}

We can now precisely state the anomaly matching condition between the 3D bulk SPT phase characterized by $[\Omega] = \{ [\nu], [\nu_i], [\nu_{ij}], [\nu_{ijk}]\} \in \coh{4}{\mathbb{Z}^{3} \times G}$ and the 2D boundary SET phase with corresponding obstruction class  $[\mathscr{O}] = \{ [o] , [o_{x}], [o_{y}] , [o_{yx}]  \} \in \coh{4}{\mathbb{Z}^{2} \times G}$ to be the requirement that
\begin{equation}
[\nu]= [o] , \quad [\nu_{x}]=[o_{x}], \quad [\nu_{y}]=[o_{y}], \quad [\nu_{yx}]=[o_{yx}]
.
\end{equation}
This assumes the 2D boundary in question is along the $xy$-plane. The remaining terms $[\nu_{z}]$, $[\nu_{xz}]$, $[\nu_{yz}]$, and $[\nu_{xyz}]$ characterizing the 3D SPT phase are not involved in matching the bulk and surface states of boundaries along the $xy$-plane, though they will be involved in constraints associated with boundaries in the other directions.

It is worth considering the different terms in the K\"unneth decomposition of the SET obstruction in more detail.
The term $[o] \in  \coh{4}{G}$ is simply the obstruction class associated with the on-site symmetries $G$.
The terms $[o_{x}], [o_{y}] \in \coh{3}{G}$ are associated with the interplay between the on-site symmetry fractionalization and
the anyonic spin orbit coupling in the $\hat{x}$ and $\hat{y}$ directions, respectively.
The term $[o_{yx}] \in \coh{2}{G}$ is associated with the interplay between the on-site symmetry fractionalization, the anyonic spin orbit coupling, and the anyonic flux per unit cell. (We note that $[o_{xy}]=[o_{yx}]^{-1}$.)

The $[o_{x}]$ and $[o_{y}]$ components of the obstruction class are automatically trivial if either the on-site symmetry fractionalization class or the anyonic spin orbit coupling class is trivial.~\footnote{This statement is easy to verify when the obstruction class is given by Eq.~(\ref{eq:obstruction_simplified}), since, in this case, we have $o_{i}({\bf g}, {\bf h}, {\bf k}) =  R^{\coho{w}({\bf g}, {\bf h}), \coho{b}(\mb{k}, T_i)} R^{\coho{b}(\mb{g}, T_i), \coho{w}({\bf h},{\bf k})}$.}
In particular, when $G$ is continuous and connected, e.g. $G= \U (1)$ or $\mathrm{SO}(3)$, the anyonic spin orbit coupling is necessarily trivial, and, hence, so is $[o_{i}]$. Conversely, nontrivial $[o_i]$ can only arise for a non-connected (e.g. discrete) on-site symmetry group $G$. Examples of such anomalous SET phases occurring at the boundary of a 3D system formed by stacking 2D topological insulators has been considered in Ref.~\onlinecite{Mross_arxiv}.

Next, we make the following observations regarding the properties of SPT phases. The boundary structure of a nontrivial $d$-dimensional strong SPT phase cannot be physically realized in a local manner as a purely $(d-1)$-dimensional system for $d \ge 2$, i.e. it cannot be realized without the corresponding $d$ dimensional bulk. On the other hand, the boundary structure of 0D and 1D strong SPT phases can be physically realized in a local manner without their corresponding $d$ dimensional bulk system. In particular, the boundary of a 0D SPT phase is trivial, while the boundary of a 1D SPT phase characterized by the index $[\omega] \in \coh{2}{G}$ gives rise to an emergent projective representation $[\omega]$ of the on-site symmetry group $G$. Projective representations can obviously exist as independent objects (i.e. without the support of a corresponding 1D bulk system) within systems of any dimensionality.

In the context of a 3D SPT with symmetry $\mathbb{Z}^{3}_{\text{trans}} \times G_{\text{int}}$, these observations imply the following properties. 1) The boundary structure corresponding to nontrivial $[\nu]$ requires the 3D bulk to exist, so it cannot be physically realized in a purely 2D system in a local manner that preserves symmetry. 2) The boundary structure corresponding to nontrivial $[\nu_i]$ requires the array of 2D SPT phases (stacked in the 3D bulk, as explained in Sec.~\ref{sec:Weak_SPT}) to exist, so it cannot be physically realized in a purely 2D system in a local manner that preserves symmetry. 3) The boundary structure corresponding to nontrivial $[\nu_{yx}]$ can be physically realized in a purely 2D system. Specifically, this class manifests emergent degrees of freedom on the boundary of the 3D system, which transform under $G$ such that each 2D unit cell on the surface transforms as the projective representation $[\nu_{yx}]$ of $G$. This structure can instead be physically realized in a purely 2D system by constructing it from physical degrees of freedom that transform under $G$ in such a way that each unit cell transforms as the projective representation $[\nu_{yx}]$ of $G$.

Finally, we make the observation that any purely 2D SET phase (i.e. one that can be physically realized in a strictly 2D system in a local manner) can be realized on the boundary of a 3D SPT.  Combining these observations with the conjectured bulk-boundary correspondence leads to the following understanding of the 2D SET obstruction class when translational symmetry is included:

\emph{The obstruction class $[\mathscr{O}] = \{ [o] , [o_{x}], [o_{y}] , [o_{yx}]  \}$ of a purely 2D SET phase with symmetry $\mathcal{G} = \mathbb{Z}^{2}_{\text{trans}} \times G_{\text{int}}$ must have trivial components $[o]=[o_{x}]=[o_{y}]=[1]$, while the component $[o_{yx}]$ must equal the projective representation of $G$ per unit cell of the 2D system. If each unit cell of the system transforms locally as a linear representation under the on-site symmetries $G$, then we must have $[o_{yx}]=[1]$.}

If these conditions on $[\mathscr{O}]$ of the 2D SET phase are not satisfied, i.e. if $[o]$, $[o_{x}]$, or $[o_{y}]$ are nontrivial or if $[o_{yx}]$ is not equal to the projective representation of $G$ per unit cell, then the SET cannot be physically realized as a strictly 2D system. We emphasize that this gives an interpretation of the obstruction class that differs from the interpretation when there is only on-site symmetry. In particular, a nontrivial obstruction class component $[o_{yx}]$ does not rule out physical realization of the system in purely 2D, but rather requires that each 2D unit cell transforms as the projective representation $[o_{yx}]$. This result provides constraints on the allowed SET order of a given 2D system.

\subsection{Projective representation per unit cell}
\label{sec:projrep}

We are finally prepared to analyze LSM-type constraints for a 2D system with a projective representation per unit cell, such as a $S=1/2$ magnet, by imposing the constraint that $[o_{yx}]$ must be equal to the projective representation per unit cell for purely 2D SET phases. For this, we consider the properties of $[o_{yx}]$ in more detail and examine several special cases.

In the following, we will focus on topological orders in which the $F$-symbols involving only Abelian anyons can be set to $1$ by an appropriate gauge choice, and symmetry $\mathcal{G} = \mathbb{Z}^{2} \times G$ that does not permute the anyon types. In this case, the obstruction class is given by Eq.~(\ref{eq:obstruction_simplified}), and we, thus, have
\begin{equation}
\label{eq:oxy_simplified}
o_{yx}({\bf g}, {\bf h}) =  M_{\cohosub{w}(\mb{g,h}), \cohosub{b}(T_y , T_x)}\frac{R^{ \cohosub{b}(T_x , \mb{g}), \cohosub{b}(T_y , \mb{h})}} {R^{\cohosub{b}(T_y , \mb{g}), \cohosub{b}(T_x , \mb{h})}}
,
\end{equation}
for ${\bf g}, {\bf h} \in G$.

For anyon models with nontrivial $F$-symbols, Eq.~\eqref{eq:oxy_simplified} acquires an additional contribution $\tilde{o}_{yx}\{F \}$, which is a combination of several $F$-symbols (see Appendix~\ref{app:F} for details). However, we have verified that if either a) $G$ is Abelian, b) $G$ is connected and continuous, or c) there is a point group symmetry relating $T_x$ to $T_y$, then $[\tilde{o}_{yx}\{F \}] = [1]$. In other words, for these cases, the modification of $o_{yx}$ is simply a 2-coboundary, and we can still use Eq.~\eqref{eq:oxy_simplified} as a representative of the cohomology class $[o_{yx}]$.

It is sometimes useful to consider quantities constructed from the cocycles, such as
\begin{align}
\frac{o_{yx}(\mathbf{g,h})}{o_{yx}(\mathbf{h,g})} =  M_{\cohosub{b}(\mathbf{g,h}),\cohosub{b}(T_y,T_x)} \frac{M_{\cohosub{b}(T_{x}, {\bf g}),\cohosub{b}(T_y,{\bf h})}}{M_{\cohosub{b}(T_{y}, {\bf g}),\cohosub{b}(T_x,{\bf h})}}
,
\end{align}
which is gauge invariant when ${\bf gh} = {\bf hg}$, and thus may readily indicate when the cohomology class is nontrivial.

\subsubsection{Trivial topological order}

A gapped topological phase with trivial topological order, for example a featureless paramagnet or insulator, will necessarily have a vanishing obstruction class $[\mathscr{O}]=[1]$. Consequently, the above interpretation of the obstruction class implies that it is forbidden for such phases to occur in a model with a nontrivial projective representation of $G$ per unit cell. This is the result of the (higher dimensional) LSM theorem, but now generalized to the case of arbitrary on-site unitary symmetries $G$, including discrete symmetries.

\subsubsection{Trivial contribution to $[o_{yx}]$ from anyonic spin orbit coupling}
\label{sec:trivialSOC}

We now consider the case when the anyonic spin orbit coupling symmetry fractionalization class gives trivial contribution to $[o_{yx}]$, so that Eq.~\eqref{eq:oxy_simplified} reduces to
\begin{equation}
\label{eqn:spinonexists}
[o_{yx}({\bf g},{\bf h})] = [M_{\cohosub{w}(\mathbf{g,h}),\cohosub{b}(T_y , T_x)}] = [\eta_{\cohosub{b}(T_y , T_x)}(\mb{g,h})]
.
\end{equation}
There are a number of scenarios in which this may occur, including: a) the anyonic spin orbit coupling class is trivial, i.e. $\coho{b}( T_{i}, \mathbf{g}) = I$ for $i=x,y$; as previously mentioned, this is automatically the case when the on-site symmetry group $G$ is continuous and connected, and b) the anyonic spin orbit coupling classes are equal in the two directions, i.e. $\coho{b}( T_{x}, \mathbf{g}) = \coho{b}( T_{y}, \mathbf{g})$ for all ${\bf g}$; this will occur when the system preserves a point group symmetry that relates the $\hat{x}$ and $\hat{y}$ directions, for example $C_3$, $C_4$, $C_6$ lattice rotations, or reflections that interchange $x$ and $y$.

Recalling the discussion from Sec.~\ref{sec:set_translation}, we know that $[\eta_{\cohosub{b}(T_y , T_x)}(\mb{g,h})]$ is the projective representation of $G$ carried by the anyon $\coho{b}(T_y , T_x)$, which is the anyonic flux per unit cell. In other words, Eq.~(\ref{eqn:spinonexists}) implies that, for purely 2D SET phases with such anyonic spin orbit coupling, the projective representation of $G$ carried by the anyonic flux per unit cell must be precisely equivalent to the projective representation describing the local symmetry action of each unit cell (i.e. of the microscopic degrees of freedom).
In this sense, the anyonic flux per unit cell, $\coho{b}(T_y,T_x)$, is a spinon which screens the projective representation per unit cell.
Indeed, this is precisely how the $S=1/2$ parton construction works, as there is a density of one parton (spinon) per $S=1/2$ moment.

We may also be interested in quasiparticles that play a role similar to that of the vison in a $\mathbb{Z}_{2}$ spin liquid. For this, we consider
\begin{align}
\frac{o_{yx}(\mathbf{g,h})}{o_{yx}(\mathbf{h,g})} =  M_{\cohosub{b}(\mathbf{g,h}),\cohosub{b}(T_y,T_x)} = \frac{\eta_{\cohosubsub{b}(\mb{g,h})}(T_y,T_x)}{\eta_{\cohosubsub{b}(\mb{g,h})}(T_x,T_y)}
,
\end{align}
which is gauge invariant when ${\bf gh} = {\bf hg}$. When this quantity is nontrivial (i.e. not equal to 1), the anyon $\coho{b}(\mathbf{g,h})$ has nontrivial mutual statistics with the anyonic flux per unit cell, i.e. the spinon, and hence transforms projectively under $T_y^{-1} T_x^{-1} T_y T_x$. In this sense, $\coho{b}(\mathbf{g,h})$ is similar to a vison excitation. In particular, for the $\mathbb{Z}_2$ spin liquid, one finds
\begin{equation}
\frac{o_{yx}(X,Z)}{o_{yx}(Z,X)} = M_{\cohosub{b}(X,Z),\cohosub{b}(T_y,T_x)}= M_{m, e / \psi} = -1
,
\end{equation}
where $\coho{b}(X,Z) = m$ is the vison.

We can understand Eq.~(\ref{eqn:spinonexists}) when the anyonic spin orbit coupling is trivial through the following heuristic flux-insertion type argument. First, create a ${\bf g}$-defect and $\bar{{\bf g}}$-defect pair from vacuum, say carrying labels $I_\mb{g}$ and $\overline{ I_{\mb{g}}}$, respectively. Next, adiabatically transport the $I_\mb{g}$ defect along a path enclosing one unit cell in a counterclockwise fashion. Finally, annihilate the pair of defects into vacuum, returning the system to the ground state.
This process can be thought of as applying the $\mb{g}$ symmetry transformation locally to a unit cell. We schematically denote the ${\bf g}$-action on the local degrees of freedom by $U_\mb{g}$. Such local actions form a projective representation of $G$, with multiplication given by
\begin{equation}
U_\mb{g}U_\mb{h}= o_{yx}(\mb{g,h})U_\mb{gh}.
\label{eqn:flux-insertion-proj-rep}
\end{equation}
In other words, the successive applications of this procedure with $I_{\bf g}$ and $I_{\bf h}$ defects is related to an application of the procedure with an $I_{\bf gh}$ by the phase $o_{yx}(\mb{g,h})$.

On the other hand,  we can arrange the processes involved in successively applying the local actions for ${\bf g}$ and ${\bf h}$ in a slightly different order. First, create both the $I_{\bf g}$ and $\overline{I_{\bf g}}$ pair from vacuum and also the $I_{\bf h}$ and $\overline{I_{\bf h}}$ pair from vacuum. Next, fuse $I_\mb{g}$ and $I_\mb{h}$, which results in $[\coho{w}(\mb{g,h})]_{\bf g h} = \coho{w}(\mb{g,h}) \times I_{\mb{g h}}$. Then, adiabatically transport $[\coho{w}(\mb{g,h})]_{\bf g h}$ around a unit cell in the counterclockwise fashion. Finally, annihilate all the defects to vacuum. Transporting $[\coho{w}(\mb{g,h})]_{\bf g h}$ around the unit cell acquires an extra phase $M_{\cohosub{w}(\mathbf{g,h}),\cohosub{b}(T_y,T_x)}$ relative to transporting $I_{\bf g h}$ around the unit cell, because of the extra topological charge $\coho{w}(\mathbf{g,h})$ that is being transported around the anyonic flux per unit cell $\coho{b}(T_y,T_x)$. Comparing the two results produces Eq.~(\ref{eqn:spinonexists}).

In summary, we have derived some sufficient conditions for the existence of a spinon:

\emph{If  i) anyons are not permuted by the symmetries and ii) either $G$ is continuous and connected or the system has a point group symmetry such as $C_3$, $C_4$, $C_6$ or a reflection that relates $x$ to $y$, then quasiparticles that carry the topological charge $\coho{b}(T_y, T_x)$ are spinons,
as they carry the same projective representation $[o_{yx}]$ of $G$ that is carried by each unit cell.}

\subsubsection{Nontrivial contribution to $[o_{yx}]$ from anyonic spin orbit coupling}

We now examine a situation where the system need not have a spinon, even though it has a nontrivial projective representation per unit cell.
Let us consider the case where the anyonic spin orbit coupling is nontrivial, i.e. $\coho{b}(T_{i} , {\bf g}) \neq I$, and the relevant point group symmetry is broken explicitly or spontaneously (e.g. by a nematic order parameter). In this case, it is possible to satisfy the condition for a purely 2D SET phase that $o_{yx}$ in Eq.~\eqref{eq:oxy_simplified} matches the projective representation per unit cell, even when the on-site symmetry fractionalization class is trivial, i.e. $[\coho{w}]=[I]$, and the anyonic flux per unit cell is trivial, i.e. $\coho{b}(T_y,T_x)=I$. In such a situation, none of the anyons carry a nontrivial projective representation of $G$ and there is no background anyonic flux. This yields
\begin{equation}
\label{eq:oxy_SOC}
o_{yx}({\bf g}, {\bf h}) = \frac{R^{ \cohosub{b}(T_x , \mb{g}), \cohosub{b}(T_y , \mb{h})}} {R^{\cohosub{b}(T_y , \mb{g}), \cohosub{b}(T_x , \mb{h})}}
,
\end{equation}
so $[o_{yx}]$ can still match a nontrivial projective representation per unit cell, but it requires a rather unconventional way of doing so.

The toric code with unconventional symmetry model described in Sec.~\ref{sec:toriccode} realizes this scenario. In particular, we see that
\begin{equation}
\frac{o_{yx}(X,Z)}{o_{yx}(Z,X)}=\frac{M_{\cohosub{b}(T_x , X),\cohosub{b}(T_y , Z)}}{M_{\cohosub{b}(T_x , Z) , \cohosub{b}(T_y , X)}}=M_{m,e}=-1
,
\end{equation}
which indicates that $[o_{yx}]$ is the nontrivial element in  $\mathcal{H}^2[\mathbb{Z}_{2}^{2}, \U (1)]=\mathbb{Z}_{2}$.
This model breaks $C_4$ symmetry, as the symmetry was defined to act nontrivially only on the vertical bonds.

\subsection{Physics of the anyonic spin-orbit coupling}

While the physics of the anyonic flux per unit cell $\coho{b}(T_y, T_x)$ is straightforward and has been understood quite well in the context of quantum Hall states and topological spin liquids, the anyonic spin-orbit coupling has received much less attention. In Sec.~\ref{sec:set_translation}, we discussed the physics of anyonic spin-orbit coupling fractionalization classes with respect to quasiparticles, which yielded the interpretation that quasiparticle string operators (Wilson lines) carry symmetry charge per directed unit length. In this section, we examine in more detail the properties of symmetry defects when there is anyonic spin orbit coupling symmetry fractionalization.

We know that the application of the translational symmetry action to an on-site symmetry ${\bf g}$-defect (symmetry flux) carrying topological charge $a_{\bf g}$ transforms the topological charge value to $\rho_{T_i}(a_{\bf g})= [\coho{b}(T_i , \mb{g}) \times a]_\mb{g}$. In other words, the translational symmetry action changes the topological charge of a ${\bf g}$-defect by $\coho{b}(T_i , \mb{g})$. Na\"ively, it might seem that the implication of such a translational symmetry action would be that ${\bf g}$-defects cannot be adiabatically transported in the $\hat{i}$-direction, since the topological charge value carried by the defect must change when it moves. Indeed, it is not possible to adiabatically transport the position of the defect in a manner where the Hamiltonian of the defect is simply translated, as this would either involve a level crossing or violate conservation of topological charge. However, it is possible to adiabatically transport a ${\bf g}$-defect by suitably changing the form of the defect Hamiltonian to favor a different value of topological charge localized at the endpoint of the defect branch line.

In particular, for a ${\bf g}$-defect carrying the energetically favored topological charge of $a_{\bf g}$, adiabatically transporting the defect by one unit length in the $\hat{i}$-direction involves extending the defect branch line by one unit length and ending with a Hamiltonian that energetically favors topological charge $[\overline{\coho{b}(T_i , \mb{g})} \times a]_\mb{g}$ at the new endpoint of the branch line.~\footnote{One might have expected the value by which the topological charge of a defect changes under translational symmetry action and under adiabatic transport to be equal. However, one must be careful not to assume that translational symmetry action and adiabatic transport are equivalent operations. Consistency dictates that these distinct operations should change a defect's topological charge by values that are the inverse of each other.} This can be interpreted to mean that ${\bf g}$-defect branch lines carry topological charge $\coho{b}(T_i , \mb{g})$ per unit length in the $\hat{i}$-direction. Thus, adiabatically transporting a $\mb{g}$ defect by a unit length in the $\hat{i}$-direction involves creating a $\coho{b}(T_i, \mb{g})$-$\overline{\coho{b}(T_i, \mb{g})}$ pair, leaving $\coho{b}(T_i, \mb{g})$ on the new segment of defect branch line, and compounding the $\overline{\coho{b}(T_i, \mb{g})}$ topological charge with the defect to change its topological charge value. An additional implication is then that we cannot adiabatically transport a ${\bf g}$-defect in the $\hat{i}$-direction while respecting ${\bf h}$-symmetry when $M_{\cohosub{b}(T_i , \mb{g}) , \cohosub{b}(T_i , \mb{h})} \neq 1$.

\begin{figure}[t!]
	\centering
	\includegraphics[width=\columnwidth]{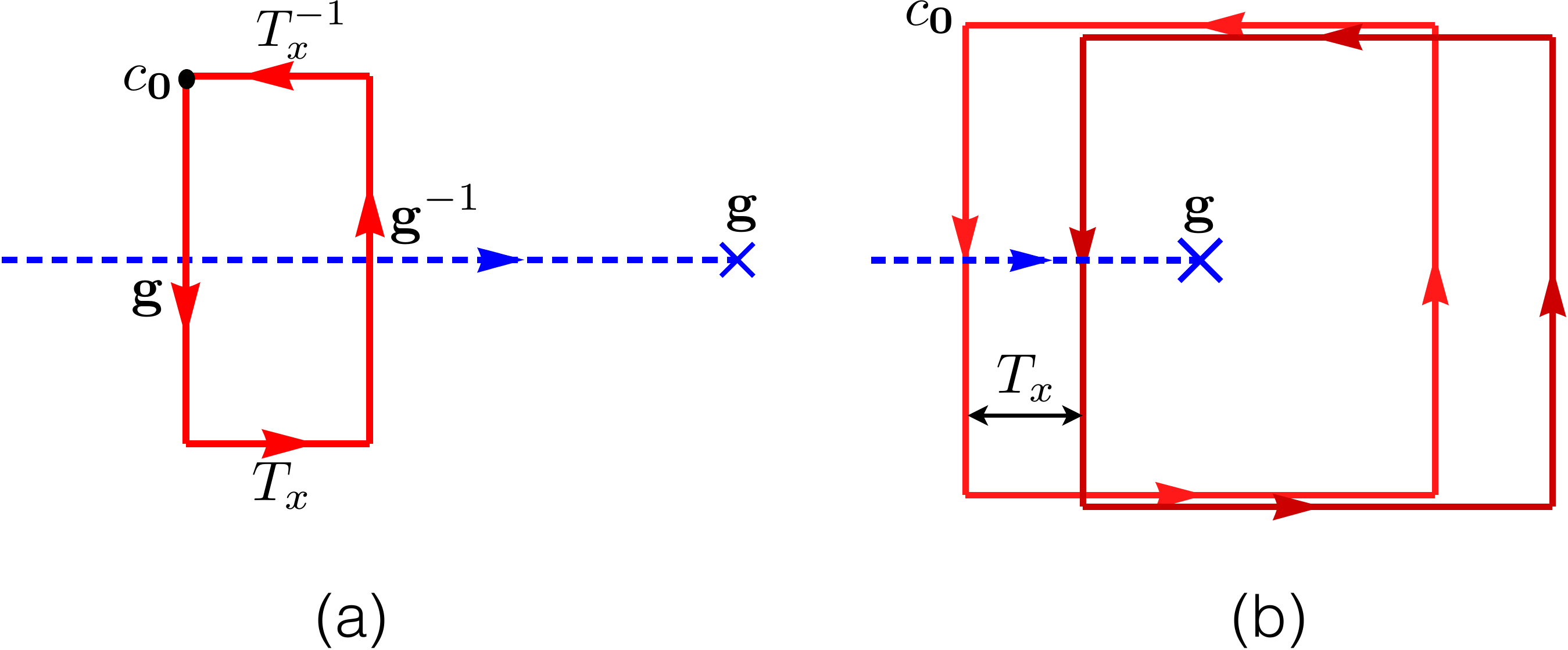}
	\caption{(a) Braiding an anyon $c_{\bf 0}$ around a unit length of ${\bf g}$-defect branch line is equivalent to the local action of $R_{T_i}^{-1} R_{\bf g}^{-1} R_{T_i} R_{\bf g}$ on the anyon (shown here for $i=x$). This implies that ${\bf g}$-defect branch lines carry topological charge $\coho{b}(T_i , \mb{g})$ per unit length in the $\hat{i}$-direction.
(b) The same result is obtained by comparing the results of braiding $c_{\bf 0}$ around two loops, each of which encloses the ${\bf g}$-defect,
and where one is translated relative to the other in the $\hat{i}$-direction. This alternative derivation allows a generalization to the case where the ${\bf g}$-defect branch line is not localized.}
	\label{fig:anyon_around_branch_line}
\end{figure}

One way to verify this interpretation is to braid a quasiparticle of topological charge $c_{\bf 0}$ around a unit segment of the defect branch line in a counterclockwise fashion, as shown in Fig.~\ref{fig:anyon_around_branch_line}(a). Since crossing the defect branch line applies the symmetry action to the object crossing the branch, this has the same effect as the local action of $R_{T_i}^{-1} R_{\bf g}^{-1} R_{T_i} R_{\bf g}$ on the anyon $c_{\bf 0}$, which is to say it generates a phase equal to $\frac{ \eta_{c_{\bf 0}}(T_i , {\bf g})} { \eta_{c_{\bf 0}}({\bf g}, T_i) } = M_{c_{\bf 0} , \cohosub{b}(T_i , \mb{g})}$. As this holds for arbitrary topological charge $c_{\bf 0}$, it implies that a unit length in the $\hat{i}$-direction of ${\bf g}$-defect branch line carries topological charge $\coho{b}(T_i , \mb{g})$.

Note that the vertical height of the loop in Fig.~\ref{fig:anyon_around_branch_line}(a) should be taken to be much larger than the
width of the ${\bf g}$-defect branch line, which is set by the correlation length of the system. In some cases, the ${\bf g}$-defect branch line is not localized, e.g. if the ${\bf g}$-defect results from a lattice dislocation~\cite{bombin2010,barkeshli2012a} or if the $G$ symmetry results from spontaneously breaking a larger
continuous symmetry. In this case, one can instead compare the result of braiding $c_{\bf 0}$ along two paths enclosing the defect, one of which is translated along
$\hat{i}$-direction relative to the other, as shown in Fig.~\ref{fig:anyon_around_branch_line}(b).

\begin{figure}[t!]
	\centering
	\includegraphics[width=0.8\columnwidth]{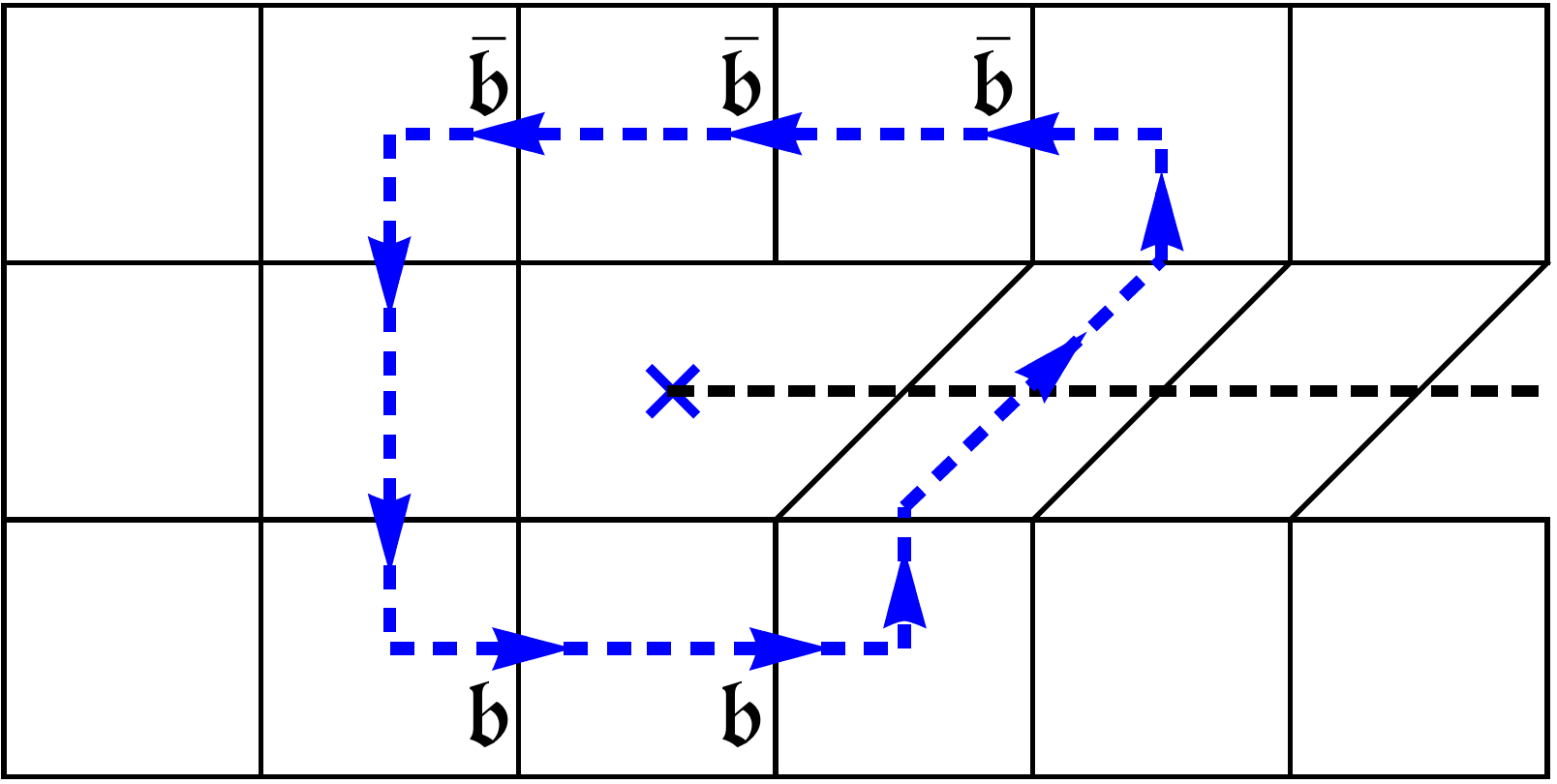}
	\caption{Braiding a ${\bf g}$-defect around an $\hat{i}$-dislocation ($T_{i}$-defect) leaves a ${\bf g}$-defect branch line loop encircling the dislocation. If ${\bf g}$-defect branch lines carry $\coho{b}(T_i , \mb{g})$ per unit length, the branch line loop carries a total topological charge of $\overline{\coho{b}(T_i , \mb{g})}$. The topological charge of the ${\bf g}$-defect will change by $\coho{b}(T_i , \mb{g})$, which matches the translational symmetry action $\rho_{T_i}(a_{\bf g})= [\coho{b}(T_i , \mb{g}) \times a]_\mb{g}$. In this figure, we display a $T_x$-defect with a ${\bf g}$-defect loop encircling it, explicitly labeling the $\coho{b}(T_x , \mb{g})$ topological charge per directed unit length of the ${\bf g}$-defect branch line ($\coho{b}(T_y , \mb{g})$ labels are left implicit).}
	\label{fig:defect_around_dislocation}
\end{figure}

Another way to verify this is to adiabatically transport a defect carrying topological charge $a_{\bf g}$ around a dislocation, i.e. a translational symmetry defect (movon). As shown in Fig.~\ref{fig:defect_around_dislocation}, if a ${\bf g}$-defect branch line carries topological charge $\coho{b}(T_i , \mb{g})$ per unit length in the $\hat{i}$-direction, then the total topological charge of ${\bf g}$-defect branch line loop left encircling the $T_{i}$-defect will equal $\overline{\coho{b}(T_i , \mb{g})}$. (This ${\bf g}$-defect branch line loop can be shrunk onto the $T_{i}$-defect and removed, leaving behind an extra topological charge of $\overline{\coho{b}(T_i , \mb{g})}$ on the $T_{i}$-defect.) By conservation of topological charge, the topological charge of the ${\bf g}$-defect must change from $a_{\bf g}$ to $[\coho{b}(T_i , \mb{g}) \times a]_\mb{g}$ as a result of the braiding process. This is consistent with the fact that transporting an object around a defect effects the symmetry action upon that object, i.e. transporting $a_{\bf g}$ around a $T_{i}$ defect yields $\rho_{T_i}(a_{\bf g})= [\coho{b}(T_i , \mb{g}) \times a]_\mb{g}$. Note that this is also consistent with the ${\bf g}$ symmetry action on $T_{i}$-defects, i.e. $\rho_{\bf g}(a_{T_i})= [\overline{\coho{b}(T_i , \mb{g})} \times a]_{T_i}$.

It is worth considering the on-site symmetry defects of the toric code with unconventional $\mathbb{Z}_2\times\mathbb{Z}_2$ symmetry (introduced in Sec.~\ref{sec:toriccode}) in detail. The defect branch lines for this example can be drawn along the links of the lattice in the $\hat{x}$-direction and across the middle of plaquettes in the $\hat{y}$-direction.~\footnote{More generally, the defect branch lines are drawn along the links of the lattice which is dual to the lattice composed of physical degrees of freedom (sites) upon which the symmetry acts nontrivially. For the toric code model on a square lattice with general symmetry (coupling to all degrees of freedom), the defect branch lines will generally run diagonally, alternatingly linking a vertex and a plaquette center. The unconventional symmetry allows us to simplify this as described.} Following the prescription described in Ref.~\onlinecite{SET}, ${\bf g}$-defects are included by modifying the Hamiltonian in the following way: for each term in the Hamiltonian that straddles the defect branch line, denote the sites on the right hand side of the directed branch line as $C_{r}$, and conjugate the corresponding term in the Hamiltonian by the operator $\prod\limits_{k \in C_{r}} R_{\bf g}^{(k)}$, i.e. apply the symmetry transformation to only the sites on one side of the branch line. The Hamiltonian at the endpoints is ambiguous under this prescription, and may possibly be chosen to favor localization of different topological charge values.

\begin{figure}[t!]
	\subfigure[]{
		\includegraphics[width=0.7\columnwidth]{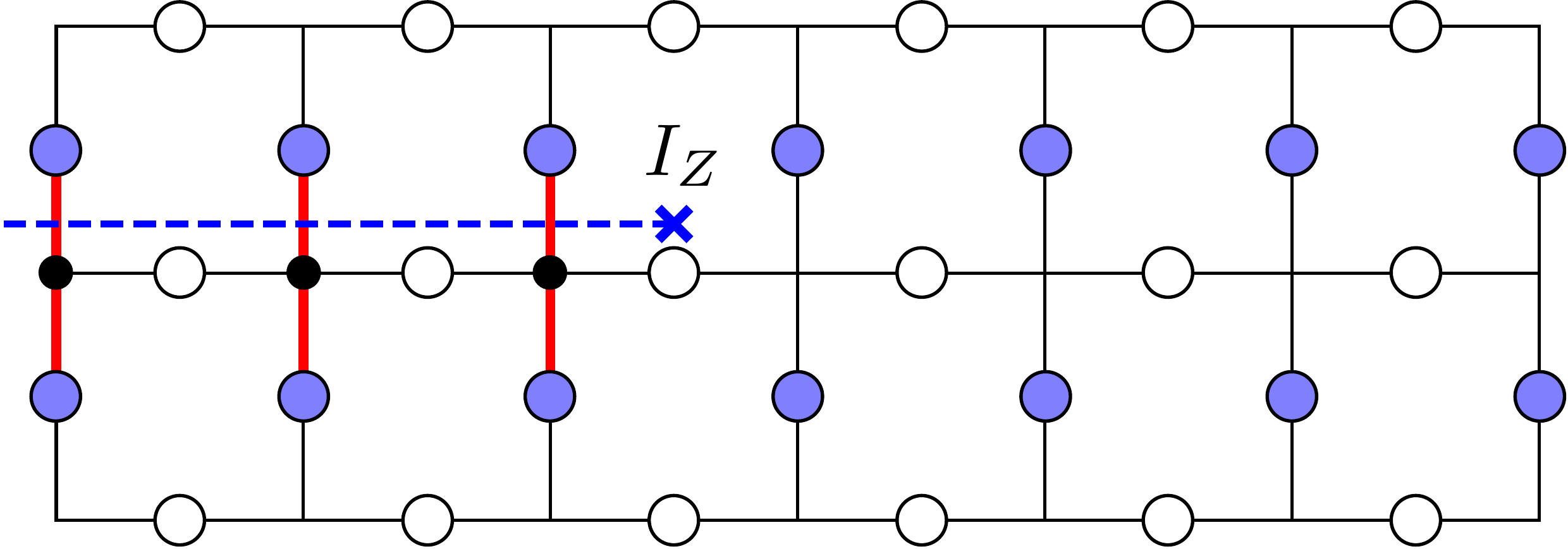}
	}
	\subfigure[]{
		\includegraphics[width=0.7\columnwidth]{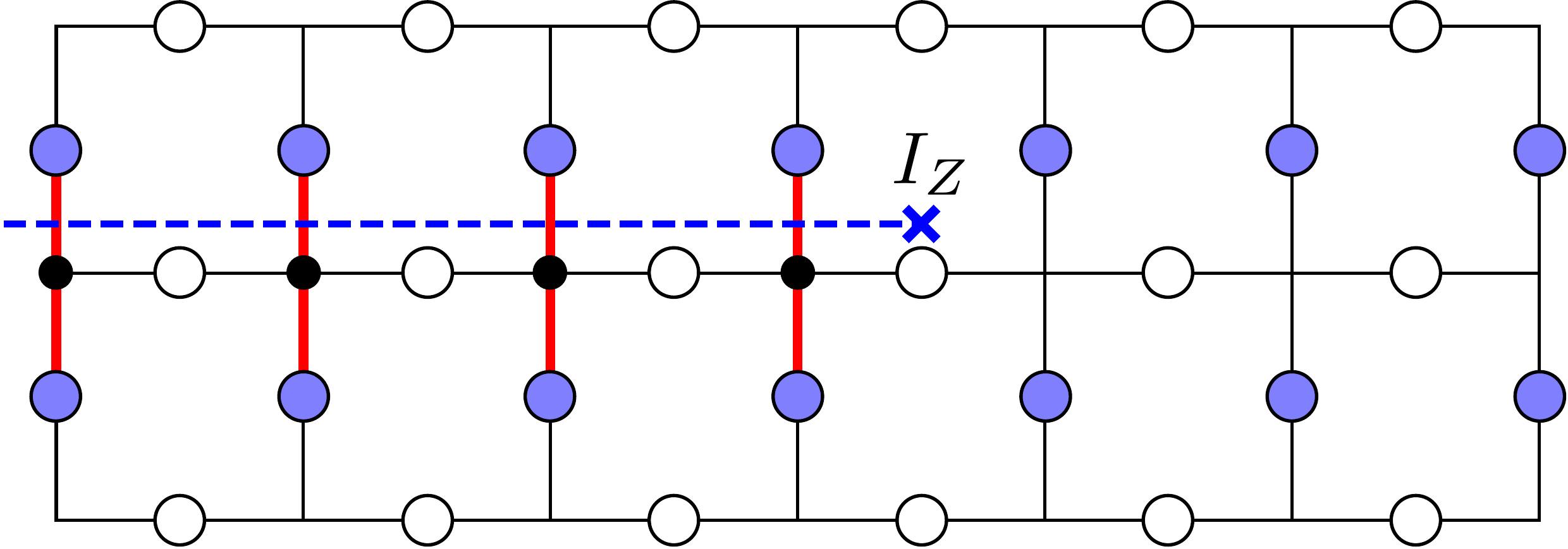}
	}
	\subfigure[]{
		\includegraphics[width=0.7\columnwidth]{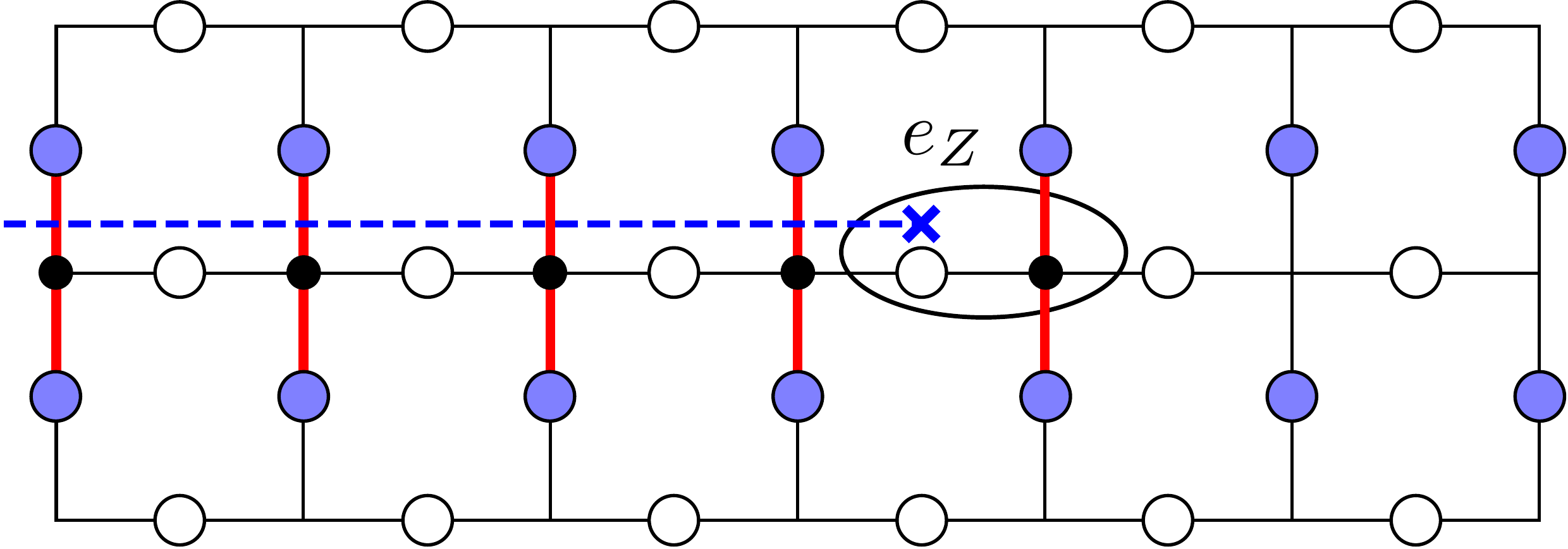}
	}
	\subfigure[]{
		\includegraphics[width=0.7\columnwidth]{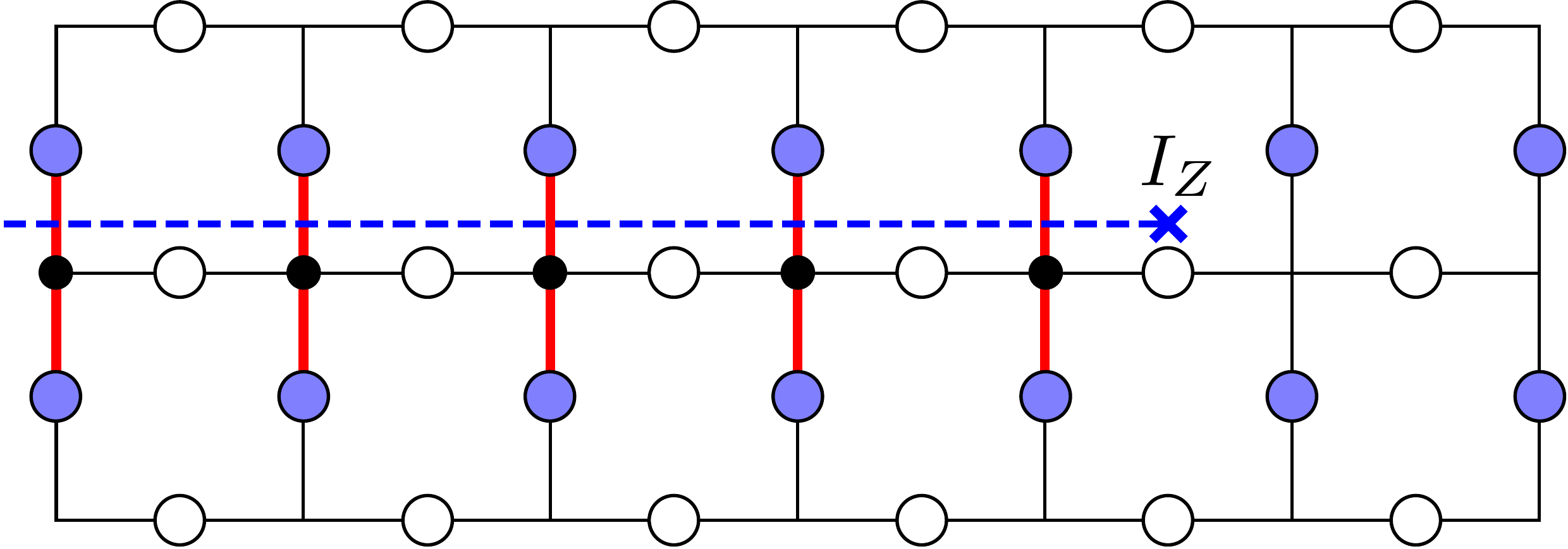}
	}
	\subfigure[]{
		\includegraphics[width=0.7\columnwidth]{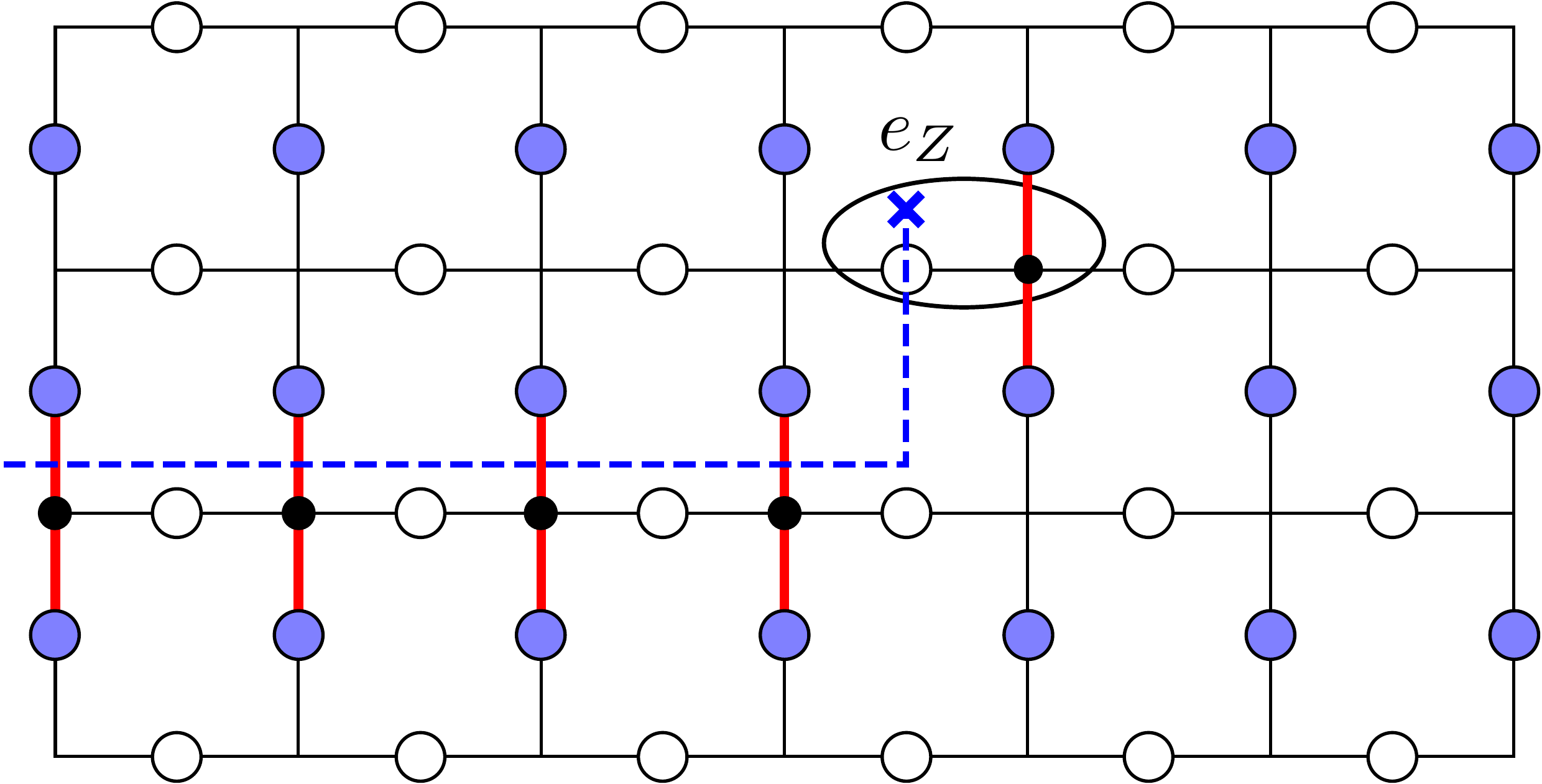}
	}
	\caption{(a) In the toric code with unconventional symmetry, $Z$ defects exist at the endpoints of a defect branch line (indicated by the dashed line), which is produced by conjugating the spins on the vertical links on one side of the branch line by $\sigma^{z}$ for terms in the Hamiltonian that straddle the branch line. This changes $-A_{v} \mapsto +A_{v}$ for vertex terms straddling the branch line, energetically favoring an $e$ quasiparticle (indicated by solid black dots) at those vertices. (b) The configuration obtained from that in (a) by acting on the system with the translation operator $R_{T_x}$. (c) The configuration obtained from (a) by adiabatically transporting the defect in the $\hat{x}$-direction by one lattice spacing. In this case, the topological charge of the defect must change from $I_{Z}$ to $e_{Z}$. (d) The configuration obtained from that in (c) by adiabatically transporting the defect in the $\hat{x}$-direction by one lattice spacing. In this case, the configuration of the physical system is unchanged, but is interpreted differently. (e) The configuration obtained from that in (c) by adiabatically transporting the defect in the $\hat{y}$-direction by one lattice spacing.}
	\label{fig:trans_defect_soc}
\end{figure}

Let us focus on a $Z$-defect branch line along the $\hat{x}$-direction in this example and apply this defect construction. The on-site symmetry operator corresponds to $R_{Z}^{(k)}= \sigma^{z}_{k}$. Conjugating plaquette terms $-B_p$ by $\sigma^{z}_{k}$ obviously leaves it invariant (which is why we are free to draw the branch line along the links in the $\hat{x}$-direction). For vertex terms $-A_{v}$ that straddle the branch line, we conjugate by $\sigma^{z}_{k}$ for the site $k$ on the vertical bond on the right hand side of the branch line, which gives the replacement $-A_{v} \mapsto +A_{v}$. In other words, we modify the Hamiltonian to energetically favor an $e$-quasiparticle at each vertex $v$ along the $Z$-defect branch line running in the $\hat{x}$-direction, as shown in Fig.~\ref{fig:trans_defect_soc}(a). This is consistent with the interpretation of the $Z$-defect branch line carrying topological charge $\coho{b}(T_x, Z)=e$ per unit length in the $\hat{x}$-direction. If we apply a translation $R_{T_x}$ to the system, the positions of the flipped vertex terms are simply moved by one lattice spacing in the $\hat{x}$-direction, as shown in Fig.~\ref{fig:trans_defect_soc}(b). Locally, the form of the Hamiltonian at the translated defect (branch line endpoint) does not change, i.e. it energetically favors the same value of topological charge and the local density matrix around the defect is related to the previous one by translation. However, we cannot reach such a configuration from the initial state by adiabatically transporting the defect, since it would violate topological charge conservation ($e$ anyons must be created or annihilated in pairs).

In order to adiabatically transport the defect by one lattice spacing in the $\hat{x}$-direction, we adiabatically change $-A_{v} \mapsto +A_{v}$ for the vertex immediately to the right of the defect, but must also modify the Hamiltonian in the vicinity of the defect (branch line endpoint) in a way that conserves topological charge. For example, if we wish to move a defect with topological charge $I_{Z}$ by one lattice spacing in the $\hat{x}$-direction, we can conserve topological charge by adiabatically changing $-A_{v} \mapsto +A_{v}$ for the \emph{two} vertices to the right of the defect, as shown in Fig.~\ref{fig:trans_defect_soc}(c). This creates two new $e$ anyons: the first $e$ anyon is interpreted as being the unit length extension of the $Z$-defect branch line in the $\hat{x}$-direction; the second $e$ anyon is interpreted as being associated with the $Z$-defect itself, modifying its topological charge from $I_{Z}$ to $e_{Z}$. If we subsequently adiabatically transport the defect a second time in the $\hat{x}$-direction, we could do so by simply leaving the Hamiltonian unchanged and subsequently interpreting the second $e$ anyon as being the second unit length extension of the $Z$-defect branch line in the $\hat{x}$-direction, as shown in Fig.~\ref{fig:trans_defect_soc}(d). If, instead, we had subsequently adiabatically transported the defect in the $\hat{y}$-direction, the second $e$ anyon would be transported in the $\hat{y}$-direction along with the defect branch line endpoint (while the first $e$ anyon remains fixed at its location), as shown in Fig.~\ref{fig:trans_defect_soc}(e).

\begin{figure}[t!]
	\centering
	\includegraphics[width=\columnwidth]{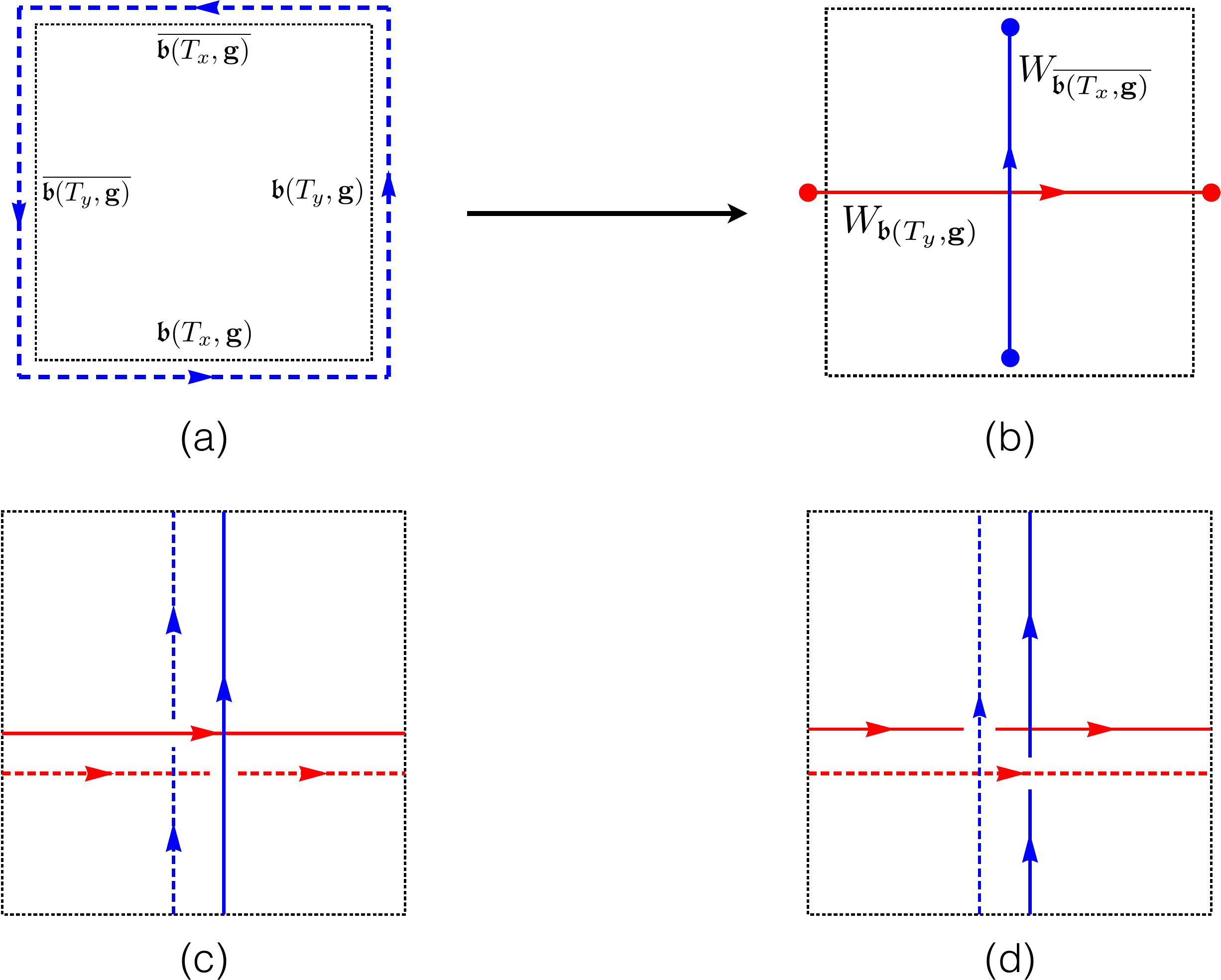}
	\caption{Physical interpretation of the contribution to the obstruction from the anyonic spin-orbit coupling from topological charge per directed unit length of defect branch lines. (a) Pair-creating a ${\bf g}$-${\bf \bar{g}}$ defect pair, adiabatically transporting the $\mb{g}$-defect counterclockwise along a path enclosing a unit cell and
then annihilating the defects generates a loop of $\mb{g}$-defect branch line encircling the unit cell (dashed blue line). This applies local ${\bf g}$-action
$U_{\bf g}$ on the unit cell. Each edge of the unit cell can then be associated with a topological charge, as indicated in the figure.
 (b) The topological charges associated with each edge of the unit cell can be obtained by applying Wilson lines in the horizontal and vertical directions, as shown. The precise manner in which the crossing point is resolved does not affect the final result.
  (c) The configuration of anyon strings associated to the process $U_\mb{g}U_\mb{h}$. Solid lines correspond to the $U_\mb{g}$ process and dashed lines to the $U_\mb{h}$ process. (d) The configuration of anyon strings associated to the process $U_\mb{h}U_\mb{g}$.}
	\label{fig:string}
\end{figure}

Returning to the general case, we can use the interpretation of the defect branch line carrying topological charge per directed unit length to re-examine the heuristic argument in Sec.~\ref{sec:trivialSOC} explaining the physical meaning of the obstruction matching condition for the case when there is anyonic spin-orbit coupling. In particular, we consider the local ${\bf g}$-action, denoted $U_\mb{g}$, on a unit cell by pair creating a ${\bf g}$-${\bf \bar{g}}$ defect pair, adiabatically transporting the ${\bf g}$-defect around a path enclosing one unit cell in a counterclockwise fashion, and then pair annihilating the defects. To be more concrete (without any loss of generality), let us choose the path to be $\mathbf{r}\rightarrow \mathbf{r}-\hat{y}\rightarrow \mathbf{r}-\hat{y}+\hat{x}\rightarrow \mathbf{r}+\hat{x}\rightarrow \mathbf{r}$.
Keeping track of the topological charge creation and annihilation due to the adiabatic transportation of the ${\bf g}$-defect creating a loop of defect branch line, we have: (1) $\mathbf{r}\rightarrow \mathbf{r}-\hat{y}$ creates topological charge $\overline{\coho{b}(T_y, \mb{g})}$ on the left segment of defect branch line; (2) $\mathbf{r}-\hat{y}\rightarrow \mathbf{r}-\hat{y}+\hat{x}$ creates topological charge $\coho{b}(T_x, \mb{g})$ on the lower segment of defect branch line; (3) $\mathbf{r}-\hat{y}+\hat{x} \rightarrow  \mathbf{r}+\hat{x}$ creates topological charge $\coho{b}(T_y, \mb{g})$ on the right segment of defect branch line; (4) $\vr+\hat{x}\rightarrow \vr$ creates topological charge $\overline{\coho{b}(T_x, \mb{g})}$ on the upper segment of defect branch line. The corresponding configuration of topological charges for this ${\bf g}$-defect branch loop is illustrated in Fig.~\ref{fig:string}(a). One can think of these topological charge values as anyons connected by anyonic string operators (Wilson lines) of the corresponding types, which can be deformed to the configuration of anyonic string operators shown in Fig.~\ref{fig:string}(b). (The intersection of anyonic string operators $W_{\overline{\cohosub{b}(T_x, \mb{g})}}$ and $W_{\cohosub{b}(T_y, \mb{g})}$ in Fig.~\ref{fig:string}(b) must be resolved to provide a proper definition, but the details of how they are resolved does not affect the quantities of interest to us, so we simply leave it drawn as an intersection.)

We can now compare the corresponding anyonic string operator configurations that arise from successive applications of ${\bf g}$ and ${\bf h}$ symmetry actions and vice-versa, i.e. $U_\mb{g}U_\mb{h}$ and $U_\mb{h}U_\mb{g}$, which are shown in Fig.~\ref{fig:string}(c) and (d), respectively. Using the relation for Abelian anyonic string operators
\begin{equation}
\includegraphics[width=0.6\columnwidth]{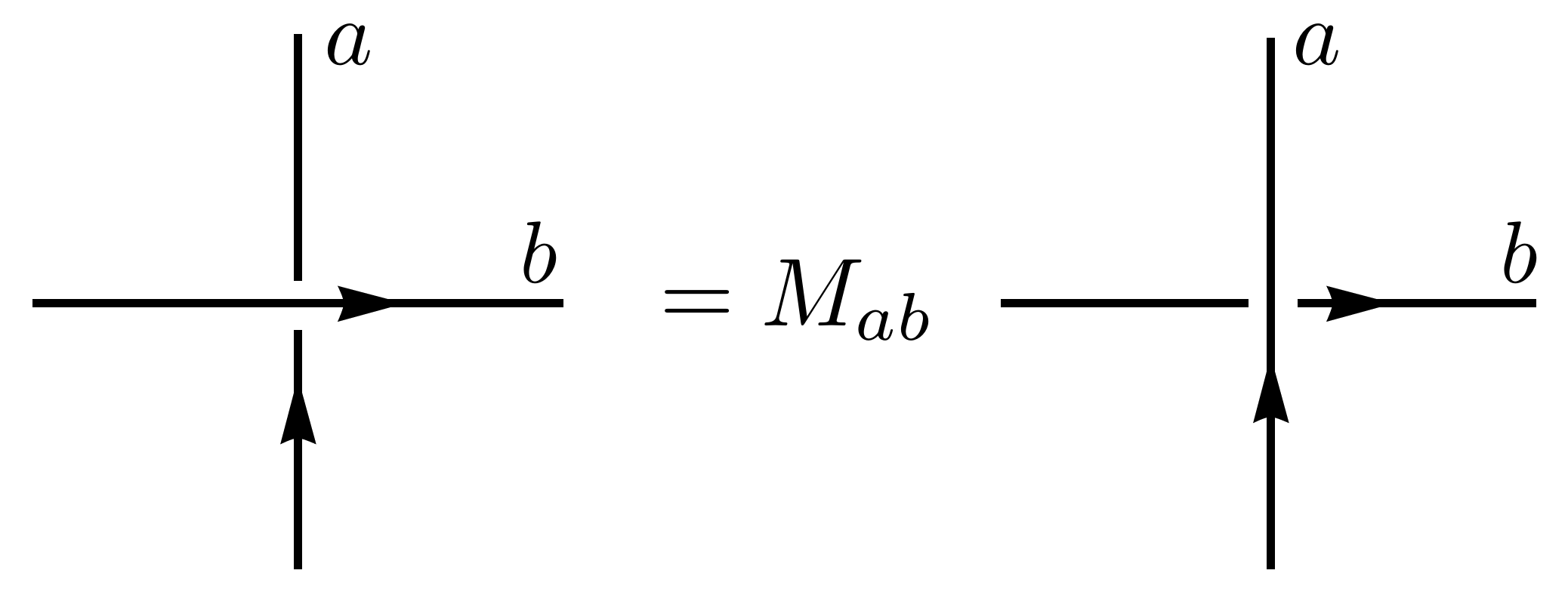}
	\label{eq:braiding}
,
\end{equation}
we find that these two configurations are equivalent up to appropriate phase factors obtained from passing the string operators for $U_{\bf g}$ through those of $U_{\bf h}$. Specifically, we find
\begin{equation}
U_\mb{g}U_\mb{h}U_\mb{g}^{-1}U_\mb{h}^{-1}=\frac{M_{\cohosub{b}(T_x, \mb{g}), \cohosub{b}(T_y, \mb{h})}}{M_{\cohosub{b}(T_y, \mb{g}), \cohosub{b}(T_x,\mb{h})}}= \frac{o_{yx}(\mb{g,h})}{o_{yx}(\mb{h,g})}.
	\label{}
\end{equation}
This is the desired obstruction-matching condition.

\begin{figure}[t!]
	\centering
	\includegraphics[width=0.8\columnwidth]{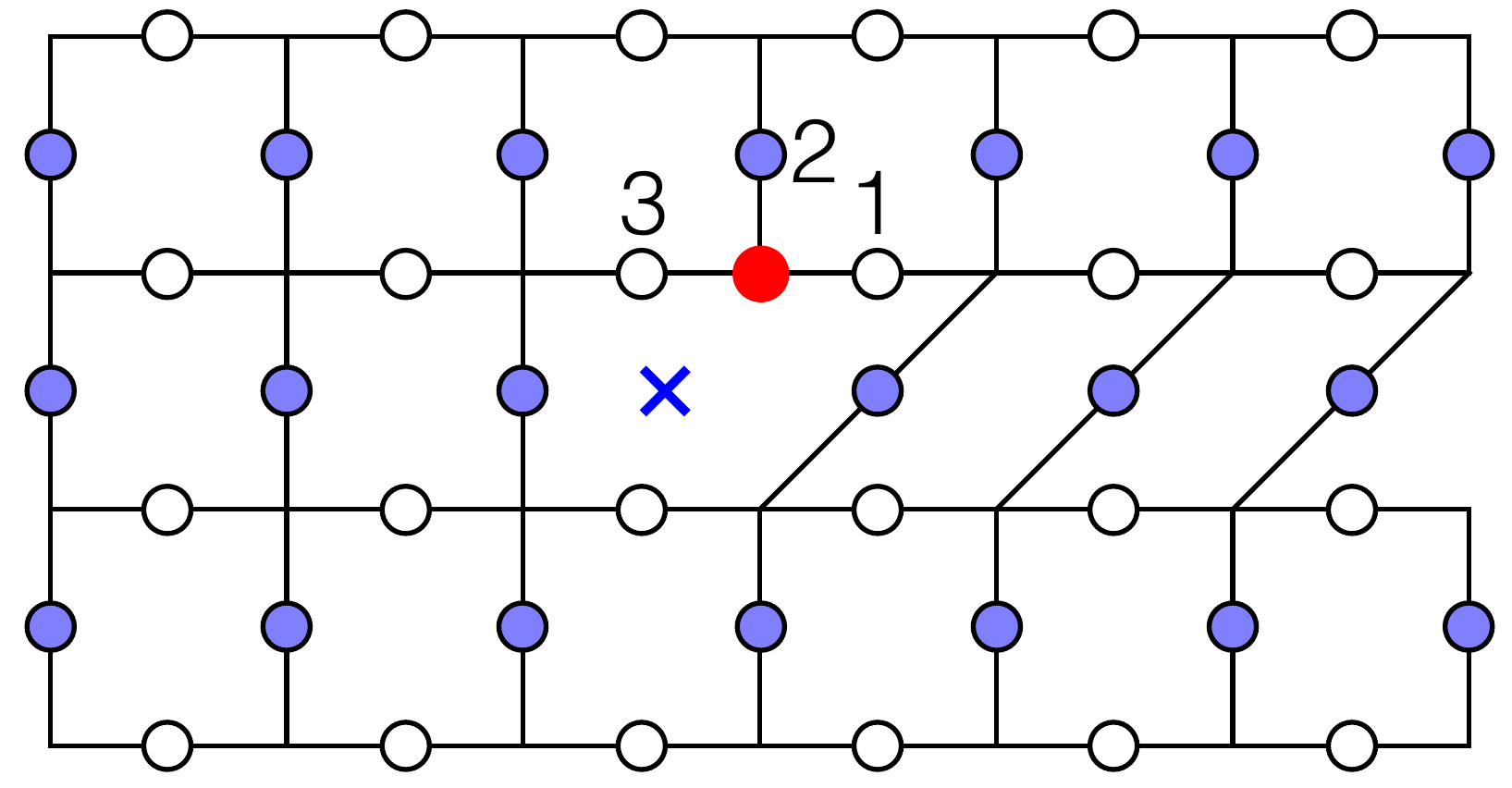}
	\caption{An $\hat{x}$-dislocation in the toric code model. When the unconventional $\mathbb{Z}_2 \times \mathbb{Z}_{2}$ symmetry is imposed, the coefficient of the vertex term $\sigma_1^x\sigma_2^x\sigma_3^x$ must be set to zero to respect the $Z$-symmetry. This results in a degeneracy between topological charge values of the translational symmetry defect that differ by $\coho{b}(T_x, Z)=e$.}
	\label{fig:dislocation}
\end{figure}

Finally, let us mention another interesting manifestation of anyonic spin-orbit coupling. Consider applying an on-site symmetry action to a translational symmetry defect (i.e. a dislocation), which gives $\rho_{\bf g}(a_{T_i})= [\overline{\coho{b}(T_i , \mb{g})} \times a]_{T_i}$, changing the topological charge of a translational symmetry defect (movon) by $\overline{\coho{b}(T_i , \mb{g})} $. Since the two types of translational symmetry defect topological charges are related by symmetry transformations, they must be energetically degenerate. We can see this explicitly in the example of the toric code model with unconventional symmetry. In Fig.~\ref{fig:dislocation}, we show an $\hat{x}$-dislocation (associated with the pentagonal plaquette) and on the dislocation there is one vertex site with three edges. The vertex term at this site $\sigma_1^x\sigma_2^x\sigma_3^x$ breaks the $Z$-symmetry and, therefore, must have its coefficient set to zero. This results in a degeneracy at the dislocation labeled by the stabilizer $\sigma_1^x\sigma_2^x\sigma_3^x=\pm 1$, which can be interpreted as the two degenerate topological charge values of the translational symmetry defect, differing from each other by $\coho{b}(T_x, Z)=e$.

\subsection{Fractional $\mathrm{U}(1)$ charge per unit cell}
\label{sec:filling}

The higher dimensional LSM theorem also holds for $\mathrm{U}(1)$-symmetric translationally invariant systems with fractional charge density $\nu = \frac{p}{q}$ per unit cell~\cite{HastingsEPL}. Since the $\mathrm{U}(1)$ group has no projective representations, there are no 3D weak SPT phases with nontrivial $\nu_{xy}$ invariants, nor 2D SET phases with nontrivial $o_{yx}$, and our previous argument does not apply.
Proceeding more heuristically than in the projective case, our goal is to establish the following
well-known claims~\cite{OshikawaLSM, HastingsLSM, OshikawaSenthil2006}, within the language of the SET formalism:

\emph{1. Adiabatically threading $2 \pi$ flux  results in an anyonic excitation called the ``vison'' $v$. The mutual statistics between $v$ and an excitation carrying topological charge $a$ determines the fractional $\U (1)$ charge $Q_{a}$ of $a$ through the relation $e^{i 2 \pi Q_{a}} = M_{v,a}$.
}

\emph{2. There is an anyon $\coho{b}(T_y, T_x)$ per unit cell whose fractional $\U (1)$ charge is equal to the filling $Q_{\cohosub{b}(T_y, T_x)} =\nu$ (mod integers). Quasiparticles that carry topological charge $\coho{b}(T_y, T_x)$ are called ``spinons.''}

\emph{3. The Hall conductance satisfies $e^{i 2 \pi \sigma_H} = M_{v, v}$.}

Consider a translationally invariant system with a non-integer filling fraction $\nu$ of some $\mathrm{U}(1)$ charge, so $\SG= \mathbb{Z}^2 \times \mathrm{U}(1)$. According to the results in Sec.~\ref{sec:set_translation},  we can completely specify the symmetry fractionalization class by the following data:
\begin{eqnarray}
\coho{b}(T_y,T_x) &=& \coho{w}(T_y,T_x) \times \overline{\coho{w}(T_x, T_y)}  \\
\coho{w}\left(\theta , \theta' \right) &=& v^{\frac{1}{2\pi}(\theta + \theta' - [\theta + \theta' ]_{2\pi})}
\label{}
\end{eqnarray}
where these are all elements in the set of Abelian anyons $\mathcal{A}$. Here we label the $\mathrm{U}(1)$ group elements by an angle $\theta \in [0,2\pi)$.
The second line serves as the definition of the vison excitation $v \in \mathcal{A}$. (As described in Sec.~\ref{sec:Laughlin_frac}, the corresponding excitation in a fractional quantum Hall state is the Laughlin-type fundamental quasihole $\phi$.)
Physically, one can argue that $v$ is the anyon nucleated by adiabatically threading $2\pi$ $\mathrm{U}(1)$ flux as follows.
Threading flux $\theta$ into the system (e.g. into a plaquette) results in a defect $I_\theta$.
If we nucleate another defect $I_{\theta'}$ nearby and then fuse them, we obtain
\begin{equation}
I_\theta \times I_{\theta'} = [\coho{w} \left(\theta, \theta' \right)]_{[\theta + \theta']_{2 \pi}} = \coho{w} \left(\theta, \theta' \right) \times I_{[\theta + \theta']_{2 \pi}}
.
\end{equation}
If we let $\theta' = 2 \pi - \theta$, this gives $I_\theta \times I_{2 \pi -\theta} = v$. Similarly, if we did this repeatedly with $\theta = 2\pi/N$, we would find $(I_{2 \pi / N})^N = v$ (where $N$ is chosen to be a multiple of $|\mathcal{A}|$). Since $\U (1)$ is continuous (and we can take the limit $N \rightarrow \infty$), these give the same result as adiabatically creating a defect $I_{\theta}$ and increasing $\theta$ from $0$ to $2\pi$. Thus, the vison $v$ is the result of $2 \pi$-flux insertion.
Note that $\mathcal{H}^2[\mathrm{U}(1), \mathcal{A}]=\mathcal{A}$, which is consistent with this physical interpretation.

Next, let us suppose that a vison $v$ is adiabatically transported around a path that encloses an anyon $a$ in a counterclockwise fashion.
On one hand, it will accumulate the mutual statistics phase $M_{v, a}$ due to $v$ encircling the anyon $a$.
On the other hand, this process is equivalent to a $2 \pi$-flux encircling the anyon $a$, and the Aharonov-Bohm effect indicates that it should acquire the phase $e^{i 2 \pi Q_a}$ due to the flux $2 \pi$ encircling a charge $Q_{a}$. Equating the two gives the relation
\begin{equation}
e^{i 2 \pi  Q_a} = M_{v, a}
.
\end{equation}

We also consider letting let $v$ be transported along a path that encloses a unit cell that is in the vacuum state.
Similar to above, transporting a $2\pi$-flux around a unit cell measures the total charge enclosed by the path, which is precisely the fractional $\U (1)$ charge $\nu$ per unit cell, giving the accumulated phase $e^{i 2\pi \nu}$. We can also view this process as the braiding between the vison $v$ and the anyon $\coho{b}(T_y, T_x)$ per unit cell, which gives the phase $M_{v , \cohosub{b}(T_y, T_x)}$. Equating the two phases, we have the relation
\begin{equation}
e^{i 2\pi \nu} = M_{v, \cohosub{b}(T_y,T_x)} = e^{i 2 \pi Q_{\cohosub{b}(T_y,T_x)}}
.
\label{}
\end{equation}
The implication is two-fold: (1) $\coho{b}(T_y, T_x)$ should carry a fractional $\U (1)$ charge $\nu$ (mod $1$). (2) The vison $v$ has nontrivial braiding statistics with $\coho{b}(T_y,T_x)$, given by $e^{i 2 \pi \nu}$. (We note that, with this definition of filling, the Laughlin states have filling $\nu=-1/m$, corresponding to a density of $1/m$ charge $-1$ electrons per unit cell.)

Finally, the Hall conductance, which is equal to the $\U (1)$ charge moved into a region by threading a $2 \pi$ flux through it, is simply equal to the charge $Q_v$ of the vison, since $v$ is the anyon that results from threading $2 \pi$ flux. Thus, we have the relation
\begin{equation}
e^{i 2 \pi \sigma_H} = e^{i 2 \pi Q_{v}} = M_{v, v}
.
\end{equation}

We again emphasize that these relations can be viewed as imposing sharp constraints on the SET orders allowed in a system. For example, a fractional Hall conductance requires the existence of anyons with matching statistics, i.e. there must be an Abelian anyon $v$ with topological twist $\theta_{v}=\pm e^{i \pi \sigma_H}$, which characterizes the $\U (1)$ symmetry fractionalization associated with winding the phase by $2 \pi$ and is the anyon created by threading a $2 \pi$ flux.

\section{Discussion}
\label{sec:discussion}

\subsection{Symmetries that permute the anyon types}

In this paper, we have focused on SET phases in which no anyon types are permuted by the symmetries.
Here we briefly discuss the case when there are permutations.

If $G$ is continuous and connected (so that $G$ cannot permute anyon types), it was argued in Ref.~\onlinecite{ZaletelPRL2015} that there must be a spinon regardless
of whether translations permute anyon types. This is not surprising, as the Oshikawa flux-threading arguments should apply.
Presumably, this result can be derived from the obstruction theory, but the calculation of $[\mathscr{O}]$ seems to be so involved that we do not attempt doing so in this paper.

If $G$ is discrete, we first consider the case where translations permute the anyon types (while $G$ does not).
In this case, the LSM constraint can be satisfied without spinons, if the on-site symmetry group $G$ is not connected and continuous.
One example is Wen's plaquette model on a square lattice, as defined in Ref.~\onlinecite{Wen2003}  and discussed in this context in Ref.~\onlinecite{ZaletelPRL2015}.
In this model, the $e$ and $m$ anyons are violations of single plaquette terms of the Hamiltonian on the two sublattices, so translations by one lattice spacing permute the two types of anyons: $\rho_{T_{i}}: e \leftrightarrow m$ for $i=x,y$.
We also define a global $\mathbb{Z}_2\times\mathbb{Z}_2$ symmetry generated by the operators $R_X=\prod_\mathbf{r}\sigma_\mathbf{r}^x$ and $R_Z=\prod_\mathbf{r}\sigma_\mathbf{r}^z$. Apparently, there is a projective representation of the $\mathbb{Z}_2\times\mathbb{Z}_2$ symmetry on each site. Therefore, we would expect that the symmetry fractionalization class should give the obstruction class consistent with the nontrivial weak SPT order in 3D.
With anyon permutation, the classification of symmetry fractionalization is now given by the 2nd cohomology group $\mathcal{H}^2_\rho(\mathcal{G}, \mathbb{Z}_2\times\mathbb{Z}_2)$.

We can determine the fractionalization class in this example. It is easy to see that the on-site symmetry does not fractionalize at all, so there are no spinons. However, the fractionalization class corresponding to non-commutativity between the on-site symmetries $G$ and the translations gives a nontrivial anyonic spin orbit coupling. In particular, a $\psi$ fermion can be created by a string of $\sigma^y$, which anti-commutes with both $\sigma^x$ and $\sigma^z$. Thus, we have
\begin{equation}
\frac{\eta_\psi(T_i , X)}{\eta_\psi(X , T_i)}=\frac{\eta_\psi(T_i , Z)}{\eta_\psi(Z , T_i)}=-1,
\label{}
\end{equation}
for $i=x,y$. This gives $\coho{b}(T_i , X)=\coho{b}(T_i , Z)=e$ or $m$ (these two values are gauge-equivalent).

On the other hand, one may consider the case when only the on-site symmetries in $G$ can permute the anyon types. When the anyonic spin-orbit fractionalization is trivial,
one can apply the heuristic argument given in Sec. \ref{sec:projrep}, since the adiabatic processes of moving on-site symmetry defects (which may be non-Abelian) remain well-defined as long as the on-site symmetry defects are not permuted under translations. The heuristic argument strongly suggests the existence of spinon per unit cell.
Again, this can presumably be resolved by calculating the obstruction in this most general case.

\subsection{Constraints from additional symmetries}

Our results can be further strengthened when combined with other symmetries.
For example, we have shown that, for $G=\mathrm{SO}(3)$, the anyonic flux per unit cell $\coho{b}(T_y, T_x)$ must be a spinon.
Now consider some other symmetry element $\mb{h}$ which commutes with all elements of $G$.
Since the anyonic flux per unit cell $\coho{b}(T_y, T_x)$ is physically well-defined and gauge invariant, it must invariant under $\mb{h}$: $\rho_\mb{h}( \coho{b}(T_y, T_x) ) = \coho{b}(T_y, T_x)$.
If $\mb{h}$ is an anti-unitary (i.e. time-reversing) or spatial parity reflecting symmetry, the topological twist is complex conjugated under the action of the symmetry, i.e. $\theta_{\rho_\mb{h}(a)} = \theta_{a}^{\ast}$.~\cite{SET}. Thus, anyons that are left invariant under $\mb{h}$ must have $\theta_{a} = \pm 1$. In particular, this would require that $\coho{b}(T_x, T_y)$ is a boson or a fermion.

This fact excludes the double-semion topological order from possessing a spinon. Recall the double semion model has four types of anyons, $\{I, s, s', b=s\times s'\}$, where $s$ and $s'$ are, respectively, semions of opposite chirality (i.e. $\theta_s = i$ and $\theta_{s'}=-i$). The only nontrivial boson or fermion is $b$, the boson formed by fusing (or forming a bound state) of $s$ and $s'$. Time-reversal interchanges $s$ and $s'$, while leaving $b$ invariant. As such,
$s$ and $s'$ must have conjugate representations of $G$ (i.e. they must map to each other under time-reversal), so the bound state $b$ must carry a trivial representation of $G$, and hence cannot be a spinon. This is the ``no-go'' theorem proved in Ref.~\onlinecite{ZaletelPRL2015}.

\subsection{Relation to cylinder arguments}

We can make connections with the arguments in Ref.~\onlinecite{ZaletelPRL2015} that demonstrated the existence of a spinon by considering the action of symmetries on the degenerate ground states of an ``infinite cylinder.'' When a topological phase is put on an infinitely long cylinder, the ground state degeneracy is the same as that of the torus, which is one ground state per topologically distinct anyon type $a$.
In the ``minimally entangled state'' (MES) basis for the ground state manifold, each basis state can thus be labeled by an anyon types as $\ket{a}$.
More generally, we can thread symmetry flux ${\bf g}$ through the cylinder by twisting the boundary condition.
It is natural to assume that the ground states are now labeled by the \emph{defect} sectors $\ket{a_{\bf g}}$.
In either case, the labeling of the MES basis is presumably ambiguous up to fusion with an Abelian anyon, just like the defect sectors, i.e. the MES basis is a torsor over $\mathcal{A}$.

To incorporate translational symmetry into this picture, we assert that a cylinder of circumference $L_y$ corresponds to defect sector $a_{{T_y}^{L_y}}$.
After all, a cylinder can be viewed as an infinite plane with a $T_y^{L_y}$ defect running along the $x$-axis; when an object crosses the defect line, it is acted on by $T_y^{L_y}$, transporting is forward, which is equivalent to periodic boundary conditions.

When a model with fractional symmetry charge per unit cell is placed on an cylinder with circumference $L_{y}$ odd, (or more generally, a circumference incommensurate with the fractional filling,) the ground states break translational symmetry in a subtle fashion.
This is most familiar in the thin-torus limit of the FQHE, where the orbital occupations become charge-density wave patterns $\cdots 0101 \cdots $, breaking $T_x$ translational symmetry.
Note that according to Eq.~\eqref{eqn:action_on_defect}, when $T_x$ acts on the defect it transforms as
\begin{align}
	\rho_{T_x}( a_{{T_y}^{L_y}} ) = \coho{b}(T_x, T_y)^{L_y}  \times a_{{T_y}^{L_y}}
\end{align}	
When $\coho{b}(T_x, T_y)^{L_y} \neq I$, this implies that the MES $\ket{a_{{T_y}^{L_y}}}$ breaks translational symmetry, in which case it is actually permuted into another MES.
This is precisely the increase in the unit cell expected of a phase at fractional filling.

It is also clear why a nontrivial anyonic spin-orbit coupling fractionalization class can result in an exception to the cylinder argument for the existence of a spinon.
The cylinder argument requires that \emph{some} MES of an odd-circumference cylinder is invariant under the on-site symmetry group $G$.
However, the defect type transforms under an on-site symmetry as
\begin{align}
\rho_{\mathbf{g}}( a_{{T_y}^{L_y}} ) = \coho{b}(\mathbf{g}, T_y)^{L_y}  \times a_{{T_y}^{L_y}}
.
\end{align}
When $\coho{b}(\mathbf{g}, T_y)^{L_y} \neq I$, the MESs are permuted by the symmetry, and it is impossible to define the symmetry properties of the entanglement spectrum as was required in the cylinder argument.

We can also understand argument why a spinon must exist when a rotation relates $T_x$ and $T_y$, even in the presence of nontrivial anyonic spin-orbit coupling fractionalization class. In this case, one can change the basis of the Bravais lattice to $\hat{x}$ and $\hat{y}'=\hat{y}-\hat{x}$. We then have $\coho{b}(\mb{g}, T_{y'})=I$, so if we compactify the lattice along the $\hat{y}'$-dierction into a cylinder (while still being extended in the $\hat{x}$-direction), the MESs are now invariant under all $\mb{g}$ and we can apply the argument to prove the existence of a spinon per unit cell.

\subsection{Possible generalizations}

Finally, we speculate on some future directions.

\subsubsection{Time-reversal symmetry}

In the presence of spin-orbit coupling, the symmetry of a spin system is broken down to time-reversal.
In a Mott insulator, each unit cell contains a Kramer's doublet with $\mathcal{T}^2 = -1$.
Thus, there is a projective representation of $\mathcal{T}$ in each unit cell, and an extension of our result should hold.
Indeed, it was argued in Ref.~\onlinecite{ZaletelPRL2015} that when translations do not permute anyons, there is a spinon excitation in each unit cell carrying $\mathcal{T}^2 = -1$. However, at present we cannot address the exotic cases, i.e.,  when anyons are permuted by translations or there is anyonic spin-orbit coupling between $\mathcal{T}$ and $T_i$.
The issue is that, while the fractionalization of $\mathcal{T}$ can easily be incorporated into the $\mathcal{H}^2_{\rho}$ classification, we do not know how to define a defect theory nor the the correct conditions for computing an obstruction $\mathscr{O}$.

\subsubsection{Non-symmorphic symmetries and magnetic translation algebras}

A series of works have shown that translations can be supplemented by non-symmorphic glide symmetries in order to get tighter LSM-type filling bounds.~\cite{Roy2012, Parameswaran2013, Watanabe_unpub}
It would be interesting to understand the corresponding generalization of this work.~\cite{parameswaran_unfinished}
However, it is not completely known how glides can be incorporated into the SET formalism.
In the least, glides will quite generically permute anyon types as they conjugate topological spin.

Another interesting case is the magnetic algebra $T^{-1}_y T^{-1}_x T_y T_x    = e^{i 2 \pi \phi \hat{N}}$, particularly in lattice type models used to generate Chern bands.
For $\phi = a / b$, we can apply our results at the cost of considering a magnetic unit cell (as we did for the example of FQH states), but there may be stronger constraints (perhaps ``$b$-times'' more powerful) that arise from considering the full algebra.

\subsubsection{Itinerant systems}

An $S=1/2$ magnet is ultimately comprised of spinful fermions, which have some residual charge fluctuations.
Thus, the true on-site symmetry group is not a projective representation of SO$(3)$, but is instead a linear representation of $\U(2)$ (a central extension of SO$(3)$ by $\U(1)$).
More generally, we can consider models in which some $\U (1)$ conserved particles at integer filling $\nu$ carry a projective representation $[\omega]$.
Intuitively, it seems that we should  be able to take a limit where charge fluctuations freeze out and consider an effective model with representation $[\omega]^\nu$ per unit cell.
In 1D, it was shown that a LSM-type result of this form indeed holds (though there are important differences in the itinerant case when LSM is further extended to include point-group symmetries).~\cite{Watanabe_unpub}

We can speculate about what version of our story we expect to hold in 2D for itinerant systems.
All local excitations carry, charge $Q=1$ and a projective representation $[\omega]$.
The relevant form of symmetry fractionalization is representation-charge separation. In other words, there may be anyons that carry projective representations $[\omega]$, but no charge $Q=0$, or anyons that carry charge $Q=1$, but only linear representations $[1]$. Such anyons may differ only by a local excitation.
These are the ``spinons'' and holons; presumably they play the same role as the spinon in our previous discussions.
We leave the full structure of the itinerant case to future work.

\subsubsection{Gapless phases and the FL$^\ast$ phase}

When a system with fractional symmetry charge per unit cell is gapless, there must also be constraints on the resulting phase.
The most obvious example is Luttinger's theorem~\cite{Luttinger}: at filling $\nu$, the volume of the Fermi sea is proportional to $\nu$.
Another example is 1D spin chains with spin-$1/2$ per unit cell, which can be viewed as the edge of a 2D AKLT state.
The 2D AKLT state is a weak SPT, and the bulk-boundary correspondence now implies that the edge spin chain cannot be gapped without breaking the spin rotation symmetry or the 1D translational symmetry. (Note that the option of topological order is not available in 1D.)
This is, of course, the content of the original LSM theorem.
Analogously, one can further examine the symmetry action in the gapless phase.
It was argued that symmetries are implemented in an anomalous fashion in the resulting conformal field theory \cite{Furuya_unpub} of the spin chain,
which is the 1D (gapless) analog of our constraint.
In 2D, there can be a combination of topological and gapless degrees of freedom.
Presumably the constraints can be accommodated in a shared fashion between the topological and gapless degrees of freedom, suggesting a generalized version of the FL$^\ast$ scenario~\cite{SenthilSachdevVojta2003}.

\begin{acknowledgments}
We thank B.~Bauer, M.~Hermele, S.~Parameswaran, C.~Wang, and Z.~Wang for helpful conversations.
In particular, we thank M.~Hermele and C.~Wang for bringing the model of Sec.~\ref{sec:toriccode} to our attention.
A.V. acknowledges support from the Templeton Foundation and a Simons Investigator award.
\end{acknowledgments}

\appendix

\section{Review of group cohomology}
\label{app:coh}

In this section, we provide a brief review of group cohomology for finite groups.
Given a finite group $G$, let $M$ be an Abelian group equipped with a $G$ action $\rho: G \times M \rightarrow M$, which is compatible with group multiplication. In particular, for any $\mathbf{g}\in G$ and $a,b \in M$, we have
\begin{equation}
\rho_\mathbf{g}(ab)=\rho_\mathbf{g}(a) \rho_\mathbf{g}(b).
\label{}
\end{equation}
(We leave the group multiplication symbols implicit.) Such an Abelian group $M$ with $G$ action $\rho$ is called a $G$-module.

Let $\omega(\mathbf{g}_1, \dots,\mathbf{g}_n)\in M$ be a function of $n$ group elements $\mathbf{g}_j \in G$ for $j=1,\dots,n$. Such a function is called a $n$-cochain and the set of all $n$-cochains is denoted as $C^n(G, M)$. They naturally form a group under multiplication,
\begin{equation}
  (\omega\cdot\omega')(\mb{g}_1, \dots, \mb{g}_n)=\omega(\mb{g}_1, \dots, \mb{g}_n)\omega'(\mb{g}_1, \dots, \mb{g}_n),
  \label{}
\end{equation}
and the identity element is the trivial cochain $\omega(\mb{g}_1,\dots,\mb{g}_n)=1$.

We now define the ``coboundary'' map $\mathrm{d}: C^n(G, M) \rightarrow C^{n+1}(G, M)$ acting on cochains to be
\begin{equation}
\begin{split}
\mathrm{d}&\omega  (\mathbf{g}_1,\dots,\mathbf{g}_{n+1})= \rho_{\mathbf{g}_1}[\omega(\mathbf{g}_2,\dots,\mathbf{g}_{n+1})] \\
	&\times \prod_{j=1}^n \omega^{(-1)^j}(\mathbf{g}_1,\dots,\mathbf{g}_{j-1},\mathbf{g}_j\mathbf{g}_{j+1},\mathbf{g}_{j+2},\dots,\mathbf{g}_{n+1}) \\
    &\times \omega^{(-1)^{n+1}}(\mathbf{g}_1,\dots,\mathbf{g}_{n})
.
\end{split}
\label{}
\end{equation}
One can directly verify that $\mathrm{d} \mathrm{d}\omega=1$ for any $\omega \in C^n(G, M)$, where $1$ is the trivial cochain in $C^{n+2}(G, M)$. This is why $\mathrm{d}$ is considered a ``boundary operator.''

With the coboundary map, we next define $\omega\in C^n(G, M)$ to be an $n$-cocycle if it satisfies the condition $\mathrm{d}\omega=1$. We denote the set of all $n$-cocycles by
\begin{equation}
\begin{split}
Z^n_{\rho}(G, M)
 = \{ \, \omega\in C^n(G, M) \,\, | \,\, \mathrm{d}\omega=1 \, \}.
\end{split}
\label{}
\end{equation}
We also define $\omega\in C^n(G, M)$ to be an $n$-coboundary if it satisfies the condition $\omega= \mathrm{d} \mu $ for some $(n-1)$-cochain $\mu \in C^{n-1}(G, M)$. We denote the set of all $n$-coboundaries by
Also we have
\begin{equation}
\begin{split}
& B^n_{\rho}(G, M) \\
&=\{ \, \omega\in C^n(G, M) \,\, | \,\, \exists \mu \in C^{n-1}(G, M) : \omega = \mathrm{d}\mu \, \}
.
\end{split}
\label{}
\end{equation}

Clearly, $B^n_{\rho}(G, M) \subset Z^n_{\rho}(G, M) \subset C^n(G, M)$. In fact, $C^n$, $Z^n$, and $B^n$ are all groups and the co-boundary maps are homomorphisms. It is easy to see that $B^n_{\rho}(G, M)$ is a normal subgroup of $Z^n_{\rho}(G, M)$. Since d is a boundary map, we think of the $n$-coboundaries as being trivial $n$-cocycles, and it is natural to consider the quotient group
\begin{equation}
	\mathcal{H}^n_{\rho}(G, M)=\frac{Z^n_{\rho}(G, M)}{B^n_{\rho}(G, M)}
,
\label{}
\end{equation}
which is called the $n$-th group cohomology. In other words, $\mathcal{H}^n_{\rho}(G, M)$ collects the equivalence classes of $n$-cocycles that only differ by $n$-coboundaries.

The algebraic definition we give for group cohomology is most convenient for discrete groups. For continuous group, formally the same definition applies but one has to impose proper continuity conditions on the cocycle functions.

Let us now consider $M$ being a $G$-module with trivial action, and let $\mb{g}\in G$ be an arbitrary element. We define the slant product $\slant_\mb{g}: C^n(G, M)\rightarrow C^{n-1}(G, M)$:
\begin{eqnarray}
&& \slant_\mathbf{g}\omega (\mb{g}_1,\ldots,\mb{g}_{n-1})
\notag \\
&& \quad = \prod_{j=0}^{n-1}\omega(\mb{g}_1,\ldots, \mb{g}_{j},\mb{g},\mb{g}_{j+1},\ldots,\mb{g}_{n-1})^{(-1)^{n-1+j}}.
\quad \label{eq:slant_product}
\end{eqnarray}
It can be shown that $\mathrm{d}(\slant_\mb{g}\omega)=\slant_\mb{g}(\mathrm{d}\omega)$. Therefore, $\slant_\mb{g}$ is in fact a group homomorphism:
\begin{equation}
	\slant_\mb{g}: \mathcal{H}^n(G, M)\rightarrow \mathcal{H}^{n-1}(G, M).
	\label{}
\end{equation}

\section{K\"unneth formula}
\label{sec:Kunneth}

The K\"unneth formula for the group cohomology of the direct product of two groups $\mathcal{H}^{d}(G\times G', M)$ reads:

\begin{equation}
	\mathcal{H}^{d}[G\times G', {M}]=\bigoplus_{k=0}^d\mathcal{H}^k[G,\mathcal{H}^{d-k}[G', M]].
	\label{eqn:kunneth}
\end{equation}
Here, $M$ is Abelian and finitely-generated (e.g. $M=\mathbb{Z}$ or a finite Abelian group), and we only consider $M$ being a trivial $G\times G'$ module. This formula has been derived in Ref.~\onlinecite{WenPRB2015} for finite $G$ and $G'$. We shall briefly review the relevant mathematical results and show that the formula generally holds for discrete groups and Lie groups.

We start from the K\"unneth formula for topological cohomology. Let $X$ and $X'$ be two topological manifolds, we have~\cite{WenPRB2015}
\begin{equation}
	H^d[X\times X', M]=\bigoplus_{k=0}^d H^k[X, H^{d-k}[X', M]].
	\label{eqn:topkunneth}
\end{equation}
Notice that $H^d[X, M]$ is the topological cohomology of chain complexes $X$ with coefficient in $M$. Now we set $X=BG$ and $X'=BG'$, i.e. the classifying spaces of principle $G$ and $G'$-bundles.  In order to derive Eq.~\eqref{eqn:kunneth} from Eq.~\eqref{eqn:topkunneth}, we need $H^d[BG, M]=\mathcal{H}^d[G, M]$. It is well-known that this is true for discrete $G$ by construction. For a compact (continuous) Lie group $G$, a theorem of Wigner~\cite{WignerCoh, Stasheff} shows that, if we define $\mathcal{H}^d[G, M]$ as the Borel cohomology (which is exactly what we need in the physical applications), the relation also holds. Therefore, Eq.~\eqref{eqn:kunneth} holds generally for discrete groups and compact Lie groups.

We also need to establish the formula for $\mathrm{U}(1)$ coefficients. This is achieved by making use of the following relation:
\begin{equation}
	\mathcal{H}^d[G, \mathrm{U}(1)]=\mathcal{H}^{d+1}[G, \mathbb{Z}].
	\label{}
\end{equation}
Again, when $G$ is continuous, we should use Borel cohomology.

We now apply the K\"unneth formula to compute the decomposition of $\mathcal{H}^n[\mathbb{Z}^d\times G, \U(1)]$. First, we can show that
\begin{equation}
\label{eqn:ZKunneth}
	\mathcal{H}^{n}[ \mathbb{Z}^{d}, \mathbb{Z}] = \left\{
\begin{matrix}
	\mathbb{Z}^{ \binom{d}{n} }  &  & \text{ for } 0 \leq n \leq d  \\
 & & \\
\mathbb{Z}_{1}  & & \text{ for } d < n
\end{matrix}
\right.
.
\end{equation}
To see this result, we notice that $B\mathbb{Z}^d=\mathbb{T}^d$, the $d$-dimensional torus, and so the right-hand side of Eq.~\eqref{eqn:ZKunneth} is the well-known (topological) cohomology of the torus. We then have
\begin{eqnarray}
\mathcal{H}^n[\mathbb{Z}^d\times G, \U(1)] &=& \mathcal{H}^{n+1}[\mathbb{Z}^d\times G, \mathbb{Z}] \notag\\
		&=& \prod_{k=0}^{n+1} \mathcal{H}^k[G, \mathcal{H}^{n+1-k}[\mathbb{Z}^d, \mathbb{Z}]] \notag \\
		&=& \prod_{k=\max\{1, n-d+1\}}^{n+1}\mathcal{H}^k[G, \mathbb{Z}]^{\binom{d}{n-k+1}} \notag \\
		&=& \prod_{k=\max\{0, n-d\}}^{n}\mathcal{H}^k[G, \U(1)]^{\binom{d}{n-k}} . \qquad
\end{eqnarray}

\onecolumngrid
\section{Explicit Representations of K\"unneth Decomposition}
\label{sec:Kunneth_rep}

We now give explicit representative $n$-cocycles for each equivalence class in $\mathcal{H}^n[G\times H, \U(1)]$, in terms of lower-dimensional cohomology classes. We write an element of $G\times H$ as $(\mb{g},\mb{h})$. A U$(1)$ $n$-cohomology class $[\omega]$ can be represented by a $n$-cocycle of the form
\begin{equation}
\omega ((\mb{g}_1,\mb{h}_1), \cdots, (\mb{g}_n,\mb{h}_n))
=\prod_{k=0}^n \nu_k(\mb{h}_1,\mb{h}_2,\cdots,\mb{h}_k; \mb{g}_{k+1},\cdots, \mb{g}_n)
,
\label{eqn:decomposition}
\end{equation}
where each $\nu_k$ represents a normalized $n$-cocycle (i.e. $\mathrm{d}\nu_k=1$). Explicitly, they must satisfy the condition
\begin{equation}
	\begin{split}
	\nu_k(\{\mb{h}_i\}_{i=2}^{k+1}; \{\mb{g}_j\}_{j=k+2}^{n+1})&\nu_k^{(-1)^{n+1}}(\{\mb{h}_i\}_{i=1}^{k}; \{\mb{g}_j\}_{j=k+1}^{n})\prod_{p=1}^k \nu_k^{(-1)^p}(\mb{h}_1, \cdots, \mb{h}_{p-1}, \mb{h}_p\mb{h}_{p+1}, \mb{h}_{p+2},\cdots, \mb{h}_{k+1}; \{\mb{g}_j\}_{j=k+2}^{n+1})\\
	&\cdot\prod_{q=k+1}^n \nu_k^{(-1)^q}(\{\mb{h}_i\}_{i=1}^{k}; \mb{g}_{k+1},\cdots, \mb{g}_{q-1},\mb{g}_q\mb{g}_{q+1},\cdots, \mb{g}_{n+1})	
	=1 .
	\end{split}
	\label{eqn:explicit-kunneth-1}
\end{equation}
In this equation, if we set $\mb{g}_{k+1}=\openone$, we get
\begin{equation}
\begin{split}
\nu_k(\{\mb{h}_i\}_{i=2}^{k+1}; \{\mb{g}_j\}_{j=k+2}^{n+1})\prod_{p=1}^k & \nu_k^{(-1)^p}(\mb{h}_1, \cdots, \mb{h}_{p-1}, \mb{h}_p\mb{h}_{p+1}, \mb{h}_{p+2},\cdots, \mb{h}_{k+1}; \{\mb{g}_j\}_{j=k+2}^{n+1})\\
&\cdot\nu_k^{(-1)^{k+1}}(\{\mb{h}_i\}_{i=1}^{k}; \{\mb{g}_j\}_{j=k+2}^{n+1})=1,
\end{split}
\label{eqn:explicit-kunneth-2}
\end{equation}
which is exactly the $k$-cocycle condition on $H$. In other words, for fixed values of $\mb{g}_{j}$, the quantity $\nu_k(\mb{h}_1,\mb{h}_2,\cdots,\mb{h}_k; \mb{g}_{k+1},\cdots, \mb{g}_n)$ is a $k$-cocycle in ${Z}^k[H, \U(1)]$.
If we modify $\nu_k$ by a $k$-coboundary of $H$ (still fixing all the $\mb{g}_{j}$), it is not difficult to see that $\nu_k$ as a $n$-cocycle of $G\times H$ is then modified by an $n$-coboundary, which is cohomologically trivial. Therefore, we can say that $\nu_k$ corresponds to a cohomology class in $\mathcal{H}^k[H, \U(1)]$. More abstractly, we can view $[\nu_k]$ as a function of $(n-k)$ elements of $G$ to $\mathcal{H}^k[H, \U(1)]$.

Now let us substitute Eq.~\eqref{eqn:explicit-kunneth-2} back into Eq. \eqref{eqn:explicit-kunneth-1}. We see that, when $\mb{h}_{i}$ are fixed,  $\nu_k(\mb{h}_1,\mb{h}_2,\cdots,\mb{h}_k; \mb{g}_{k+1},\cdots, \mb{g}_n)$  is a $(n-k)$-cocycle in $Z^{n-k}[G, \U(1)]$. Combined with the previous results, it follows that $\nu_k$ corresponds to a $(n-k)$-cocycle in $\mathcal{H}^{n-k}[G, \mathcal{H}^k[H, \U(1)]]$.

We also notice that this parametrization provides a concrete recipe to write down a representative $n$-cocycle for the cohomology classes in $\mathcal{H}^n[G\times H, \U(1)]$, given a cohomology class in $\mathcal{H}^{n-k}[G, \mathcal{H}^k[H, \U(1)]]$.

Furthermore, given the decomposition of Eq.~\eqref{eqn:decomposition}, one can extract $\nu_k$ by applying slant products:
\begin{equation}
	i_{(\openone,\mb{h}_k)}i_{(\openone,\mb{h}_{k-1})}\cdots i_{(\openone,\mb{h}_1)}\omega\big( (\mb{g}_{k+1},\openone),\cdots, (\mb{g}_{n},\openone)\big)=\prod_p\nu(\mb{h}_{p_1},\dots, \mb{h}_{p_k};\mb{g}_{k+1},\cdots, \mb{g}_n)^{\sigma(p)}.
	\label{}
\end{equation}
This follows from a straightforward application of the definition of the slant product, so we will omit the details.

\section{Simplifying $[o_{yx}]$ when the $F$-symbols are nontrivial}
\label{app:F}

We will simplify the expression of the obstruction class involving nontrivial $F$-symbols. First let us display the full expression of the obstruction:
\begin{equation}
\mathscr{O} ({\bf g},{\bf h},{\bf k},{\bf l})
=
\frac{ F^{\cohosub{w}({\bf g},{\bf h}) \cohosub{w}({\bf k},{\bf l}) \cohosub{w}({\bf gh},{\bf kl})}
F^{ \cohosub{w}({\bf k},{\bf l})  \cohosub{w}({\bf h},{\bf kl})  \cohosub{w}({\bf g},{\bf hkl}) }
F^{\cohosub{w}({\bf h},{\bf k})  \cohosub{w}({\bf g},{\bf hk})  \cohosub{w}({\bf ghk},{\bf l}) }}
{F^{\cohosub{w}({\bf g},{\bf h})  \cohosub{w}({\bf gh},{\bf k})  \cohosub{w}({\bf ghk},{\bf l}) }
F^{\cohosub{w}({\bf k},{\bf l})  \cohosub{w}({\bf g},{\bf h})  \cohosub{w}({\bf gh},{\bf kl}) }
F^{ \cohosub{w}({\bf h},{\bf k})  \cohosub{w}({\bf hk},{\bf l})  \cohosub{w}({\bf g},{\bf hkl}) }}
R^{\cohosub{w}({\bf g},{\bf h}), \cohosub{w}({\bf k},{\bf l})}
.
\end{equation}
For a derivation of the result we refer the readers to \Ref{chen2014} and \Ref{SET}.

We will first need to have an explicit general expression of the $F$-symbols. The Abelian subcategory $\mathcal{A}$ can always be decomposed into cyclic groups as $\mathcal{A}=\prod_j \mathbb{Z}_{N_j}$, i.e. each $a\in \mathcal{A}$ can be uniquely represented as a tuple $a=(a_1, a_2, \dots )$, where $a_j \in \mathbb{Z}_{N_j}$. For Abelian theories, the possible $F$-symbols that satisfy the pentagon consistency equation are classified by $\coh{3}{\mathcal{A}}$. These are further constrained by the hexagon consistency equations, which must be satisfied by a braided theory of anyons. The allowed $F$-symbols can then be brought to the form
\begin{equation}
F^{abc}=\exp\Big[i 2\pi \sum_{j\leq k}\frac{ p_{jk}}{N_j N_k}[a_j]_{N_j}\big([b_k]+[c_k]-[b_k+c_k]\big)_{N_k}\Big]
,
\label{eqn:Flookslikethis}
\end{equation}
where $p_{jk}\in \mathbb{Z}$. In the nomenclature of \Ref{Propitius1995}, this only involves 3-cocycles of type I and type II, as the type III cocycles are excluded by the hexagon equations.

To shorten the expressions we define
\begin{equation}
f(a,b)=\exp\Big(i 2\pi \sum_{j\leq k}\frac{ p_{jk}}{N_j N_k}[a_j]_{N_j}[b_k]_{N_k}\Big).
	\label{}
\end{equation}
So the $F$-symbols can be written as
\begin{equation}
F^{abc}=\frac{f(a,b)f(a,c)}{f(a,b\times c)}.
\label{eqn:Fasf}
\end{equation}

We also define for convenience
\begin{equation}
	\beta\equiv\coho{b}(T_x,T_y), x_\mb{g}=\coho{b}(\mb{g}, T_x), y_\mb{g}=\coho{b}(\mb{g}, T_y).
	\label{}
\end{equation}

We now define $\tilde{o}(\mb{g,h})$ formally as the contribution to the slant product $o_{yx}$
from the $F$-symbols, which was neglected in the main text where we assumed all $F$-symbols can be made trivial.
We will not bother to write down the general expression since it is too complicated and not very enlightening.
However, before we proceed to calculate $\tilde{o}(\mb{g,h})$, we first discuss how it is affected by the gauge transformations on $F$ and $R$ symbols.
Recall that $F$-symbols have the following gauge degrees of freedom:
\begin{equation}
	F^{abc}\rightarrow \frac{u(a,b)u(a\times b,c)}{u(a,b\times c)u(b,c)}F^{abc}.
	\label{}
\end{equation}
Here $u(a,b)$ is an arbitrary $\mathrm{U}(1)$-valued function.
Mathematically, the gauge transformations modify the $F$-symbols by a coboundary in $B^3(\mathcal{A}, \mathrm{U}(1))$. Under such a gauge transformation, one finds that
\begin{equation}
	\tilde{o}(\mb{g,h})\rightarrow \tilde{o}(\mb{g,h})\frac{u(x_\mb{g}, y_\mb{h})u(x_\mb{h}, y_\mb{g})}{u(y_\mb{g}, x_\mb{h})u(y_\mb{h}, x_\mb{g})}\times (2-\text{coboundary}).
	\label{eqn:2coboundary}
\end{equation}
When combining $\tilde{o}(\mb{g,h})$ with the definition of $o_{yx}$, the extra piece $\frac{u(x_\mb{g}, y_\mb{h})u(x_\mb{h}, y_\mb{g})}{u(y_\mb{g}, x_\mb{h})u(y_\mb{h}, x_\mb{g})}$, which is not generally a coboundary, will be cancelled by a similar factor coming from the $R$ symbols, so that $o(\mb{g,h})$ indeed acquires just a coboundary.

Now let us compute $\tilde{o}$ using the gauge choice of Eq.~\eqref{eqn:Fasf}:
\begin{equation}
		\tilde{o}(\mb{g,h})=\frac{ f(\beta, x_\mb{gh}y_\mb{gh})}{f(\beta, x_\mb{g}y_\mb{g})f(\beta, x_\mb{h}y_\mb{h})}\frac{f(x_\mb{g}, \beta y_\mb{g})f(x_\mb{h},\beta y_\mb{h})}{f(x_\mb{g},\beta y_\mb{gh})f(x_\mb{h},\beta y_\mb{gh})} \frac{f(y_\mb{g},x_\mb{gh})f(y_\mb{h}, x_\mb{gh})}{f(y_\mb{g},x_\mb{g})f(y_\mb{h},x_\mb{h})} \frac{f(x_\mb{g},y_\mb{h})f(x_\mb{h},y_\mb{g})}{f(y_\mb{g}, x_\mb{h})f(y_\mb{h},x_\mb{g})}
\end{equation}
Note that $x_\mb{gh} = x_\mb{hg}$ since $\coho{b}(\cdot, T_x) \in \mathcal{H}^1[G, \mathcal{A}]$.
It follows that
\begin{equation}
	\frac{\tilde{o}(\mb{g,h})}{\tilde{o}(\mb{h,g})}=1
	\label{eq:tilden_sym}
\end{equation}
For Abelian $G$, Eq.~\eqref{eq:tilden_sym} is the necessary and sufficient condition to show $\tilde{o}(\mb{g,h})$ is a coboundary.
One may wonder whether this statement depends on gauge choices for $F$ and $R$ symbols.
But from Eq.~\eqref{eqn:2coboundary}, we can be assured that $\tilde{o}(\mb{g,h})/\tilde{o}(\mb{g,h})$ does not depend on the gauge choice of $F$ and $R$ symbols.

We now assume lattice rotational symmetry so $x_\mb{g}=y_\mb{g}$. Note that if $G$ is continuous, $x_\mb{g}=y_\mb{g} = I$ and the analysis also holds.
Again we can show $\tilde{o}(\mb{g,h})$ is a coboundary:
\begin{equation}
	\tilde{o}(\mb{g,h})=\frac{ f(\beta, x_\mb{gh}^2)}{f(\beta, x_\mb{g}^2)f(\beta, x_\mb{h}^2)}\frac{f(x_\mb{g}, \beta x_\mb{g})f(x_\mb{h},\beta x_\mb{h})}{f(x_\mb{gh},\beta x_\mb{gh})} \frac{f(x_\mb{gh},x_\mb{gh})}{f(x_\mb{g},x_\mb{g})f(x_\mb{h},x_\mb{h})} \frac{f(x_\mb{gh}, \beta x_\mb{gh})}{f(x_\mb{g},\beta x_\mb{gh})f(x_\mb{h},\beta x_\mb{gh})} \frac{f(x_\mb{g},x_\mb{gh})f(x_\mb{h}, x_\mb{gh})}{f(x_\mb{gh}, x_\mb{gh})}
		\label{}
	\end{equation}
The first three factors are exactly in the form of coboundaries. We now write out the last two more explicitly:
\begin{equation}
	\begin{split}
	&\frac{f(x_\mb{gh}, \beta x_\mb{gh})}{f(x_\mb{g},\beta x_\mb{gh})f(x_\mb{h},\beta x_\mb{gh})} \frac{f(x_\mb{g},x_\mb{gh})f(x_\mb{h}, x_\mb{gh})}{f(x_\mb{gh}, x_\mb{gh})}\\
	&=\exp\Big[i 2\pi \sum_{j\leq k}\frac{ p_{jk}}{N_j N_k}([x^j_\mb{gh}]-[x^j_\mb{g}]-[x^j_\mb{h}])_{N_j}([\beta^k+x_\mb{gh}^k]-[x_\mb{gh}^k])_{N_k}\Big]	\\
	&=\exp\Big[i 2\pi \sum_{j\leq k}\frac{ p_{jk}}{N_j N_k}([x^j_\mb{gh}]-[x^j_\mb{g}]-[x^j_\mb{h}])_{N_j}([\beta^k+x_\mb{gh}^k]-[x_\mb{gh}^k]-[\beta^k])_{N_k}\Big] \\
    & \quad \times  \exp\Big[i 2\pi \sum_{j\leq k}\frac{ p_{jk}}{N_j N_k}([x^j_\mb{gh}]-[x^j_\mb{g}]-[x^j_\mb{h}])_{N_j}[\beta^k]_{N_k}\Big]\\
&= \exp\Big[i 2\pi \sum_{j\leq k}\frac{ p_{jk}}{N_j N_k}([x^j_\mb{gh}]-[x^j_\mb{g}]-[x^j_\mb{h}])_{N_j}[\beta^k]_{N_k}\Big],
	\end{split}
	\label{}
\end{equation}
which is also a coboundary.
From Eq.~\eqref{eqn:2coboundary} we see that when $x_\mb{g}=y_\mb{g}$, under the gauge transformation of $F$ and $R$ symbols $\tilde{o}(\mb{g,h})$ is only modified by a coboundary, so although we have chosen a particular gauge Eq.~\eqref{eqn:Flookslikethis} for our computation our conclusion holds generally.

In summary, we have shown that if either $G$ is Abelian, $G$ is continuous, or there is a rotation symmetry, then $[\tilde{o}] = [1]$ and the analysis of the main text holds.

\twocolumngrid

\bibliography{refs}

\end{document}